\documentclass[ALICE,manyauthors,nocleardouble]{cernphprep}
\usepackage{graphicx}  
\usepackage{dcolumn}   
\usepackage{bm}        
\usepackage{amssymb}   
\usepackage{amsfonts}
\usepackage{graphics}
\usepackage{epsfig}
\usepackage[normalem]{ulem}
\usepackage[usenames]{color}
\usepackage{multirow}
\usepackage{hyperref}
\usepackage{booktabs}
\usepackage{placeins}
\usepackage[version=3]{mhchem}

\newcommand{\dphi}{\ensuremath{\Delta\varphi}}
\newcommand{\pT}{\ensuremath{p_{\mathrm{T}}}}
\newcommand{\pTassoc}{\ensuremath{p_{\mathrm{T,\,assoc}}}}
\newcommand{\pTtrig}{\ensuremath{p_{\mathrm{T,\,trig}}}}
\newcommand{\sqrts}{\ensuremath{\sqrt{s}}}
\newcommand{\Nch}{\ensuremath{N_{\mathrm{charged}}}}
\newcommand{\NchLong}{\ensuremath{N_{\mathrm{charged},\,|\eta|\, <\, 0.9,\ p_{\mathrm{T}}\, >\, 0.2\,\mathrm{GeV}/c }}}
\newcommand{\Nrec}{\ensuremath{N_{\mathrm{rec,\, charged}}}}
\newcommand{\Nmpi}{\ensuremath{N_{\mathrm{MPI}}}}
\newcommand{\Nuncorr}{\ensuremath{N_{\mathrm{uncorrelated\ seeds}}}}
\newcommand{\Nnear}{\ensuremath{\langle N_{\mathrm{assoc,\, near-side}} \rangle}}
\newcommand{\Naway}{\ensuremath{\langle N_{\mathrm{assoc,\, away-side}} \rangle}}
\newcommand{\Nisotrop}{\ensuremath{\langle N_{\mathrm{isotrop}} \rangle}}
\newcommand{\Ntrigger}{\ensuremath{\langle N_{\mathrm{trigger}} \rangle}}
\newcommand{\NuncorrAv}{\ensuremath{\langle N_{\mathrm{uncorrelated\, seeds}} \rangle}}
\newcommand{\gmom}{\ensuremath{\mathrm{GeV}/c}}
\newcommand{\plmi}[1] {{$\pm #1$}}

\begin{document}
%
%
\begin{titlepage}
\PHnumber{2013-107}              
\PHdate{24 June 2013}                  

  \title{Multiplicity dependence of two-particle azimuthal correlations  \\in pp collisions at the LHC}
\ShortTitle{Multiplicity dependence of two-particle azimuthal correlations}   

\Collaboration{ALICE Collaboration \thanks{See appendix~\ref{app:collab} for the list of collaboration members}}
\ShortAuthor{ALICE Collaboration}      

\begin{abstract}
We present the measurements of particle pair yields per trigger particle obtained from di-hadron azimuthal correlations
in pp collisions at $\sqrt{s}=0.9$, 2.76, and 7\,TeV recorded with the ALICE detector.
The yields are studied as a function of the charged particle multiplicity.
Taken together with the single particle yields the pair yields provide information 
about parton fragmentation at low transverse momenta, 
as well as on the contribution of multiple parton interactions to particle production.
Data are compared to calculations using the PYTHIA6, PYTHIA8, and PHOJET event generators.

\end{abstract}
\end{titlepage}
\setcounter{page}{2}

\section{Introduction}
The multiplicity distribution of particles produced in proton-proton (pp) collisions 
and the multiplicity dependence of other global event characteristics represent fundamental 
observables reflecting the properties of the underlying particle production mechanisms. In the Feynman
picture, the strongly interacting hadrons can be seen as bunches of point-like partons producing particles in
interactions with small (soft) and large (hard) momentum transfer. As expected from
Feynman scaling \cite{Feynman:1969}, 
at low centre-of-mass energies ($\sqrt{s}$), where particle production is dominated by soft interactions,
the mean number of particles $\langle M \rangle$ was found to rise logarithmically with $\sqrt{s}$. Moreover,
the evolution of the charged particle multiplicity distribution $P(M)$ as a function of $\sqrt{s}$ follows
the Koba-Nielsen-Oleson (KNO) scaling \cite{Koba:1972ng} 
with scaling variable $z = M/\langle M \rangle$ and $P(M) \langle M
\rangle = \psi(z)$, where $\psi(z)$ is an energy independent function. Experimentally one finds that KNO
scaling is violated for $\sqrt{s} > 200 \, {\rm GeV}$ \cite{Alner:1985wj}. This scaling violation which increases with 
$\sqrt{s}$ has been interpreted as a consequence of particle production through multiple parton-parton interactions (MPI) \cite{Dremin, GrosseOetringhaus:2009kz}. 
Further, at the LHC, already at a transverse momentum transfer of a few $\gmom$ the cross section for leading order (LO) parton-parton 
scatterings exceeds the total pp inelastic cross section. This apparent inconsistency can be resolved by
aggregating several quasi independent scatterings in the same pp collision \cite{Sjostrand:1987su, Bahr:2008wk}. 
If multiple semi-hard scatterings play a dominant role in the production of high multiplicity events, this should lead to distinct
experimentally observable effects. 
The search for these is the aim of the present analysis of pp collisions recorded with the ALICE detector at the LHC.

Each parton-parton scattering produces partons almost back-to-back in azimuth, $\varphi$. 
They fragment producing two correlated bundles of particles.
With increasing multiplicity we expect that both the 
number of sources of correlated particles and the number of correlated particles per source increase. Thus, we have 
designed our analysis methods in a way that the two effects can be separated as much as possible. 
Since many of the bundles of particles
(low transverse-momentum jets) overlap in the same event, they can not be identified and separated event-by-event. 
An alternative method, pursued in this analysis, is to study two-particle angular correlations as a function of the event multiplicity \cite{Wang:1993}.

Such studies involve measuring the distributions of the relative angle $\Delta \varphi$
between particle pairs consisting of a ``trigger'' particle in a certain transverse momentum
$\pTtrig$ interval and an ``associated'' particle in a $\pTassoc$ interval, where $\Delta \varphi
$ is the difference in azimuth $\varphi$ between the two particles.
The $p_{\rm T}$ ranges chosen for the analysis ($\pTtrig > 0.7 \, \gmom$ and $\pTassoc > 0.4 (0.7) \, \gmom$) are a 
compromise between being high enough to decrease the sensitivity to low energy phenomena such as the
breaking of individual strings ($p_{\rm T}^{\mathrm{min}} \gg \Lambda_{\mathrm{QCD}}$) and sufficiently low such
that the correlations are sensitive to the bulk of the particle production. These cuts have been also used by the CDF collaboration
to define so-called track-clusters: a track with $\pT > 0.7 \, \gmom$ with at least one other track with $\pT > 0.4 \, \gmom$
in a cone of radius $\sqrt{\Delta \varphi^2 + \Delta \eta^2} <  0.7$, where $\Delta \eta $ is the pseudo-rapidity difference \cite{cdf:2002}.
In the CDF analysis, the presence of a track-cluster has been used  for an event-by-event identification of hard events. 
In the present correlation analysis the $\Delta \varphi$ distributions are averaged over all events of a given sample.
This has the advantage that random correlations, which become dominant at high multiplicities, can be subtracted.
The mean number of trigger particles per event and the correlated pair-yield per trigger are measured and combined in a way that they can provide information about the number of semi-hard scatterings 
in the event of a given charged particle multiplicity as well as the fragmentation properties of low-$p_{\rm T}$ partons
biased by the multiplicity selection.

Although the full final state of pp collisions cannot be calculated  in perturbative QCD, pQCD-inspired models
based on multiple parton interactions provide a consistent way to describe high multiplicity pp
collisions, and have been implemented in recent Monte Carlo (MC) generators like PYTHIA6 \cite{Sjostrand:1987su, Sjostrand:2006za}, PYTHIA8 \cite{Sjostrand:2007gs}, PHOJET \cite{Engel:1995sb} and HERWIG \cite{Bahr:2008pv}.
Using the QCD factorisation theorem \cite{factorisation}
cross sections are calculated from a convolution of the short-distance  parton-parton cross section and the long-distance parton distribution function ({\em pdf}) of the proton. Approaching zero momentum transfer, the leading order short distance cross sections
diverge and the models
have to implement regularisation mechanisms to control 
this divergence. Moreover, parton distribution functions are only
known for single parton scatterings and, hence, extensions for multiple interactions are needed. 
Furthermore, each partonic interaction produces coloured strings between the final state partons 
which overlap in the case of many interactions. It is possible that in this case partons do not hadronise independently and phenomenological
models have been developed to account for so-called colour connections and reconnections.
Measurements that can provide information on multiple parton interactions and fragmentation 
properties are important to constrain such models. 
Consequently, we compare our results for pp collisions at  $\sqrt{s} =$ 0.9, 2.76 and 7~TeV  
among each other and to the outcome of Monte Carlo simulations on generator level with different PYTHIA6 tunes (Perugia-0 and Perugia-2011 \cite{Skands:2010ak}), PYTHIA8 and PHOJET. 
We have chosen this set of generators and tunes since they have been already compared to previous ALICE measurements based on azimuthal correlations: 
the underlying event \cite{ALICE:2011ac} and  transverse sphericity \cite{Abelev:2012sk}; 
Ref.\ \cite{ALICE:2011ac} contains a short description of them. 
   
The paper is organised in the following way: the ALICE sub-systems used in the analysis are described in
section~\ref{sec:alice} and the data samples, event and track selection in section~\ref{sec:eventtrack}. 
Section \ref{sec:method} introduces the analysis strategy. In sections~\ref{sec:correction} and \ref{sec:systematics} we focus on the data correction
procedure and systematic uncertainties, respectively. Final results are presented in section~\ref{sec:results} and in section~\ref{sec:conclusion} we draw conclusions.
\section{Experimental setup}\label{sec:alice}
The pp collision data used for this analysis were recorded by the ALICE detector at the LHC.
The detector is described in detail in Ref.~\cite{Aamodt:2008zz}.
In the following, only the sub-detectors used in this analysis are described in detail. 
These are the VZERO detector, the Inner Tracking System~(ITS) including the Silicon Pixel Detector~(SPD), 
the Silicon Drift Detector~(SDD), and the Silicon Strip Detector~(SSD), as well as the Time Projection Chamber~(TPC). 
The VZERO detector and the SPD are used to trigger on minimum bias events.
The track reconstruction of charged particles is performed with the combined information from the ITS and the TPC. 

The VZERO scintillator hodoscope is divided into two arrays of counters, VZERO-A and VZERO-C
located at 3.4\,m and -0.9\,m from the nominal interaction point along the beam axis, respectively. 
VZERO-A covers the pseudorapidity range of $2.8 < \eta < 5.1$ and VZERO-C $-3.7 < \eta < -1.7$. 

The Inner Tracking System (ITS) comprises 6 cylindrical layers of silicon detectors of three different detector types, each contributing with two layers. 
The Silicon Pixel Detector constitutes the first two layers of the ITS. 
The sensitive part of the detector is made of high granularity 250\, $\mu$m-thick
hybrid silicon pixels consisting of a 2-dimensional matrix 
of reversed-biased silicon detector diodes with $10^7$ read-out channels. 
The pseudorapidity coverage is $\arrowvert \eta \arrowvert<1.98$ for the first layer and $\arrowvert \eta \arrowvert<1.4$ for the second layer. 
The SPD contributes to the minimum bias trigger as well as to the reconstruction of tracks left by 
charged particles, and the vertex reconstruction.
The Silicon Drift Detector comprises the two intermediate layers of the ITS. 
The sensitive part consists of homogeneous high-resistivity 300\,$\mu$m-thick n-type silicon wafers with 133000 read-out channels.
The SDD contributes to the reconstruction of tracks of charged particles as well as to the particle identification using energy loss information.
The Silicon Strip Detector composes the two outermost layers of the ITS. 
The double-sided SSD has 2.6~million read-out channels and contributes like the SDD to the track reconstruction and the particle identification. 
Furthermore, it is optimised for track matching between the ITS and the Time Projection Chamber. 
The total material budget of the ITS traversed by straight tracks  perpendicular to the detector surface amounts to 
$7.2\% \, X_0$.

The main tracking detector of the ALICE central barrel is the Time Projection Chamber.
It is a cylindrical detector filled with 90\,$\mathrm{m}^3$ of gaseous $\ce{Ne}/\ce{CO}_2/\ce{N}_2$ at a mixing ratio of (85.7/9.5/4.8). 
High-voltage is applied to the central membrane, resulting in an electric field between the central electrode and the end caps, which are each equipped with multi-wire proportional chambers. 
The TPC provides full azimuthal acceptance for particles produced in the pseudo-rapidity interval $\arrowvert \eta \arrowvert<0.9$. 
It is used to perform charged-particle momentum measurements with a good two-track separation 
adequate to cope with the extreme particle densities present in central heavy-ion collisions.
Hence, in pp collisions, two-particle reconstruction effects like track merging and track splitting 
are small and manageable.
The ITS and TPC cover the full azimuth and a combined pseudo-rapidity interval $\arrowvert \eta \arrowvert < 0.9$. 
All detectors are operated inside the L3 magnet which generates a homogeneous magnetic field 
of $B=0.5$\,T in the detector region.

\section{Event and track selection}\label{sec:eventtrack}
The present analysis uses 
pp collisions collected with ALICE minimum bias triggers
at the collision energies $\sqrts~=~0.9$, 2.76, and 7\,TeV. 
In May 2010, 7 million events were collected at $\sqrts=0.9$\,TeV, 
in March 2011, 27 million events were collected at 2.76\,TeV, 
and from April to August 2010, 204~million events were collected at 7\,TeV.
The probability for pile-up events is negligible for the  $\sqrts=0.9$\,TeV data taking period but sizeable for the $\sqrts=2.76$ and 7\,TeV data taking periods. 
The impact of pile-up events on the final analysis results has been tested and quantified using a high pile-up data set 
as well as by performing a comparison of results obtained with sub-sets of the nominal data sets at relatively high and relatively low pile-up probability.

ALICE data are compared to model predictions of PYTHIA6.4~\cite{Sjostrand:1987su, Sjostrand:2006za} (tune Perugia-0~\cite{Skands:2010ak} and tune Perugia-2011~\cite{Skands:2010ak}), PYTHIA8.1~\cite{Sjostrand:2007gs} (tune 4C~\cite{Corke:2010yf}), and PHOJET~\cite{Engel:1995sb} (version~1.12). 
The detector response in full detector simulations has been modeled using GEANT3~\cite{Brun:1994zzo} as well as GEANT4~\cite{Agostinelli:2002hh,Allison:2006ve}. 

\subsection{Trigger and offline event selection}\label{sec:event}
Minimum bias events were selected using the following trigger requirements: 
at least one charged particle needs to be detected in either the SPD or in one of the two VZERO detectors in coincidence with signals 
from the two BPTX beam pick-up counters indicating the presence of two intersecting proton bunches~\cite{Aamodt:2009aa}.
In addition to the online trigger selection, the trigger decision is reprocessed offline using the same selection criteria;
however, the reconstructed information are used instead of the online signals.\\
Only events having exactly one good quality reconstructed primary collision vertex are used in the analysis.
Collision vertices are reconstructed using either reconstructed tracks or so-called tracklets~\cite{Aamodt:2009aa}
based on correlated hits measured in the two SPD layers.
A vertex passes the quality selection if it is located within $\arrowvert z_{\mathrm{vertex}} \arrowvert <10$\,cm with respect to the nominal interaction point 
in beam direction and if at least one track contributes to the reconstruction of the vertex. 
Pile-up events with more than one reconstructed collision vertex are rejected from the analysis. 
Furthermore, we require at least one reconstructed high-quality track (see section~\ref{sec:track}) in the combined ITS-TPC acceptance of $\pT>0.2$\,GeV/$c$ and $|\eta|<0.9$.
The discussed event selection cuts efficiently suppress events from beam-gas and beam-halo interactions as well as from cosmic rays. 
Table \ref{tab:EventCutFlow} shows the number of recorded minimum bias events that pass the event selection cuts. 
The vertex-cut efficiency is dominated by the vertex quality requirements. The single vertex requirement after vertex quality cuts removes up to 0.5\,\% additional events.

\begin{table}[t]
\centering
\begin{tabular}{lrr}
\toprule
         & Events (million) & Fraction of all (\%) \\
\midrule 
\multicolumn{3}{l}{pp @ $\sqrts=0.9$\,TeV}\\
\midrule 
Triggered            & 6.96 & 100.0  \\ 
Vertex cuts & 4.91 &  70.6 \\
Track in acceptance  & 4.64 &  66.7 \\
\midrule
\multicolumn{3}{l}{pp @ $\sqrts=2.76$\,TeV}\\
\midrule 
Triggered            & 26.65 & 100.0  \\ 
Vertex cuts & 19.42 & 72.9  \\
Track in acceptance  & 18.49 & 69.4 \\
\midrule
\multicolumn{3}{l}{pp @ $\sqrts=7$\,TeV}\\
\midrule 
Triggered            & 203.96 & 100.0  \\ 
Vertex cuts & 157.89 & 77.4  \\
Track in acceptance  & 152.02 & 74.5 \\
\bottomrule
\end{tabular}
\caption{Number of pp minimum bias events after event selection for the data sets at $\sqrts=0.9$, 2.76, and 7\,TeV. The track selection used in the last event selection step is described in section \ref{sec:track}.} 
\label{tab:EventCutFlow}
\end{table}

\newpage
\subsection{Track cuts}\label{sec:track}
In the analysis, we consider only charged primary particles which are defined as prompt particles produced in the collision and their decay products except products of weak decays of strange particles.
The data analysis is performed using track selection cuts optimised for a uniform azimuth ($\varphi$) 
acceptance and 
for a minimal contamination of tracks by particles originating 
from secondary vertices (secondary particles) \cite{ALICE:2011ac}. 
The track selection comprises the following cuts: 
tracks are required to have at least three associated hits in the ITS, 
one of which has to be located in the first three ITS layers. 
Furthermore, each track needs to have at least 70 associated TPC clusters measured in the 159 TPC pad rows.
The quality of the track parameter fitting is measured by the $\chi^2$ per TPC cluster and 
tracks passing our selection have $\chi^2$ per cluster $<4$.
No tracks with a kink topology indicating a particle decay are accepted.
A \pT-dependent $\mathrm{DCA}_{xy}$-cut corresponding to 7 times the $\sigma $ of the expected primary track distribution  
($\mathrm{DCA}_{xy,\ \mathrm{max}} \approx 0.2$\,cm)
assures that the tracks passing the selection criteria are
predominantly those from the primary vertex. 
In addition, a cut on the distance of closest approach in the $z$-direction of maximal $\mathrm{DCA}_z=2$\,cm improves the selection 
of primary particles and rejects particles from secondary vertices originating from, for example, the decay
of long-lived particles or hadronic interaction in the detector material. 
Moreover, this cut removes tracks originating from displaced pile-up vertices. 
Out of the selected high quality tracks, the data analysis accepts tracks within
the ITS-TPC acceptance $|\eta|<0.9$ and with $\pT>0.2$\,GeV/$c$. 
The track selection cuts are summarised in table~\ref{tab:TrackCuts}.

\begin{table}[t]
\centering
\begin{tabular}{lr}
\toprule
Criterion & Value \\
\midrule
Minimum number of ITS hits & 3 \\
Minimum number of ITS hits in first 3 layers & 1 \\
Minimum number of TPC clusters & 70 \\
Maximum $\chi^2$ per TPC cluster & 4\\
Maximum $\mathrm{DCA}_{xy}(\pT)$ & $7\sigma$ ($\mathrm{DCA}_{xy,\ \mathrm{max}} \approx 0.2$\,cm) \\
Maximum $\mathrm{DCA}_{z}$ & 2\,cm\\
\bottomrule
\end{tabular}
\caption{Track selection criteria.} 
\label{tab:TrackCuts}
\end{table}

\section{Analysis method}\label{sec:method}
\subsection{Definitions}\label{subsec:defintions}
We are analysing the sample-averaged probability distribution 
of the azimuthal difference $\dphi = \varphi_{\mathrm{trig}}-\varphi_{\mathrm{assoc}}$ 
between trigger particles ($\pTtrig>\pTtrig^{\mathrm{min}}$,  $\arrowvert \eta \arrowvert<0.9$) 
and associated particles  ($\pTassoc>\pTassoc^{\mathrm{min}}$, $\arrowvert \eta \arrowvert<0.9$). 
The {\em pair-yield per trigger} as a function of $\dphi $ is defined as 
\begin{equation}
\frac{\mathrm{d}N}{\mathrm{d}\dphi} = \frac{1}{N_{\mathrm{trig}}} \frac{\mathrm{d}N_{\mathrm{assoc}}}{\mathrm{d}\dphi},
\end{equation}
where $N_{\mathrm{trig}}$ is the number of trigger particles and $N_{\mathrm{assoc}}$ is the number of associated particles. 
We study the pair-yield per trigger as a function of the charged particle multiplicity $\NchLong$, 
as well as for different transverse momentum thresholds $\pTtrig^{\mathrm{min}}$ and $\pTassoc^{\mathrm{min}}$. \\

\begin{figure}[t]
\centering
\includegraphics[width=0.49\textwidth]{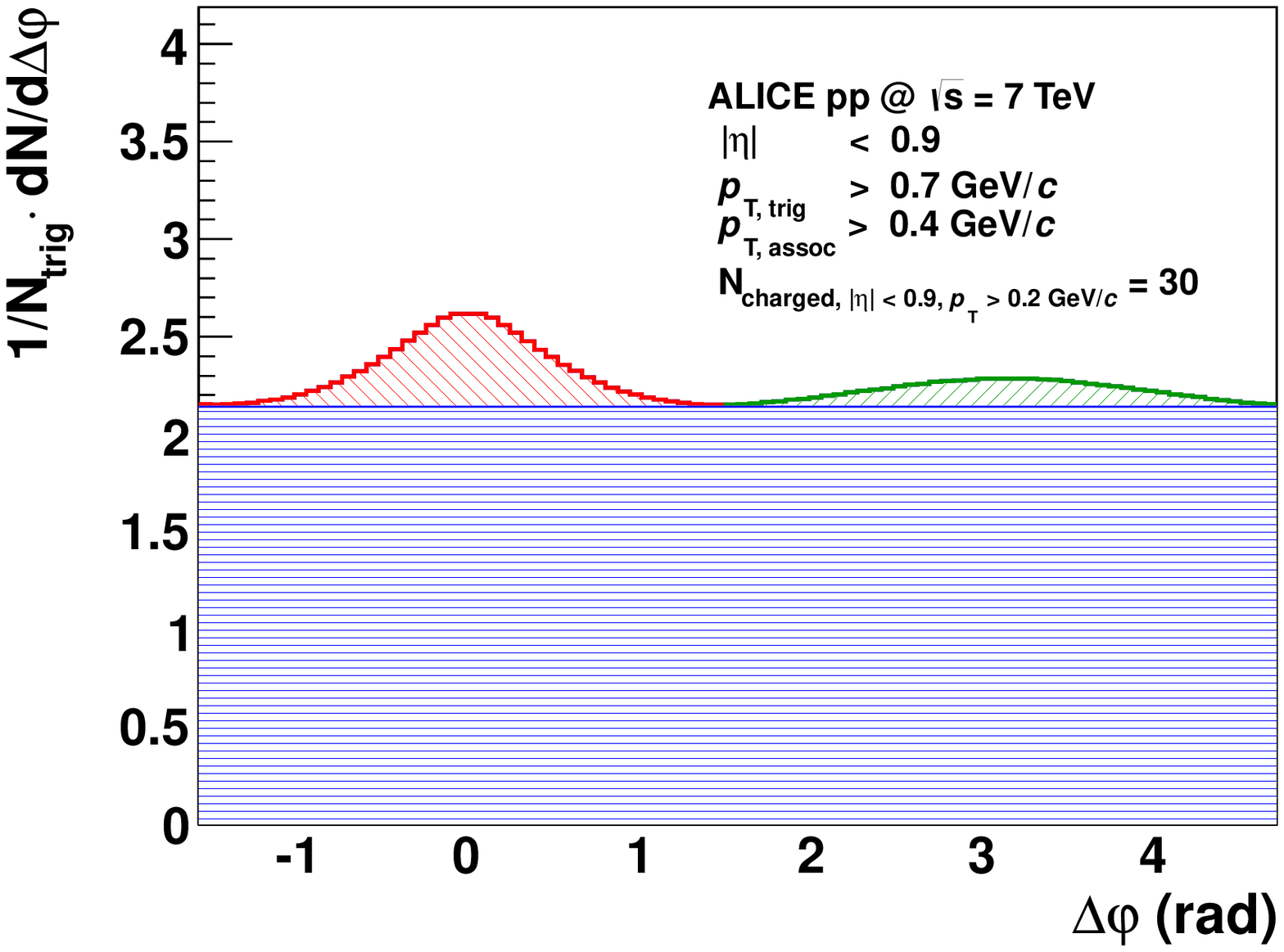}
\includegraphics[width=0.49\textwidth]{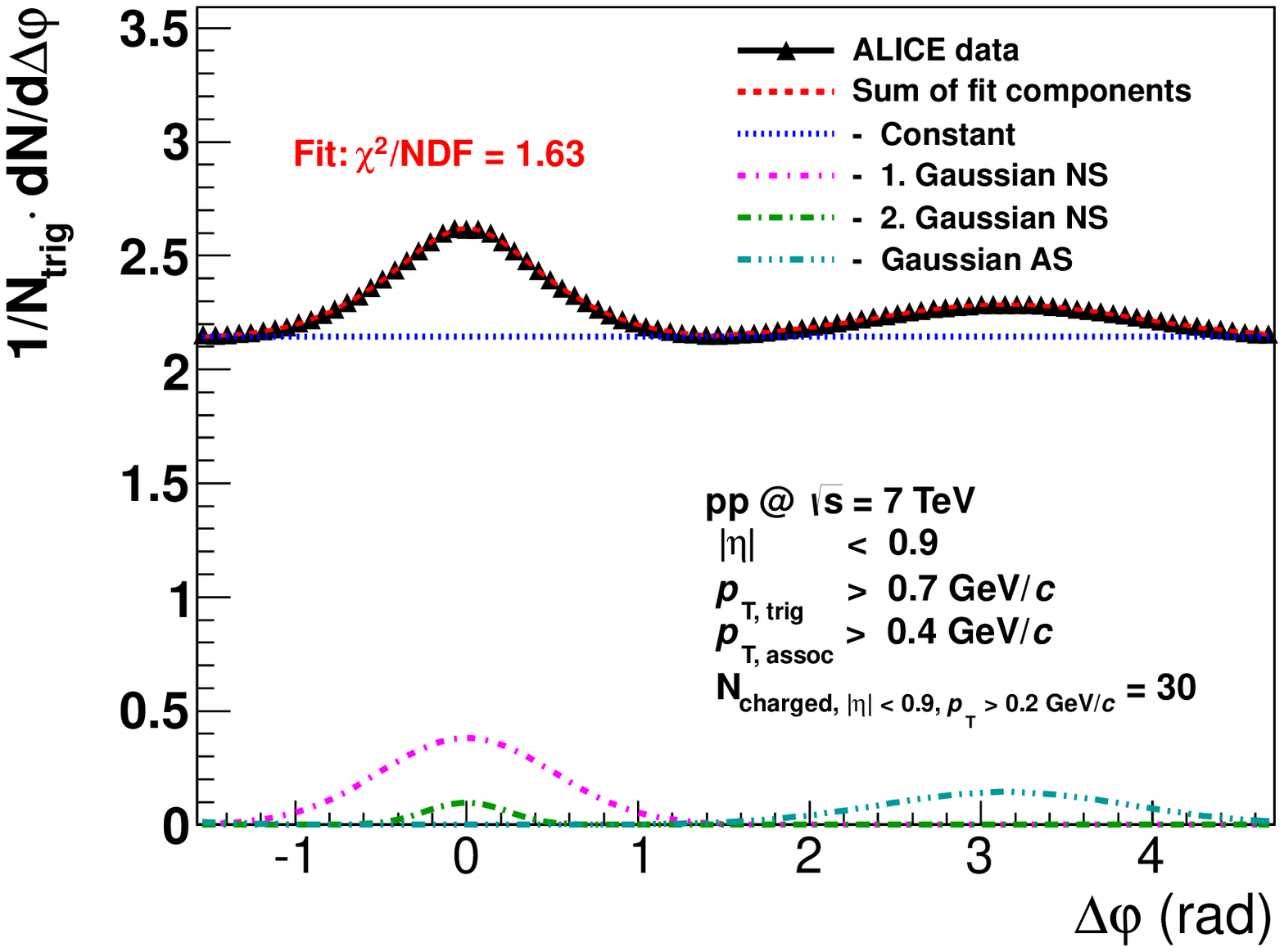}
\caption{Left panel: illustration of the contributions to the per-trigger pair yield as a function of $\dphi$.
Right panel: the per-trigger pair yield as a function of $\dphi$ described by the fit function and its sub-components
(see text).}
\label{fig:dphiExtract}
\end{figure}

The left panel of figure~\ref{fig:dphiExtract} shows an example of the measured
per-trigger pair yield as a function of $\dphi$ for $\pTtrig > 0.7 \, \gmom$
and $\pTassoc > 0.4 \, \gmom$ and \\$\NchLong = 30$. 
The two structures at the near-side ($\dphi \approx0$) and away-side ($\dphi \approx\pi$) of the trigger particle are dominantly induced by the fragmentation of back-to-back parton pairs.
In order to extract the per-trigger pair-yields for all multiplicity and $\pT$-cut classes, a fit function is introduced which allows us to decompose the azimuthal correlation into its main components. 
Whereas the away-side peak can be fitted using a single Gaussian, the near-side peak shows an enhanced tail-region and 
needs the superposition of two Gaussians with different widths. 
Including a constant $C$ to describe the combinatorial background, we obtained the fitting function

\begin{align} \label{eq:fit}
    f(\dphi) = C 
           + A_1 \exp\left(- \frac{\dphi^2}{2 \cdot \sigma_1^2} \right) 
           + A_2 \exp\left(- \frac{\dphi^2}{2 \cdot \sigma_2^2} \right) 
	  + A_3 \exp\left(- \frac{(\dphi-\pi)^2}{2 \cdot \sigma_3^2} \right). 
\end{align}
\vspace{0.2cm}
To increase the stability of the fit, the first near-side Gaussian and the away-side Gaussian are restricted to $-\pi/2 < \dphi < \pi/2 $ and $\pi/2 < \dphi < 3\pi/2 $, respectively. The second near-side Gaussian is fitted in the region $-\pi/5 < \dphi < \pi/5 $.

The right panel of figure~\ref{fig:dphiExtract} shows the measured azimuthal correlation, the parametrisation of the correlation based on the fit function, and the sub-components of the fit function. The $\chi^2$ per degree of freedom for this fit is 1.63.

\paragraph{Pair yield for asymmetric and symmetric $\pT$-bins} 
In the case of
non-overlapping $\pT$ intervals for the trigger and associated particles, the pair yield per
trigger measures the conditional yield 
of associated particles under the trigger condition.
Beside non-overlapping $\pT$ intervals (asymmetric bins), we are using symmetric bins for 
which the two intervals are identical. In this case, for $n$ trigger particles, $n(n-1)/2$ unique pairs with $\pTtrig > \pTassoc$
can be formed and, hence, the pair yield per trigger particle measures: 

\begin{equation}\label{eq:pair}
\frac{\langle N_{\mathrm{pair}} \rangle} {\langle N_{\mathrm{trig}} \rangle}   =\frac{1}{\langle n \rangle} \cdot \frac{\langle n(n-1)\rangle}{2} = \frac{1}{2} \left(\frac{\langle n^2\rangle}{\langle n\rangle} -1 \right).
\end{equation}

In general, without the knowledge of the second moment $\langle n^2 \rangle$ of the number distribution function $P_n$, the
mean number of correlated particles $\langle n \rangle $ cannot be determined. However, for small $\langle n \rangle$ and 
monotonically falling $P_n$, the expression has a well defined limit: 
\begin{equation}\label{eq:relation}
 \frac{1}{2} \left(\frac{\langle n^2\rangle}{\langle n\rangle} -1 \right) \approx \frac{\langle n \rangle}{1-P_0} -1.
\end{equation}
Since ${\langle n \rangle}/(1-P_0)$ is the mean value of the distribution $P_n$ under the condition 
that at least one particle has been produced, 
the right-hand side represents the number of particles associated with a trigger particle.
Note that for jet-like  self-similar particle emission (geometric series) the approximation is exact.

\paragraph{Pair yield extraction} 
Based on the fit function of equation \ref{eq:fit}, five observables can be derived. 
Three of the observables are directly related to the decomposed pair yield per trigger:
\begin{itemize}
\item{Per-trigger pair yield in the combinatorial background}
\begin{eqnarray} \label{eq:iso}
\Nisotrop &=& \frac{1}{N_{\mathrm{trigger}}} \cdot C,
\end{eqnarray}
\item{Per-trigger pair yield in the near-side peak}
\begin{eqnarray}\label{eq:near}
\Nnear &=& \frac{\sqrt{2\pi}}{N_{\mathrm{trigger}}} (A_{1}\cdot \sigma_1 +A_{2}\cdot \sigma_2), 
\end{eqnarray}
\item{Per-trigger pair yield in the away-side peak}
\begin{eqnarray}\label{eq:away}
\Naway &=& \frac{\sqrt{2\pi}}{N_{\mathrm{trigger}}} (A_{3}\cdot \sigma_3 ).
\end{eqnarray}
\end{itemize}
The yields in the near-side and away-side peaks measure fragmentation properties of low-$\pT$ partons.
In addition, the average number of trigger particles $\Ntrigger$ is determined:
\begin{eqnarray}\label{eq:trigger}
\Ntrigger &=& \frac{N_{\mathrm{trigger}}}{N_{\mathrm{events}}}.
\end{eqnarray}
The average number of trigger particles depends on the number of semi-hard scatterings per event and the fragmentation
properties of partons. With the aim to reduce the fragmentation dependence and to increase the sensitivity
to the number of scatterings per event we define for symmetric $\pT$-bins a new observable, {\em average number of uncorrelated seeds},
by combining the average number of trigger particles with the near-side and away-side yield of trigger particles  
($\pT>p_{\mathrm{T,\; trig}}$).
\begin{eqnarray}\label{eq:uncorr}
\NuncorrAv = \frac{\langle N_{\mathrm{trigger}} \rangle}{\langle 1 + N_{\mathrm{assoc, \; near+away}, \;p_{\mathrm{T}}>p_{\mathrm{T,\; trig}}}\rangle}.
\end{eqnarray}
where 
\begin{eqnarray}
\langle N_{\mathrm{assoc, \; near+away}, \;p_{\mathrm{T}}>p_{\mathrm{T,\; trig}}} \rangle = \Nnear + \Naway
\end{eqnarray} 
and also the associated particles have $\pT>p_{\mathrm{T,\; trig}}$.

Model studies (see section \ref{sec:methodMPI}) show that
the ratio effectively corrects for the multiplicative effect of fragmentation 
so that the obtained 
quantity provides information about the number of uncorrelated sources of particle production. 

\subsection{Relation between experimental observables and the PYTHIA MPI model}
\label{sec:methodMPI}

\begin{figure}[t]
\centering
\begin{minipage}{0.49\textwidth}
\includegraphics[width=\textwidth]{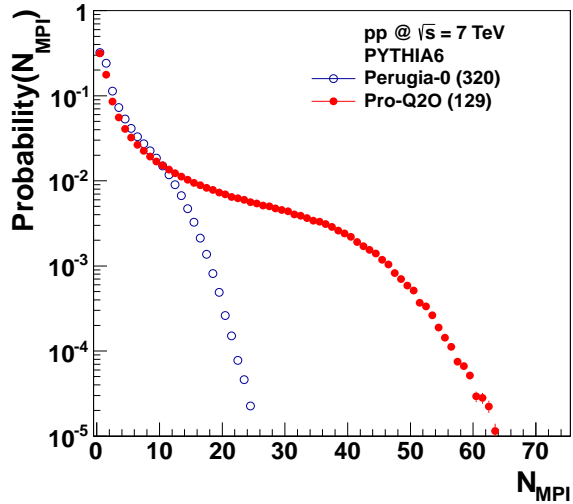} 
\end{minipage}
\vspace{0.2cm}
\caption{Number of multiple parton interactions \Nmpi\ in PYTHIA6 tune  Perugia-0 and tune Pro-Q2O.}
\label{fig:mpi}
\end{figure}

The PYTHIA MC for pp collisions includes a model for multiple parton interactions. 
Within the PYTHIA model, the dependence between the number of uncorrelated seeds $\NuncorrAv$ and the average number of multiple parton interactions $\langle \Nmpi \rangle$ can be studied. 
Here, the number of multiple parton interactions $\Nmpi$ is defined 
as the number of hard or semi-hard scatterings that occurred in a single pp collision \cite{Sjostrand:2006za}.
The number of multiple parton interactions $\Nmpi$ is  shown in figure \ref{fig:mpi} for the PYTHIA6 tunes Perugia-0 and Pro-Q2O~\cite{Skands:2010ak}.
Both MC tunes predate LHC data and give a good description of Tevatron (${\rm p\bar p}$ at $\sqrts =$~2~TeV)
results. However, they have very different probability distributions for $\Nmpi$. 
Whereas Pro-Q2O features a wide plateau, that of Perugia-0 is much narrower. 

\begin{figure}[t]
\centering
\begin{minipage}{0.49\textwidth}
\includegraphics[width=\textwidth]{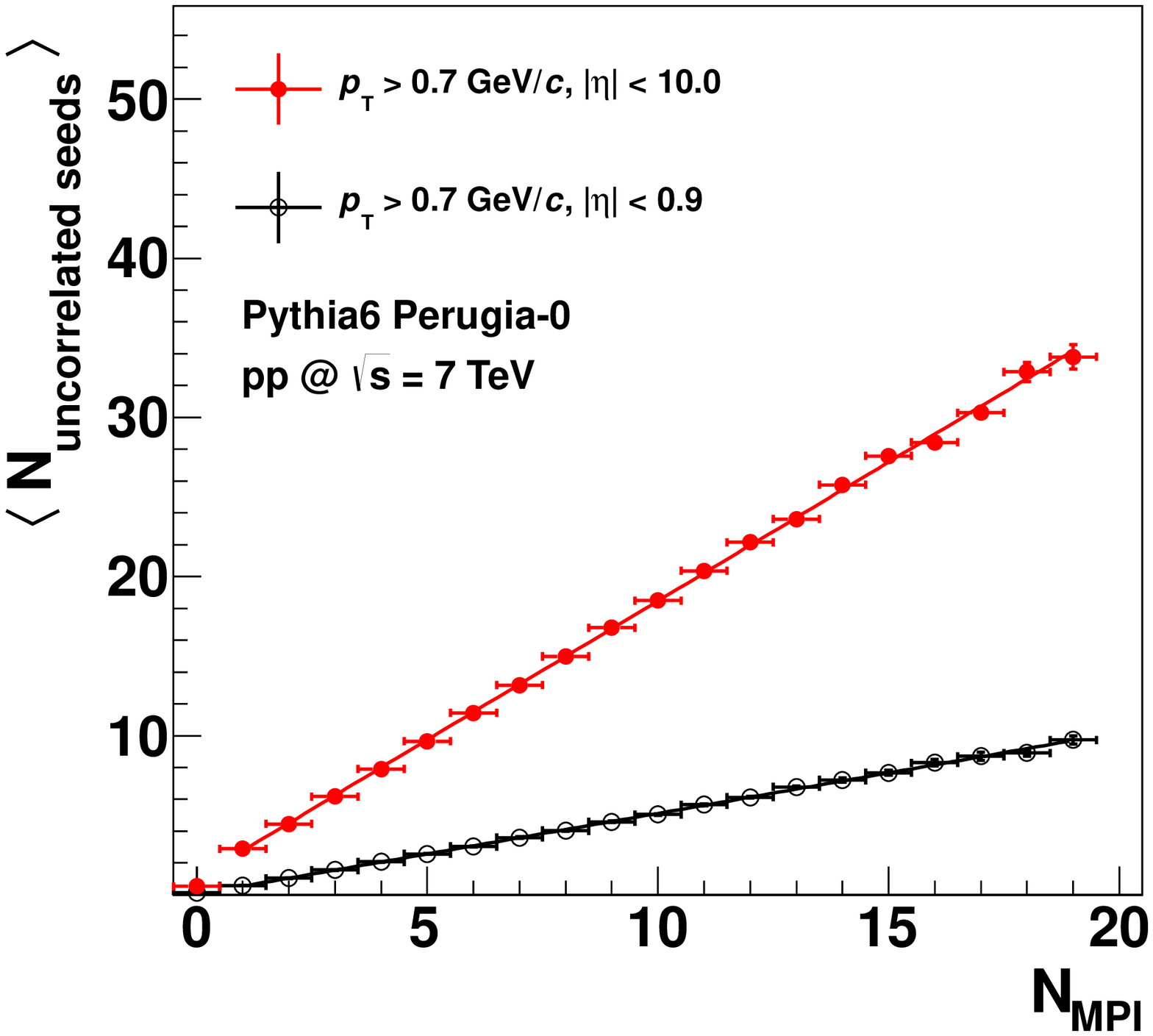} 
\end{minipage}
\begin{minipage}{0.49\textwidth}
\includegraphics[width=\textwidth]{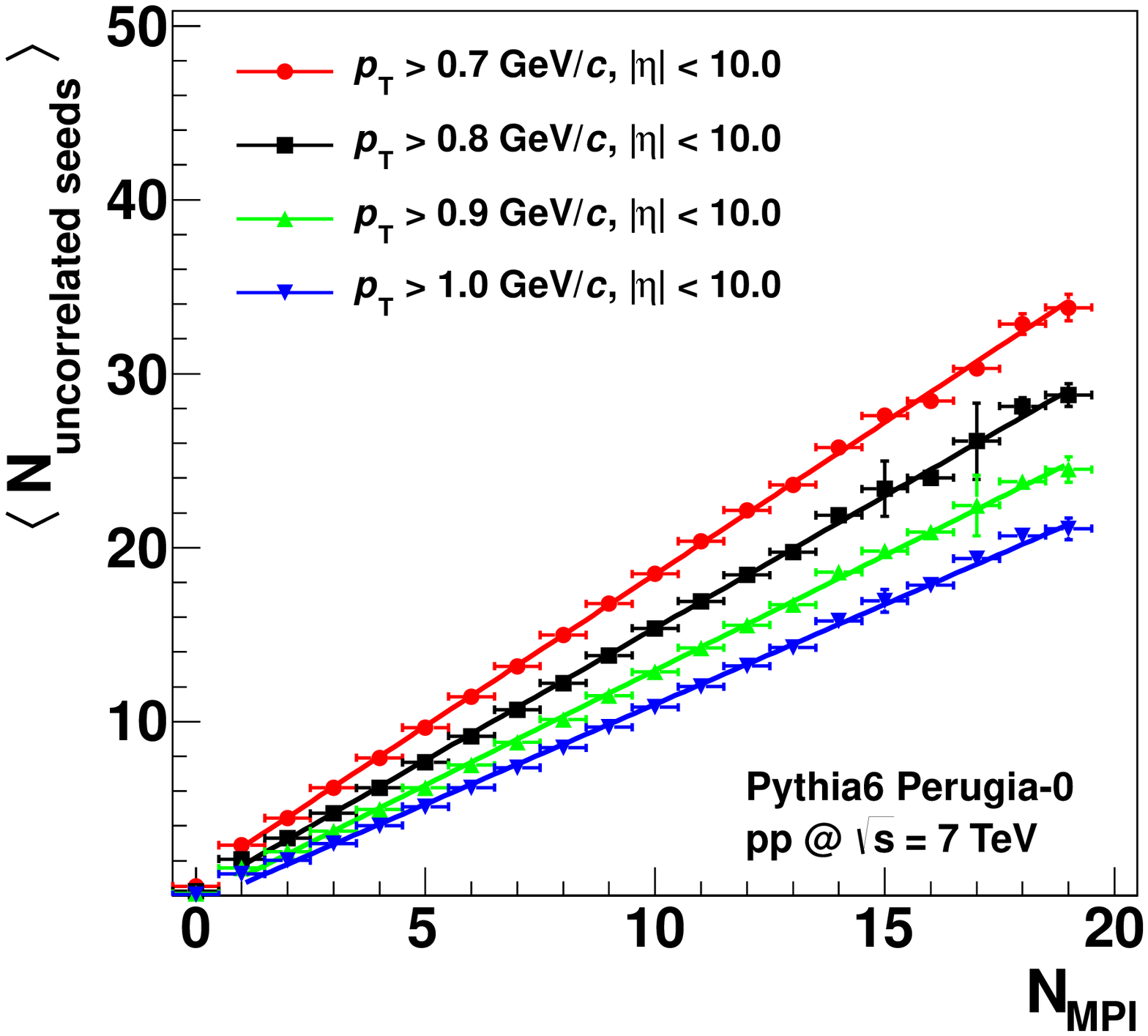} 
\end{minipage}
\caption{Linear dependence between $\Nuncorr$ and $\Nmpi$ in PYTHIA6 Perugia-0 simulations on generator level. Left panel: $\Nuncorr$ versus $\Nmpi$ for different $|\eta|$-ranges. Right panel: $\Nuncorr$ versus $\Nmpi$ for different \pTtrig-thresholds.} 
\label{fig:NuncorrSeedsNmpi}
\end{figure}

\begin{figure}[t]
\centering
\begin{minipage}{0.49\textwidth}
\includegraphics[width=\textwidth]{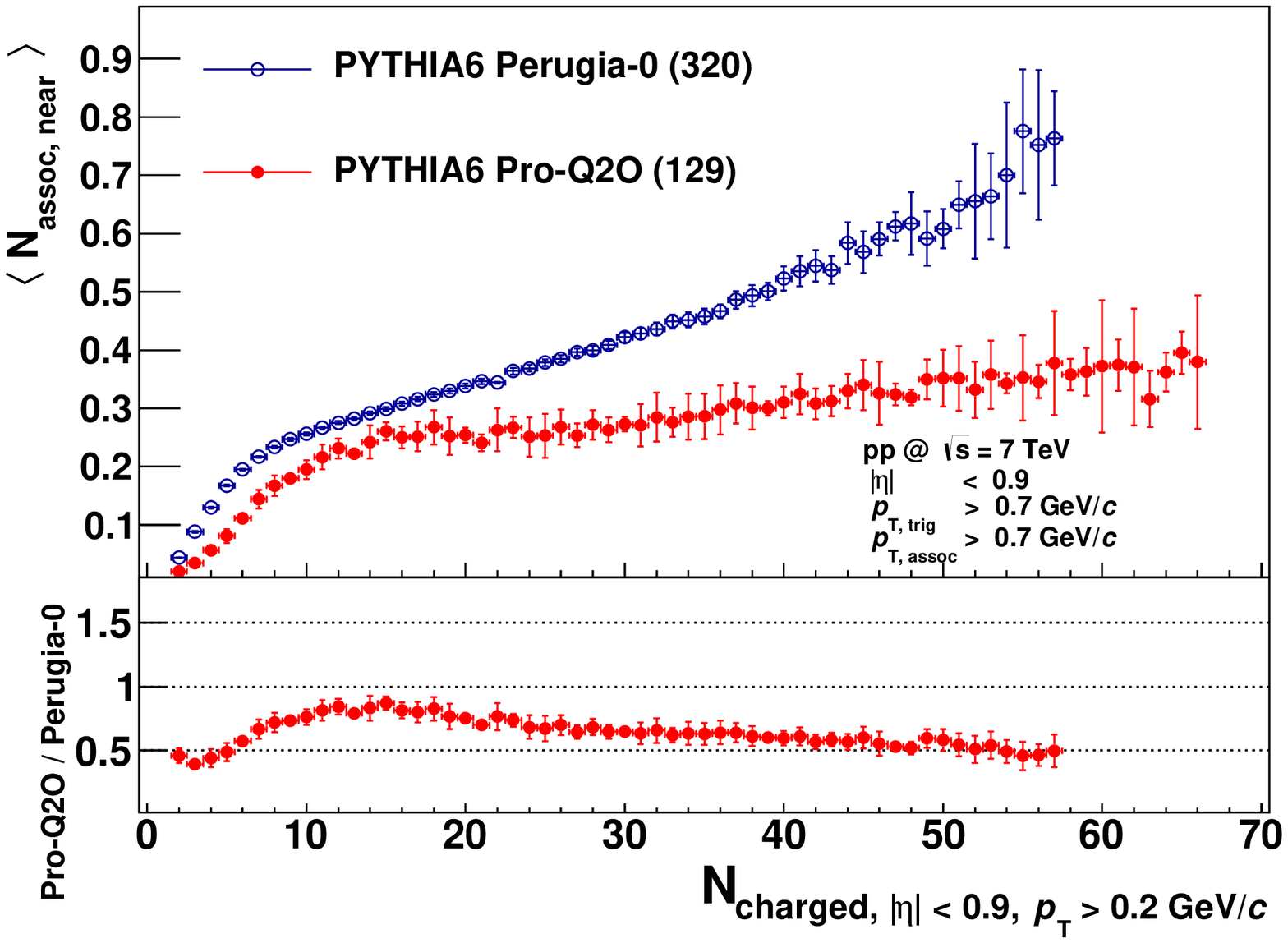} 
\end{minipage}
\begin{minipage}{0.49\textwidth}
\includegraphics[width=\textwidth]{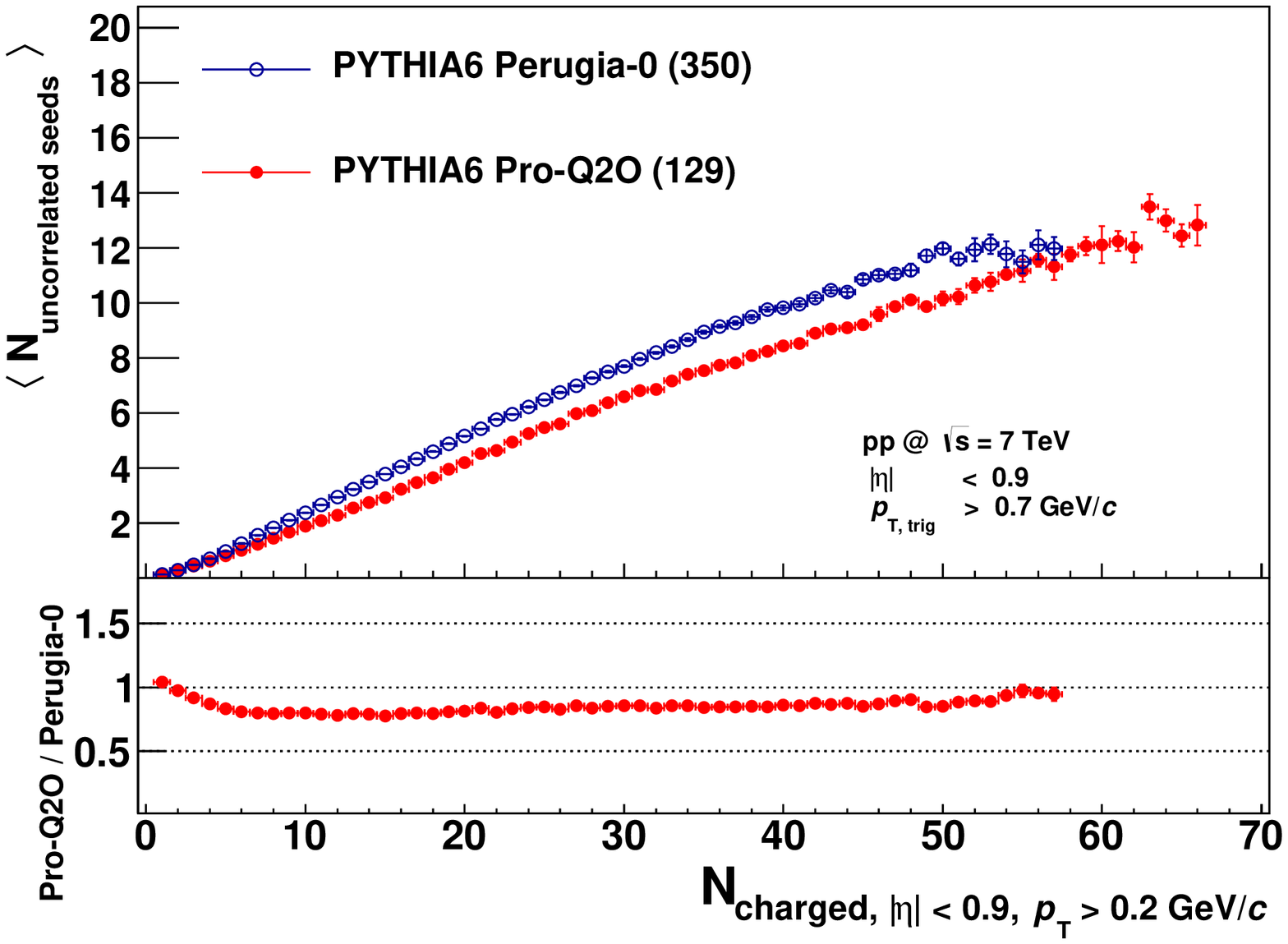} 
\end{minipage}
\caption{Comparison of PYTHIA6 tunes Perugia-0 and Pro-Q2O for near-side pair-yield per trigger particle (left panel)
and number of uncorrelated seeds (right panel).} 
\label{fig:p0vsproq20}
\end{figure}

The dependence between the number of uncorrelated seeds 
and the number of multiple parton interactions in PYTHIA6 tune Perugia-0 simulations on generator level is shown in figure \ref{fig:NuncorrSeedsNmpi} for different $|\eta|$-ranges and $\pTtrig$-thresholds.  
For all cases, we see a linear dependence. The same is observed for the tune Pro-Q2O (not shown).
However, the difference in width of the MPI distributions has direct consequences for the experimental observables defined in the previous subsection~\ref{subsec:defintions} demonstrating their sensitivity to MPI and fragmentation properties.
 Figure \ref{fig:p0vsproq20} (left panel) shows the near-side pair-yield per trigger as a function of multiplicity. 
In the case of tune Pro-Q2O the yield reaches a plateau at $N_{\rm ch} > 15$ after which it rises only very slowly. In contrast, tune Perugia-0 shows a rather steep rise with a change to an even steeper slope at $N_{\rm ch}  \approx 50$.  The reason is the limited 
$\Nmpi$ in this tune. In order to reach high multiplicities the number of fragments per parton has to increase together with $\Nmpi$.
This can also be observed in figure \ref{fig:p0vsproq20} (right panel) where the number of uncorrelated seeds as a function of charged multiplicity is shown. For the tune Pro-Q2O an almost linear rise as  a function of charged multiplicity is observed up to the highest multiplicities, whereas for the tune Perugia-0, it
starts to level off at about $N_{\rm ch}  \approx 50$.

\section{Correction procedure}\label{sec:correction}
We corrected for the relevant inefficiencies such as detector acceptance, reconstruction, two-track 
and vertex reconstruction efficiency. In addition, the contamination of the sample of primary tracks 
by secondary particles was also corrected for.
The trigger inefficiency is not part of these corrections as it is negligible for events with at least one track in the considered acceptance $|\eta| < 0.9$. 
In the following paragraphs, the correction steps are discussed in detail. 
Table \ref{tab:correction} shows a breakdown of the main correction steps and corresponding efficiencies or contamination
for the different collision energies. They have been estimated from full transport 
and detector response simulations of PHOJET and PYTHIA6 tune Perugia-0 events using GEANT3 and a data driven correction procedure.
We show the efficiencies for the lowest \pT\ -cut used in the analysis ($\pT > 0.2$\,GeV/$c$ for the charged particle multiplicity),
because it corresponds to the largest inefficiency and contamination. 
\begin{table}[t]
\centering
\begin{tabular}{lrrr}
\toprule
Correction & $\sqrts=0.9$\,TeV & $\sqrts=2.76$\,TeV & $\sqrts=7$\,TeV \\
\midrule 
Tracking efficiency & $76.4\,\%$  & $75.5\,\%$  & $76.8\,\%$ \\ 
Contamination (MC based)      & $5.0\,\%$& $5.2\,\%$& $4.9\,\%$ \\
Contamination (data-driven)       & $1.1\,\%$ & $1.0\,\%$ & $1.1\,\%$ \\
Two-track and detector effects   & $0.5\,\%$ & $0.6\,\%$ & $0.5\,\%$ \\
Vertex reconstruction efficiency   & $97.5\,\%$ & $98.3\,\%$ & $98.8\,\%$ \\
\bottomrule
\end{tabular}
\caption{Main contributions to the track-to-particle correction averaged over $\pT>0.2$\,GeV/$c$, $|\eta|<0.9$, and charged particle multiplicities $\Nch$.} 
\label{tab:correction}
\end{table}

\begin{figure}[t]
\begin{minipage}{0.5\textwidth}
\includegraphics[width=\textwidth]{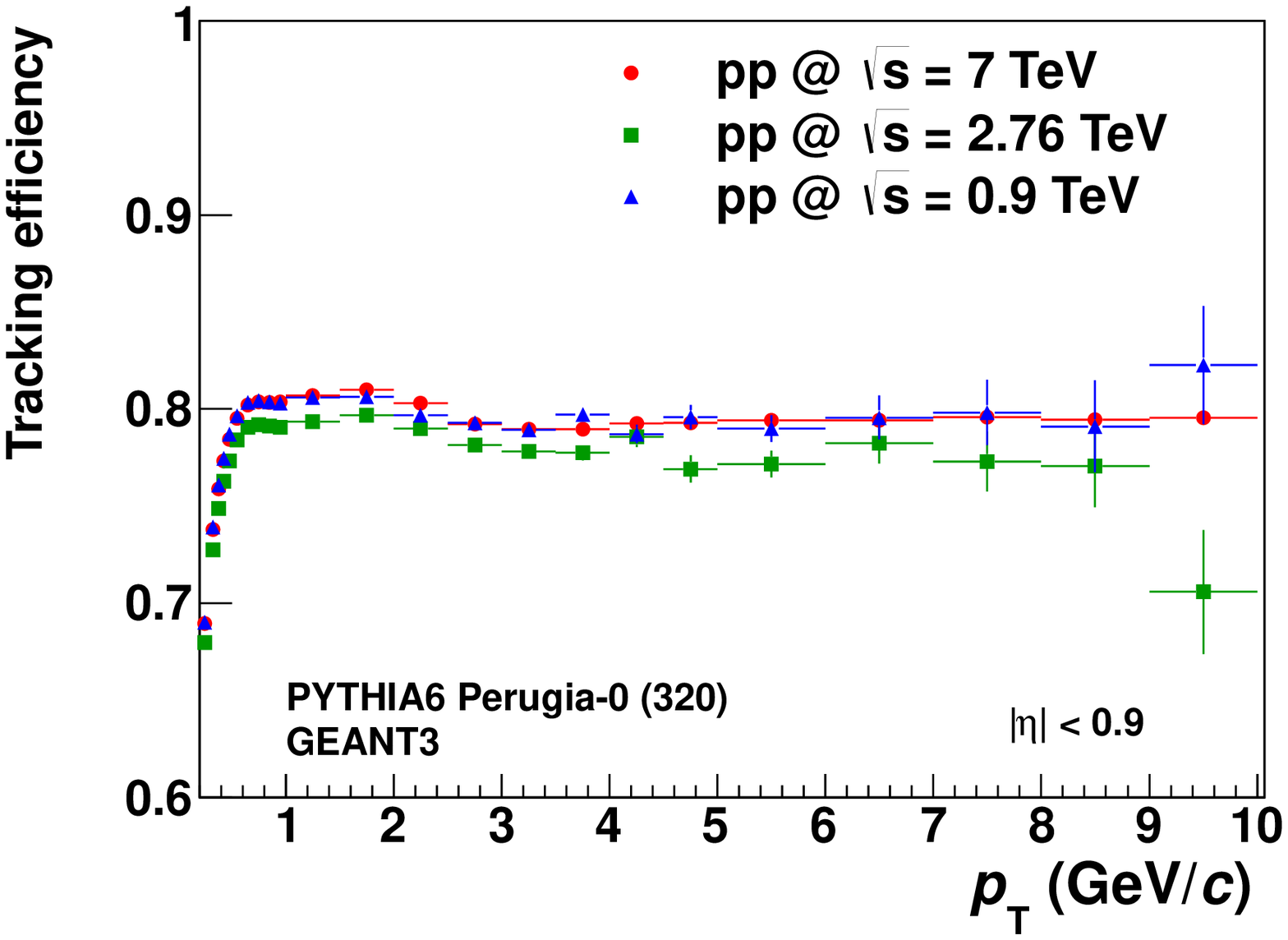} 
\end{minipage}
\begin{minipage}{0.5\textwidth}
\includegraphics[width=\textwidth]{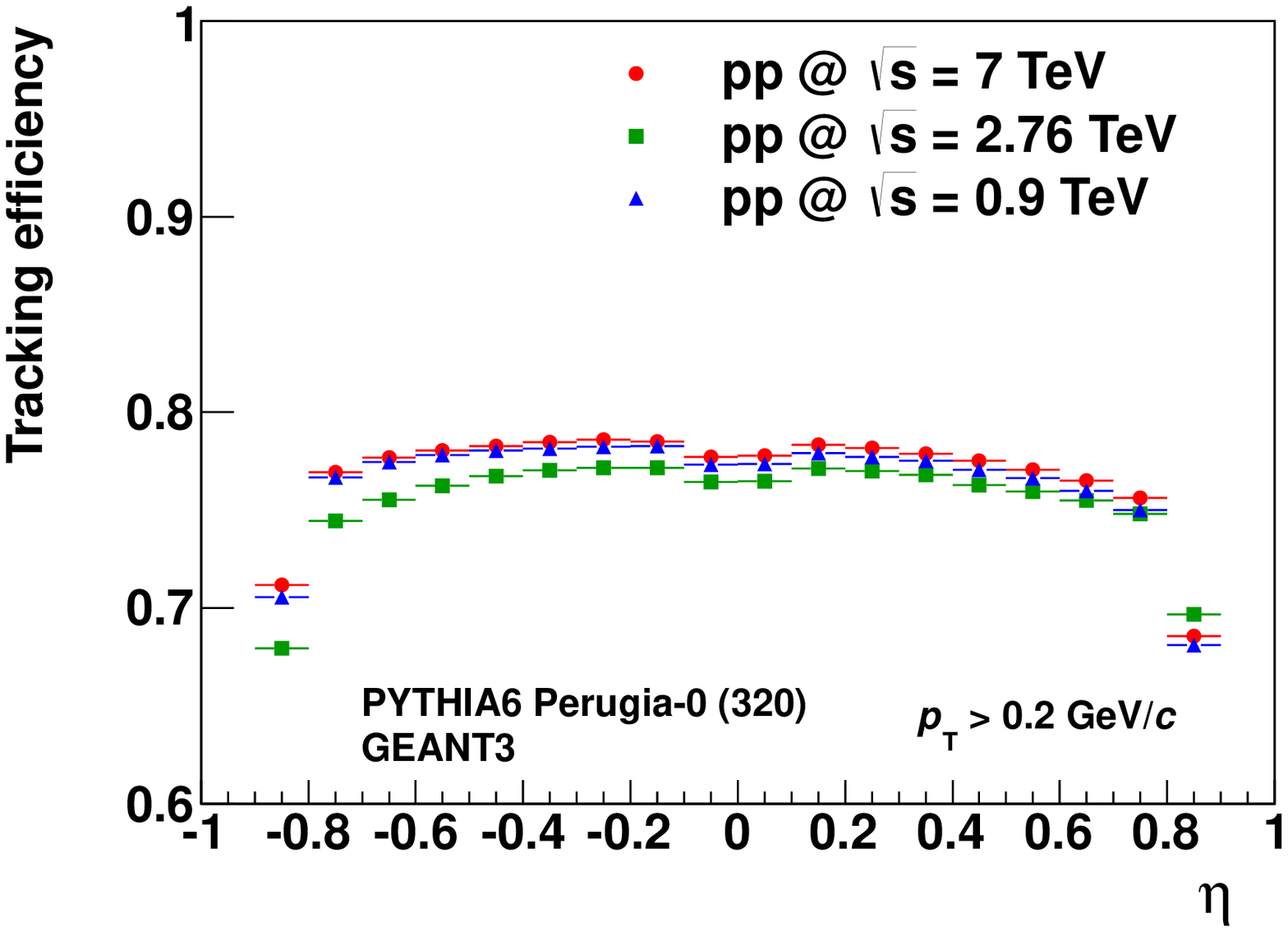} 
\end{minipage}
\caption{Reconstruction efficiency for primary particles. 
Left panel: reconstruction efficiency versus transverse momentum ($|\eta|<0.9$). 
Right panel: reconstruction efficiency versus pseudorapidity ($p_{\mathrm{T}}>0.2$ GeV/$c$). } 
\label{fig:trackEff}
\end{figure}

\paragraph{Tracking efficiency}
The tracking efficiency is given by the ratio of the number of reconstructed tracks from primary particles after track quality cuts 
to the number of primary particles. 
The tracking efficiency depends on
the kinematic properties of the particle ($\pT$, $\eta$, $\varphi$) and is 
influenced by the detector geometry, the probability of particle absorption in the detector material and particle decays. 
Figure \ref{fig:trackEff} shows the tracking efficiency for the different centre-of-mass energies obtained by projecting 2-dimensional $\eta-\pT$-correction maps
and integrating over $\varphi$. 
For the analysed data sets, the integrated tracking efficiency lies in the range 76\,\% to 77\,\%.

\begin{figure}[t]
\begin{minipage}{0.5\textwidth}
\includegraphics[width=\textwidth]{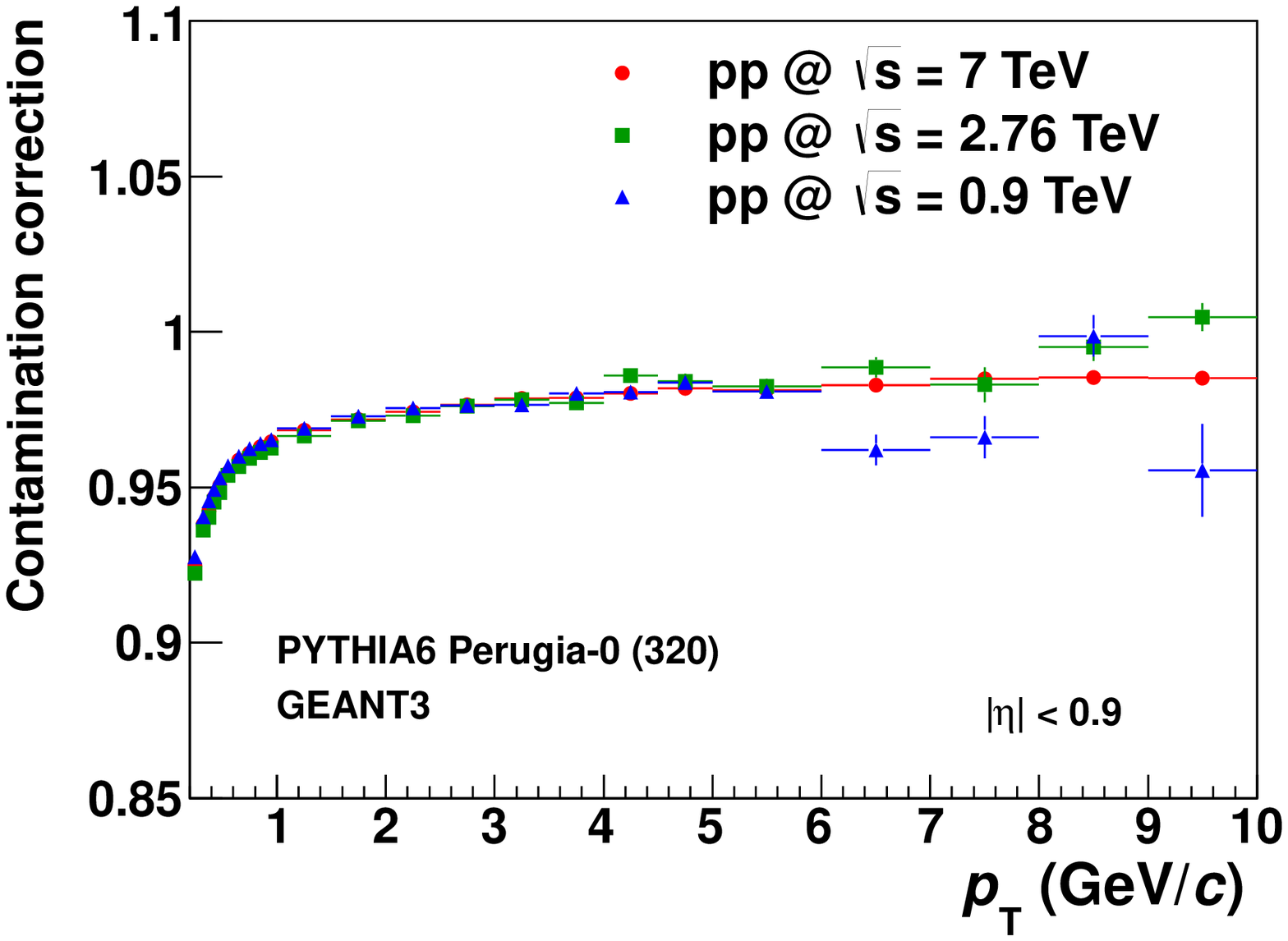} 
\end{minipage}
\begin{minipage}{0.5\textwidth}
\includegraphics[width=\textwidth]{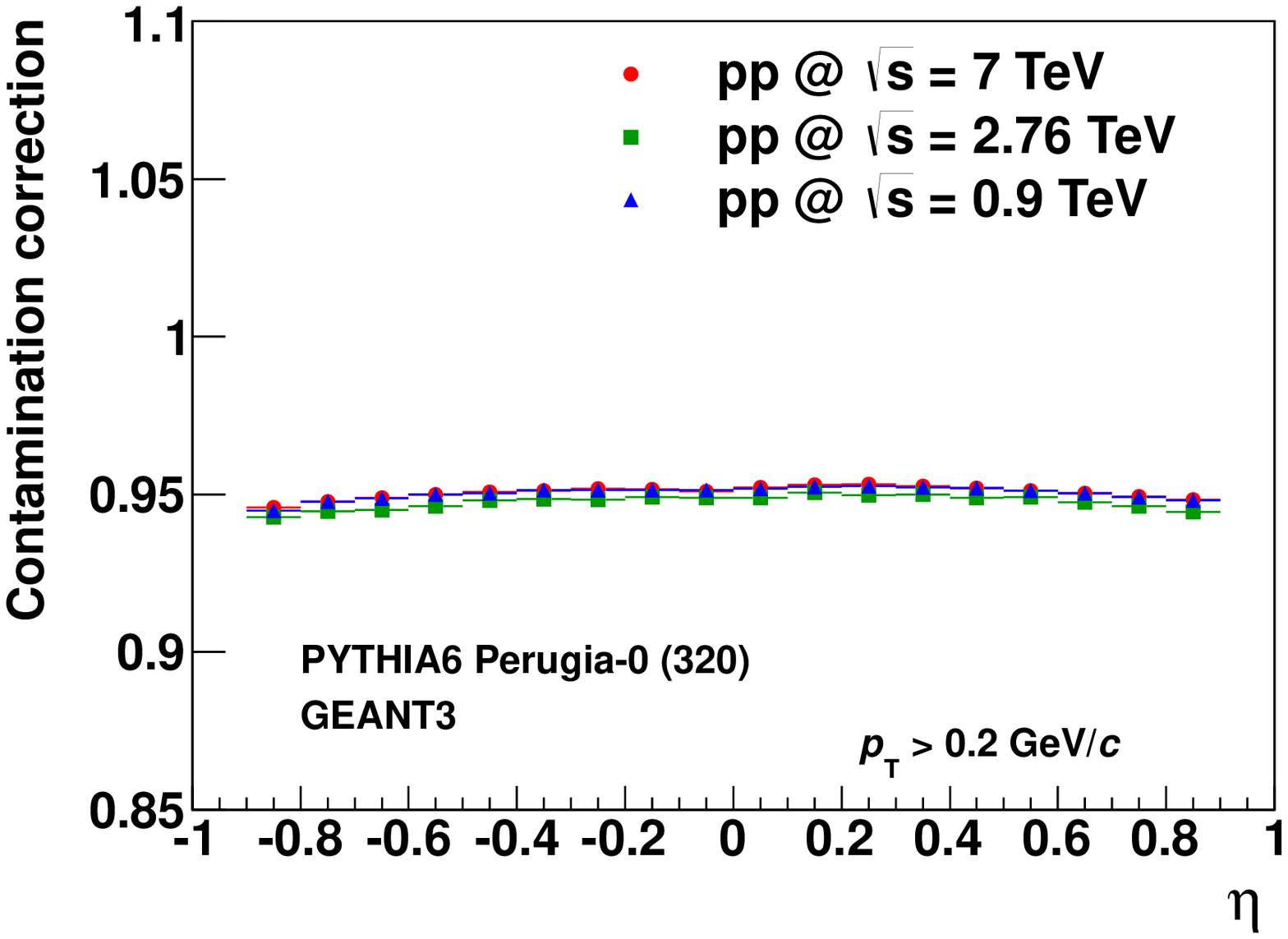} 
\end{minipage}
\caption{Contamination correction. Left panel: contamination correction versus transverse momentum ($|\eta|<0.9$). Right panel: contamination correction versus pseudorapidity ($p_{\mathrm{T}}>0.2$~GeV/$c$).} 
\label{fig:contamination}
\end{figure}

\paragraph{Secondary particle contamination}\label{sec:corr:contamination}
The standard Monte Carlo based contamination correction is given by the ratio of the number of reconstructed tracks after track quality cuts to the number of reconstructed tracks of primary particles. 
The contamination of the reconstructed tracks passing the quality cuts is mainly due to decay products from strange particles, 
photon conversions, and hadronic interactions with the detector material. 
Figure \ref{fig:contamination} shows the contamination correction as a function of the transverse momentum and the pseudorapidity. 
For the analysed data sets, the integrated contamination correction amounts to approximately 5\,\%.

In addition to the Monte Carlo based contamination correction, a data driven correction has been applied. 
This correction is based on the results of Ref.~\cite{Aamodt:2011zj, Aamodt:2011zza} which show that 
the generators PHOJET and PYTHIA6 tune Perugia-0 used in the correction procedure strongly underestimate
strange particle yields. This underestimation leads to an incomplete correction of the contamination in ALICE data when using Monte Carlo based correction maps only.
Based on the measured yields of strange particles, 
an additional correction factor of approximately 1\,\% has been added to the 5\,\% obtained from the standard MC contamination correction. 

\begin{figure}[t]
\centering
\includegraphics[width=0.6\textwidth]{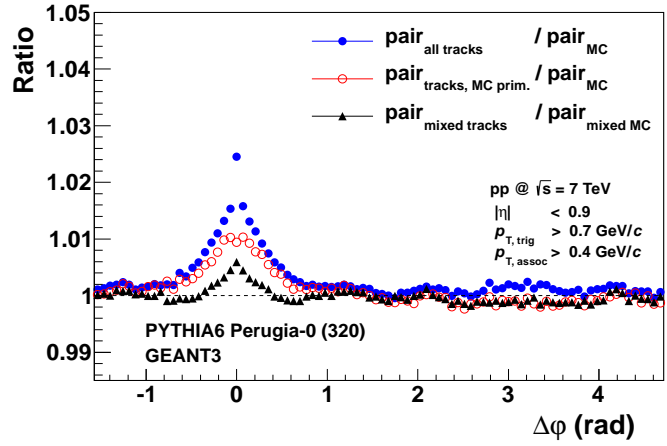}
\caption{Ratio between the track pair distribution of reconstructed and corrected tracks using single track corrections and the pair distribution of MC primary particles ($p_{\mathrm{T,\; trig }}>0.7$\,GeV/$c$,  $p_{\mathrm{T,\; assoc }}>0.4$\,GeV/$c$, and $\arrowvert \eta \arrowvert < 0.9$) as a function of the difference in azimuthal angle $\varphi_{\mathrm{track1}}-\varphi_{\mathrm{track2}}=\dphi$. 
The full detector simulations have been performed for $\sqrts=7$\,TeV.}
\label{fig:npair}
\end{figure}

\paragraph{Two-track and detector effects}
Effects such as track splitting, track merging, decay of long-lived particles, hadronic interactions with the detector material, gamma conversions 
as well as a non-uniform $\varphi$-acceptance induce modulations of the $\dphi$-distributions that have to be taken into account. 
These modifications can not be corrected in single-track-corrections only.
Figure \ref{fig:npair} shows the ratio
$$
 \frac{{\mathrm d}N}{\dphi}(\mathrm{pair}_{\mathrm{corrected \ tracks}})/ \frac{{\mathrm d}N}{\dphi}(\mathrm{pair}_{\mathrm{MC\ particles}}).
$$
The ratio is presented for all tracks, for tracks from primary particles only, 
and for tracks of mixed events each after single track correction. 
An enhanced number of particle pairs peaked around $\dphi=0$ is found after single track correction for the three cases. 
The ratio of corrected pairs to Monte Carlo particle pairs including secondary particles also shows a small enhancement around $\dphi=\pi$.
To correct for this effect, a two-track post-correction is performed after the single track correction, 
using Monte Carlo based correction factors which depend on  $\dphi$, $\pTtrig^{\mathrm{min}}$, and $\pTassoc^{\mathrm{min}}$. 
The correction decreases with increasing transverse momentum thresholds.
For the analysed data sets, 
the maximum effect from this correction (5\,\%) is observed for the lowest values of $\pTtrig^{\mathrm{min}}=0.7$\,GeV/$c$ and $\pTassoc^{\mathrm{min}}=0.4$\,GeV/$c$ and at the highest $\Nch$, where the ratio near-side yield over combinatorial background is lowest.

\paragraph{Vertex reconstruction efficiency}
The vertex reconstruction efficiency is the ratio between the number of triggered events with a reconstructed accepted vertex of good quality and the number of triggered events. 
The vertex reconstruction efficiency has not only an impact on the number of events but also on the total number of particles entering the data sample. 
The effect of the vertex reconstruction efficiency contributes with $1.2$\,\% to $2.5$\,\% to the multiplicity integrated track-to-particle correction 
for the analysed data sets.
The impact of the vertex reconstruction efficiency depends strongly on the charged particle multiplicity effecting only the low $\Nch$ bins. 
For $\Nch>10$, the vertex reconstruction efficiency is consistent with unity. 

\paragraph{Trigger efficiency}
The correction of the  trigger efficiency takes into account the fact 
that the number of triggered events is only a subset of the produced events of a given event class. 
However, the trigger is fully efficient for events with at least one charged track in the considered ITS-TPC acceptance. 
Hence, no correction for the trigger efficiency is applied.

\begin{figure}[t]
\centering
\begin{minipage}{0.47\textwidth}
\includegraphics[width=\textwidth]{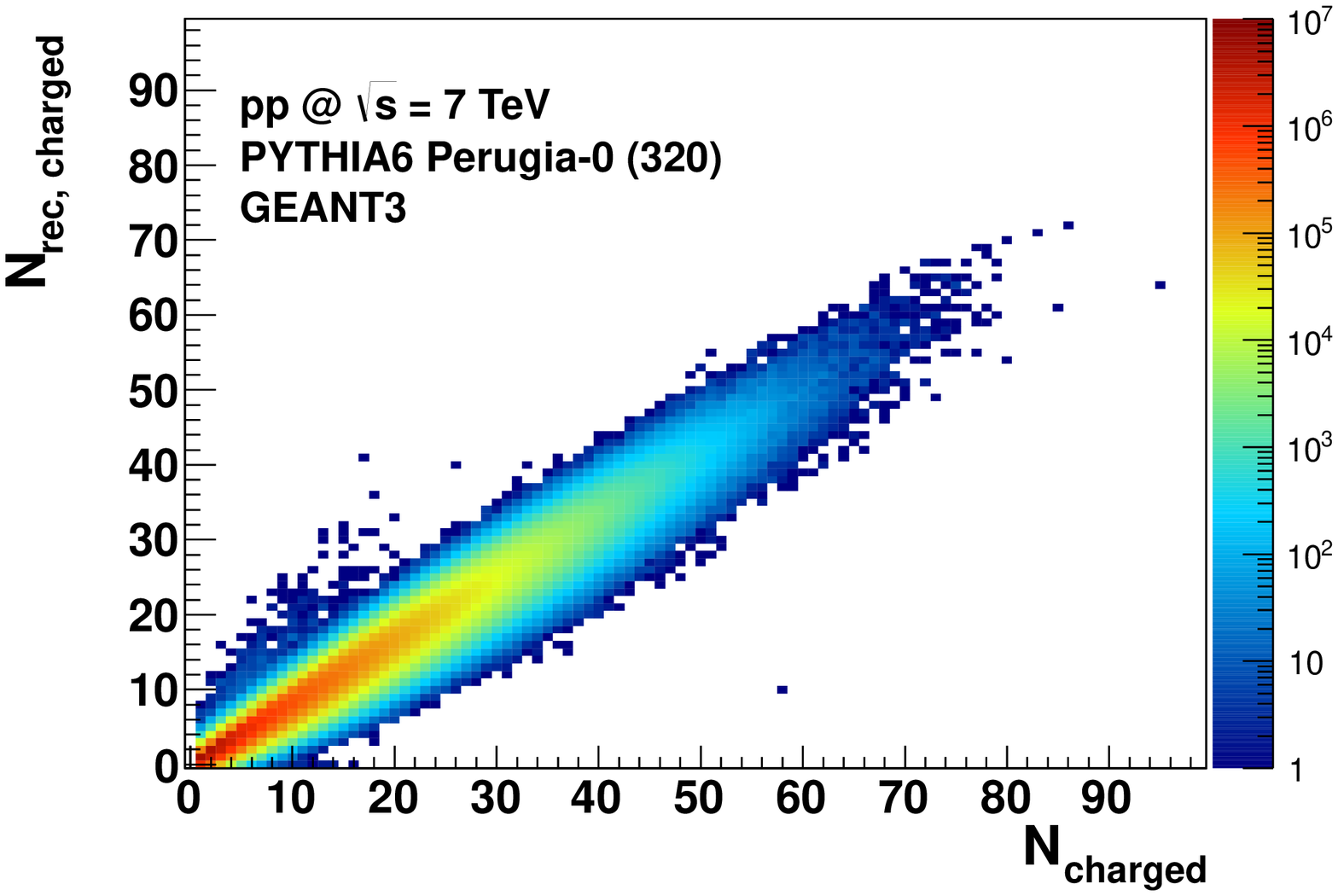}
\end{minipage}
\begin{minipage}{0.47\textwidth}
\includegraphics[width=\textwidth]{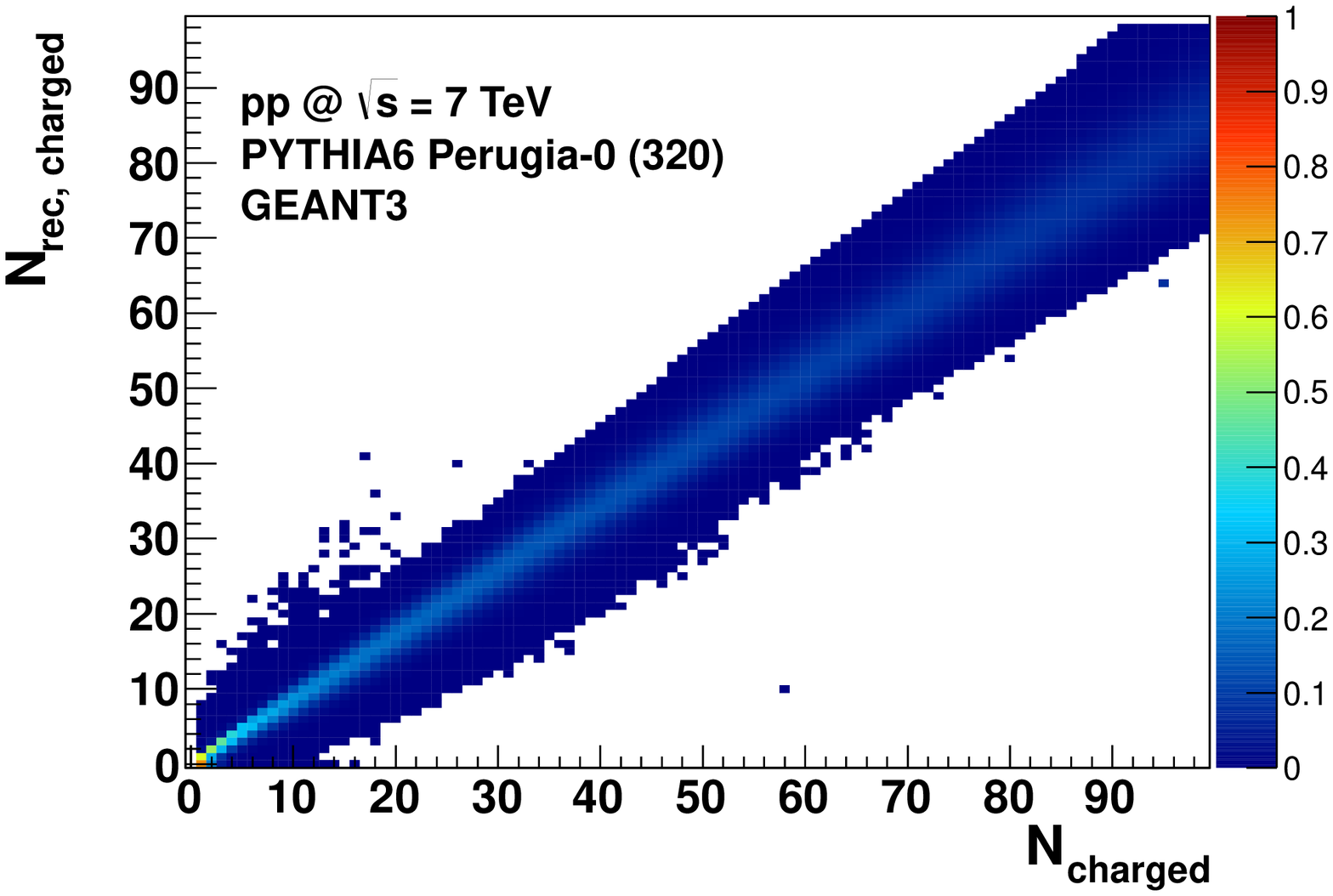}
\end{minipage}
\vspace{0.2cm}
\caption{Left panel: simulated correlation matrix. 
Right panel: normalized and extended correlation matrix. 
Input for the extension are Gaussian distributions with extrapolated $\langle \Nch \rangle$ and $\sigma_{\langle \Nch \rangle}$.} 
\label{fig:correlationExtend}
\end{figure}

\paragraph{Charged particle multiplicity correction}
The present analysis studies the evolution of the integrated yields of the azimuthal correlation as a function 
of the true charged particle multiplicity $\Nch$ in the range $\pT>0.2$\,GeV/$c$ and $\arrowvert\eta\arrowvert<0.9$. 
Our approach to a full correction of detector effects on the multiplicity is a two-step procedure: 
first, the correction of the raw two-particle correlation observables $O_{unc}$ from $O_{\mathrm{unc}}(\Nrec)$ 
to its corrected value $O_{\mathrm{corr}}(\Nrec)$ is performed as a function of the reconstructed uncorrected multiplicty $\Nrec$.
Then, the correction of the charged particle multiplicity from $O_{\mathrm{corr}}(\Nrec)$ to $O_{\mathrm{corr}}(\Nch)$ is carried out to 
obtain the corresponding observable at the corrected charged particle multiplicity $\Nch$.
The same procedure has also been used for the measurement of the mean transverse momentum 
and the transverse sphericity as a function of the true multiplicity as described in Refs.~\cite{Aamodt:2010my,Abelev:2012sk}.
The correction employs the correlation matrices $R(\Nch,\, \Nrec)$ which are proportional to the probability of reconstructing
$\Nrec$ particles under the condition that $\Nch$ particles have been produced.
They are obtained from full detector simulations quantifying the relation between the number of charged primary particles and the number 
of reconstructed tracks both in $\pT>0.2$\,GeV/$c$ and $\arrowvert\eta\arrowvert<0.9$ as shown in the left panel of figure~\ref{fig:correlationExtend}.

The columns of the correlation matrix have to be normalised to one 
\begin{equation}
  \forall \Nch : \sum_{\Nrec}R_{1}(\Nch,\, \Nrec)=1.
\end{equation}
The normalised correlation matrix represents the conditional probability for measuring an event of a given true multiplicity, $\Nch$, 
for a given reconstructed track multiplicity of $\Nrec$.
In a second step, the correlation matrix is extrapolated to the highest multiplicities not covered in the detector simulation 
due to the limited number of simulated events. 
To this end, the distribution of each matrix column at low multiplicities is fitted 
with a Gaussian function.
As expected, the width of the Gaussian functions grows approximately as $\sigma \propto \sqrt{\Nrec}$ and the mean grows as $\langle\Nch\rangle \propto \Nrec$. 
This scaling is used to extrapolate the correlation matrix to higher multiplicities.
An extrapolated correlation matrix with normalised columns is shown in the right panel of figure~\ref{fig:correlationExtend}.
Based on the normalised and extended correlation matrix, the observable $O_{\mathrm{corr}}(\Nrec)$  can be converted to $O_{\mathrm{corr}}(\Nch)$ using
\begin{equation}
O(\Nch) = \sum_{\Nrec} O(\Nrec) \cdot R_{1}(\Nch, \Nrec).
\end{equation}

\section{Systematic uncertainties}\label{sec:systematics}
A comprehensive study of the systematic uncertainties of the final analysis results has been performed. 
In the following, the sources of systematic uncertainties and their impact on the analysis results are described.
Representative for all final analysis results, the systematic uncertainties of the per-trigger near-side pair yield measured using $\pTtrig>0.7$\,GeV/$c$ and $\pTassoc>0.4$\,GeV/$c$ as a function of the charged particle multiplicity are discussed in the text and summarised in table~\ref{tab:systSum}.

\begin{table}[t]
\begin{center}
\begin{tabular}{lcccccc}
\toprule
\multicolumn{1}{c}{} & \multicolumn{2}{c}{$\sqrts = 0.9$\,TeV} & \multicolumn{2}{c}{$\sqrts = 2.76$\,TeV} & \multicolumn{2}{c}{$\sqrts = 7$\,TeV}\\
      & $N$=$2$  &  $N$=$2\langle N \rangle$ & $N$=$2$  &  $N$=$2\langle N \rangle$& $N$=$2$  &  $N$=$2\langle N \rangle$\\
\midrule
Signal extraction & \plmi{0.3}\,\%& \plmi{0.1}\,\% & \plmi{0.2}\,\%& \plmi{0.1}\,\% & \plmi{0.5}\,\%& \plmi{0.1}\,\%\\
Bin width & \plmi{0.2}\,\% & \plmi{0.1}\,\% & \plmi{0.2}\,\% & \plmi{0.1}\,\%& \plmi{0.2}\,\% & \plmi{0.1}\,\%\\
Correction procedure & \plmi{1.9}\%& \plmi{0.9}\,\% & \plmi{5.0}\,\%& \plmi{3.0}\,\% & \plmi{12.8}\,\%& \plmi{1.2}\,\%\\
Event generator & \plmi{1.1}\,\%& \plmi{1.8}\,\%& \plmi{1.8}\,\%& \plmi{2.0}\,\%& \plmi{1.9}\,\%& \plmi{0.1}\,\%  \\
Transport MC  &\plmi{0.3}\,\%& \plmi{0.1}\,\%&\plmi{0.3}\,\%& \plmi{0.1}\,\%&\plmi{0.3}\,\%& \plmi{0.1}\,\% \\
Track cut & \plmi{15.0}\,\%& \plmi{2.5}\,\%& \plmi{16.9}\,\%& \plmi{2.3}\,\%& \plmi{10.6}\,\%& \plmi{2.0}\,\%\\
Vertex cut & \plmi{2.7}\,\%& \plmi{0.5}\,\% & \plmi{1.5}\,\%& - & \plmi{2.1}\,\%& - \\
Detector efficiency & \plmi{3.0}\,\% & \plmi{3.0}\,\%& \plmi{4.1}\,\%& \plmi{4.1}\,\% & \plmi{4.1}\,\%& \plmi{4.1}\,\%\\
Material budget & \plmi{0.4}\,\%& \plmi{0.3}\,\%& \plmi{0.4}\,\%& \plmi{0.3}\,\%& \plmi{0.4}\,\%& \plmi{0.3}\,\%\\
Particle composition & \plmi{2.0}\,\%& \plmi{1.0}\,\%& \plmi{2.1}\,\%& \plmi{1.3}\,\% & \plmi{2.0}\,\%& \plmi{1.5}\,\%\\
Pileup &- &- &- &- &\plmi{5.0}\,\%& \plmi{1.0}\,\%\\
Extrapol. of S.-Corr.& - & - & \plmi{2}\,\% & - & \plmi{2}\,\% & -\\
\bottomrule
\end{tabular}
\end{center}
\caption{Systematic uncertainties for the per-trigger near-side pair yield measured using $\pTtrig>0.7$\,GeV/$c$, $\pTassoc>0.4$\,GeV/$c$, and $\arrowvert\eta\arrowvert<0.9$ exemplary for all final analysis results for two charged particle multiplicity bins. The full charged particle multiplicity dependence can be found in Ref. \cite{Sicking:2012}.} 
\label{tab:systSum}
\end{table}

\paragraph{Per-trigger pair yield measurement based on a fit function}
The per-trigger pair yield of the azimuthal correlation is extracted utilizing the fit function of equation~\ref{eq:fit}. 
A good agreement between the data distribution and the fit function has been found using residuals as well as the $\chi^2/\mathrm{NDF}$ test.
The stability of the fit results has been verified based on various tests. 
For example, it has been checked that a modification of the combinatorial background of the azimuthal correlation does not change the extracted yields of the near and the away-side peaks. 
Moreover, it has been verified that the combination of events results in the expected modification of the per-trigger pair yield components. 
In addition, the minimum number of events needed for a stable fit result as well as the  optimised resolution of the $\dphi$-distribution in terms of the bin-size have been determined. 

\paragraph{Correction procedure}
In section \ref{sec:correction}, a full correction procedure of detector effects has been introduced.
When correcting event generator data after full detector simulations with correction maps obtained with the same event generator, it is expected to recover the Monte Carlo input. 
A remaining disagreement between the corrected results and the input Monte Carlo results represents the systematic uncertainty of the correction procedure.
As an example, the per-trigger near-side pair yield obtained from the MC input and the corrected results differ from each other by up to 12.8\,\% for the first charged particle multiplicity bin and by less than 3.0\,\% for higher charged particle multiplicity bins. 

Correction maps can be estimated with different Monte Carlo generators.
When using correction maps of one Monte Carlo generator for the correction of data of a second Monte Carlo generator, further discrepancies can emerge. 
The per-trigger near-side pair yields of corrected data obtained using PYTHIA6 tune Perugia-0 correction maps and using PHOJET correction maps differ by less than 2\,\% for all charged particle multiplicities.

We have estimated the impact of the transport Monte Carlo choice on the final analysis results. 
For this purpose, in addition to the default GEANT3~\cite{Brun:1994zzo} detector simulations, a sample of pp events has been simulated using GEANT4~\cite{Agostinelli:2002hh, Allison:2006ve}. 
The results obtained with the GEANT3 and the GEANT4 based correction maps are in very good agreement. The results differ from each other by a maximum of  0.3\,\% for all charged particle multiplicities.

\paragraph{Track and vertex selection}
The systematic uncertainty related to the choice of the track selection cuts introduced in section~\ref{sec:track} is estimated by performing a full correction and analysis chain using varying track selection cuts. 
For this purpose, the default ITS-TPC track cuts have been loosened and tightened within reasonable limits. 
In addition, tracks measured exclusively with the TPC have been analysed.
The per-trigger near-side pair yield shows a sizable difference when using the different track cuts of up to 16\,\% for the first charged particle multiplicity bin, however, the impact decreases to less than 2.5\,\% for higher charged particle multiplicities.

The impact of the vertex selection choice is tested by varying the vertex quality cuts. 
Instead of requiring at least one track associated to the collision vertex, two tracks are required.
The impact of this modification on the per-trigger near-side pair yield is 2\,\% for the lowest charged particle multiplicity bin and compatible with zero for charged particle multiplicities above $\Nch>10$.

\paragraph{Tracking efficiency}
The ITS-TPC tracking efficiency uncertainty has been estimated by comparing the track matching efficiency between ITS and TPC and vice versa for simulated data and real data \cite{Aamodt:2010my, Aamodt:2010jd}. 
The disagreement between the matching efficiencies is then converted into a transverse momentum dependent reconstruction efficiency uncertainty. 
By varying the reconstruction efficiency accordingly, the systematic uncertainty on the final analysis results can be estimated. 
The impact of this uncertainty on the per-trigger near-side pair yield is about 4\,\% for all charged particle multiplicities.

The material budget of ALICE has been measured with the help of photon conversions in the detector material. 
The remaining uncertainty in the knowledge of the material budget can be converted into a transverse momentum dependent uncertainty of the tracking efficiency. The effect of this uncertainty results in a small variation of analysis results. For example, the per-trigger near-side pair yield is modified by below 0.4\,\% for all charged particle multiplicities.

The ITS-TPC tracking efficiency estimated in full detector simulations depends to some extent on the composition of the particle yields. This is due to the fact that the particle decay length and the probability to be absorbed in the detector material depends on the particle type. 
The systematic uncertainty related to the particle composition has already been studied in Ref.~\cite{ALICE:2011ac}. 
Motivated by a disagreement of measured particle yields to predictions of PYTHIA6 and PHOJET \cite{Aamodt:2011zj, Aamodt:2011zza}, the yields of pions, kaons, and protons used in the calculation of correction maps of detector effects have been modified by $\pm 30$\,\% \cite{ALICE:2011ac}. 
The effect of this modification accounts for a variation of the final results of at most 2.0\,\%.

\paragraph{Pile-up events}
The impact of pile-up events on the analysis results has been tested by analysing high pile-up data sets. 
A quantitative estimation of the systematic uncertainty related to pile-up events in the data analysis has been performed by splitting the default data sets into sub-sets of relatively low and relatively high pileup-probability. 
The difference between the analysis results of the two sub-sets accounts for about 5\,\% for the lowest charged particle multiplicity bin and below 1\,\% for all higher charged particle multiplicities.

\paragraph{Extrapolation of strangeness correction}
As part of the contamination correction procedure described in section~\ref{sec:corr:contamination}, a data driven contamination correction has been performed accounting for the underestimated strangeness yield in the Monte Carlo generators.
This correction is based on ALICE measurements at $\sqrts=0.9$\,TeV \cite{Aamodt:2011zj,Aamodt:2011zza}, however, these corrections were also used to correct collision data measured at $\sqrts=2.76$ and 7\,TeV.
The uncertainty related to the extrapolation of this correction to higher centre-of-mass energies can be estimated using measurements of strange particle yields performed by the CMS experiment at $\sqrts~=~0.9$ and 7\,TeV \cite{Khachatryan:2011tm}. 
When performing the same data driven contamination correction based on the CMS measurements, small modification of the final results can be observed. 
The systematic uncertainty of the per-trigger near-side pair yield related to the extrapolation of the strangeness correction is below 2\,\% for the first charged particle multiplicity and compatible with zero for charged particle multiplicities above $\Nch>8$. 

\section{Results}\label{sec:results}
The two-particle correlation analysis are now presented,
after having included the corresponding corrections described in the previous sections.
Results are discussed for the three different centre-of-mass energies and two sets of 
$\pT$-cuts: $\pTtrig > 0.7 \, \gmom, \, \pTassoc > 0.4 \, \gmom$  and 
$\pT$-cuts: $\pTtrig > 0.7 \, \gmom, \, \pTassoc > 0.7 \, \gmom$.
The second, symmetric, bin is used to analyse the number of uncorrelated seeds.

ALICE data are presented as black points and the results of Monte Carlo calculations as coloured symbols.
The error bars represent  the statistical errors and the boxes the systematic uncertainties. 
The horizontal error bars  correspond to the bin-width. For measurements as a function of the charged particle 
multiplicity, the upper part of the figures shows the analysis results and the lower part shows the ratio between data and
the Monte Carlo calculations.

\begin{figure}[t]
\centering
\includegraphics[width=0.6\textwidth]{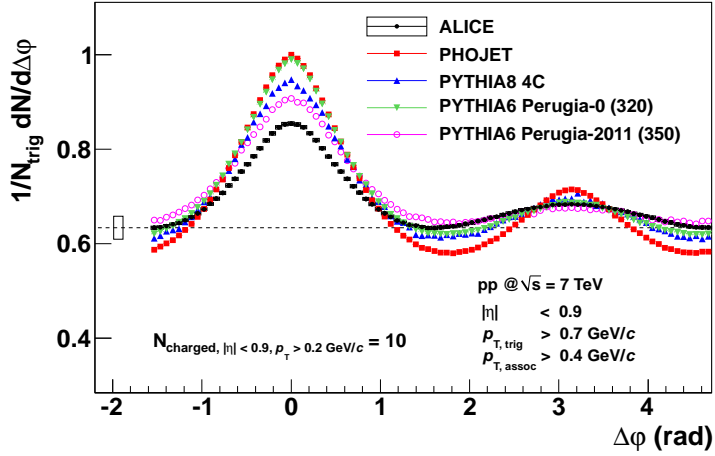}
\caption{Azimuthal correlation for events with $\Nch=10$ measured at $\sqrts=7$\,TeV. }
\label{fig:azimuthal}
\end{figure}

Before discussing in detail the multiplicity and centre-of-mass energy dependence and their implications for 
multiple parton interactions, we present in figure \ref{fig:azimuthal} an example of 
a measured  azimuthal correlation
function. In the figure, the data are compared to various MC simulations on generator level for the charged particle multiplicity bin $\Nch=10$ at $\sqrts=7$\,TeV.
The part of the systematic uncertainty that has the same relative contribution for all $\dphi$-bins is presented as a box
on the left side of the data points. 
The height of the box corresponds to the value of the leftmost data point (at $\dphi=-\pi/2$) and must be scaled for all other data points according to their absolute values.

Within the systematic uncertainties, the constant combinatorial background is of the same height for data and all
PYTHIA tunes. 
PHOJET shows a lower combinatorial background.
The near-side peak centred around $\dphi=0$ is overestimated by all Monte Carlo generators in terms of its height and
 its integral above the combinatorial background. Here, PYTHIA6 tune Perugia-2011 shows the best agreement with
 data.
The width of the near-side peak is roughly reproduced by the Monte Carlo generators.
PHOJET and PYTHIA8 tune 4C produce an away-side peak ($\dphi=\pi$) with a higher absolute height than in data.
The PYTHIA6 tunes Perugia-0 and Perugia-2011 both agree with data in terms of the height of the away-side peak.
PYTHIA8 4C and PHOJET overestimate the integral of the away-side peak above the constant combinatorial 
background. Here, PYTHIA6 Perugia-0 agrees with
data, and PYTHIA6 tune Perugia-2011 underestimates the data slightly.
The width of the away-side peak is much narrower in PHOJET than in data  while the PYTHIA tunes give only a slightly
narrower away-side peak.

\subsection{Yields}
\begin{figure}[p]
\centering
\vspace{-0.3cm}
\begin{minipage}{0.49\textwidth}
\includegraphics[width=\textwidth]{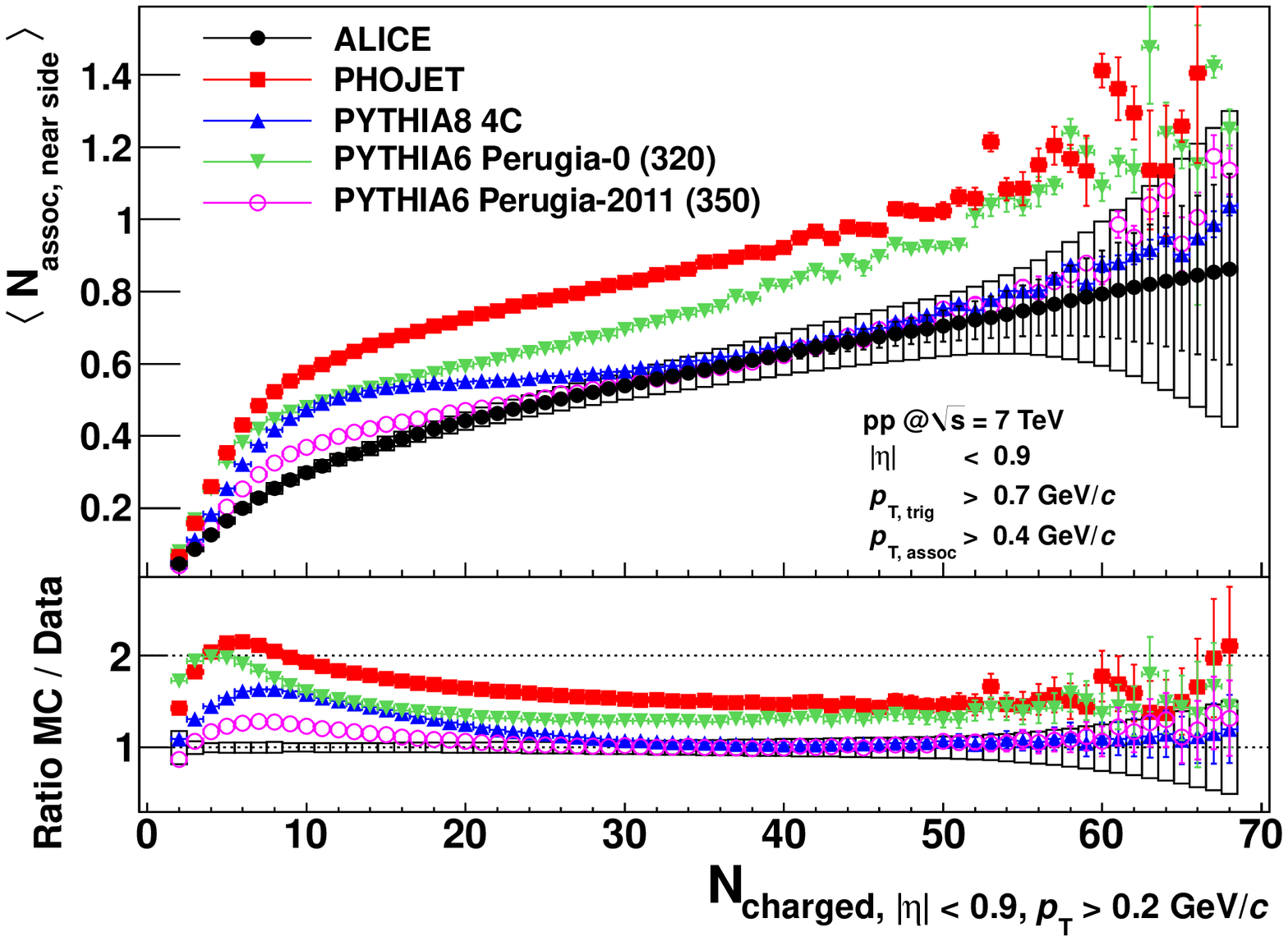} 
\end{minipage}
\begin{minipage}{0.49\textwidth}
\includegraphics[width=\textwidth]{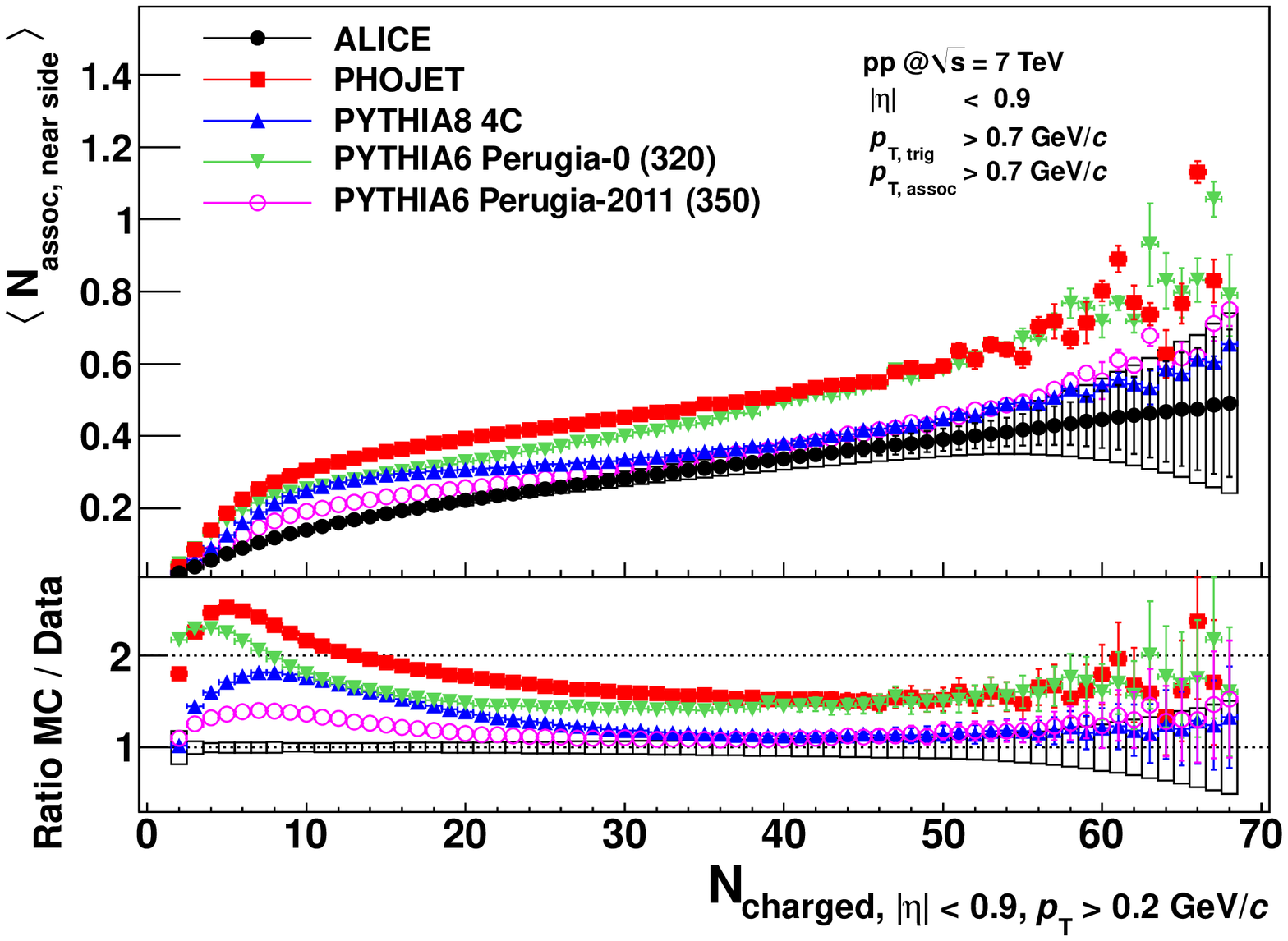} 
\end{minipage}
\vspace{-0.1cm}
\begin{minipage}{0.49\textwidth}
\includegraphics[width=\textwidth]{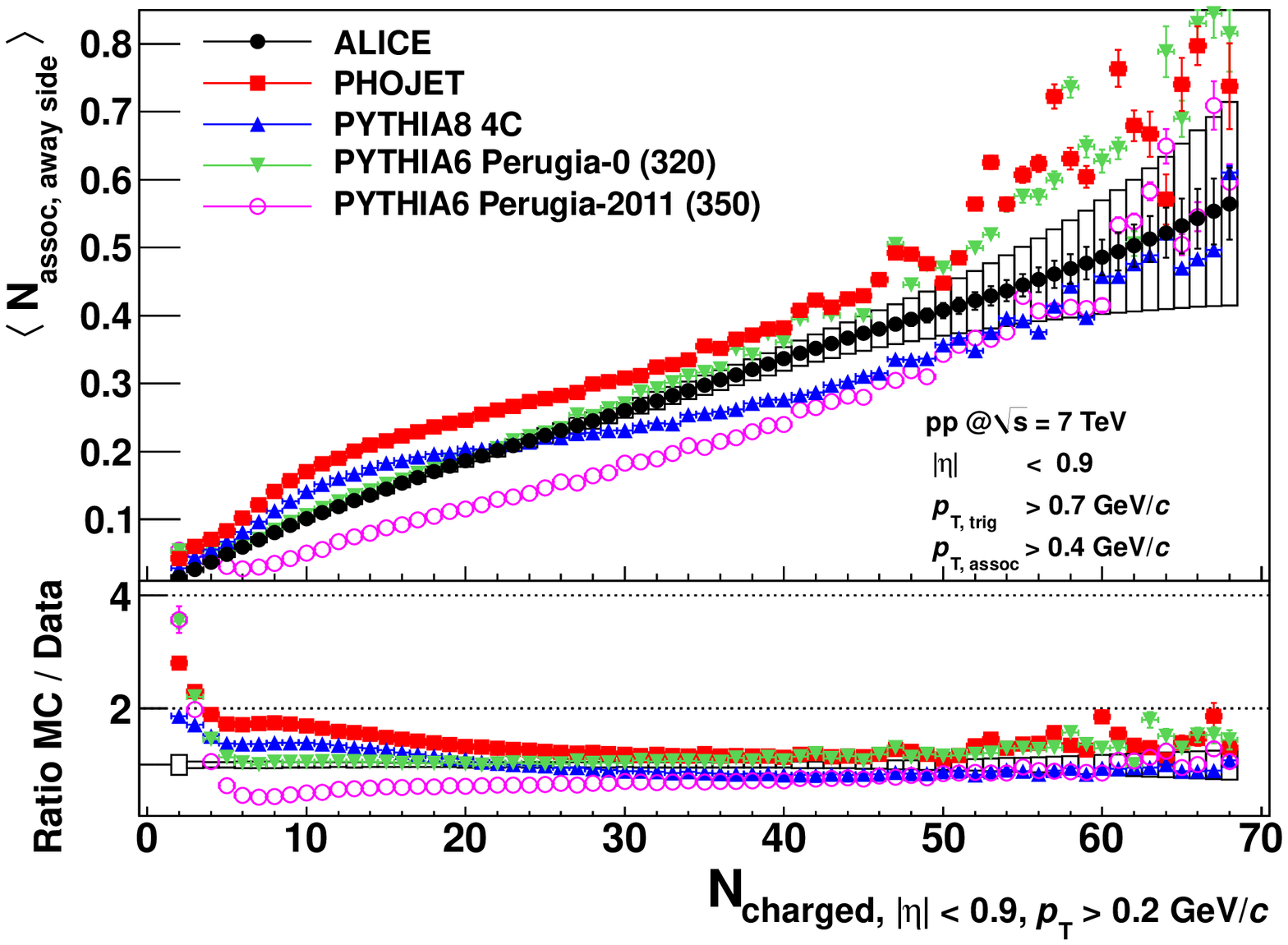} 
\end{minipage}
\begin{minipage}{0.49\textwidth}
\includegraphics[width=\textwidth]{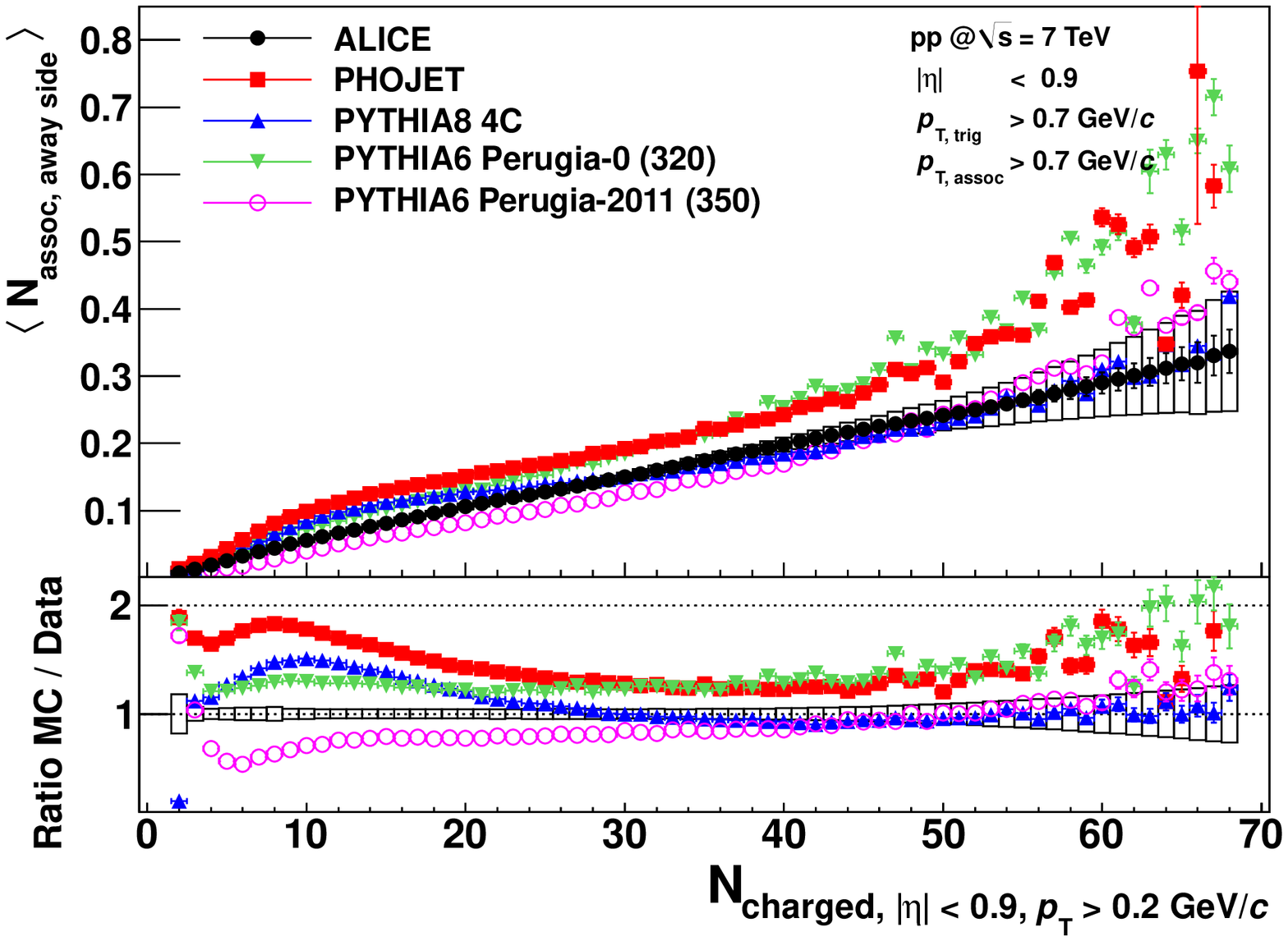} 
\end{minipage}
\vspace{-0.1cm}
\begin{minipage}{0.49\textwidth}
\includegraphics[width=\textwidth]{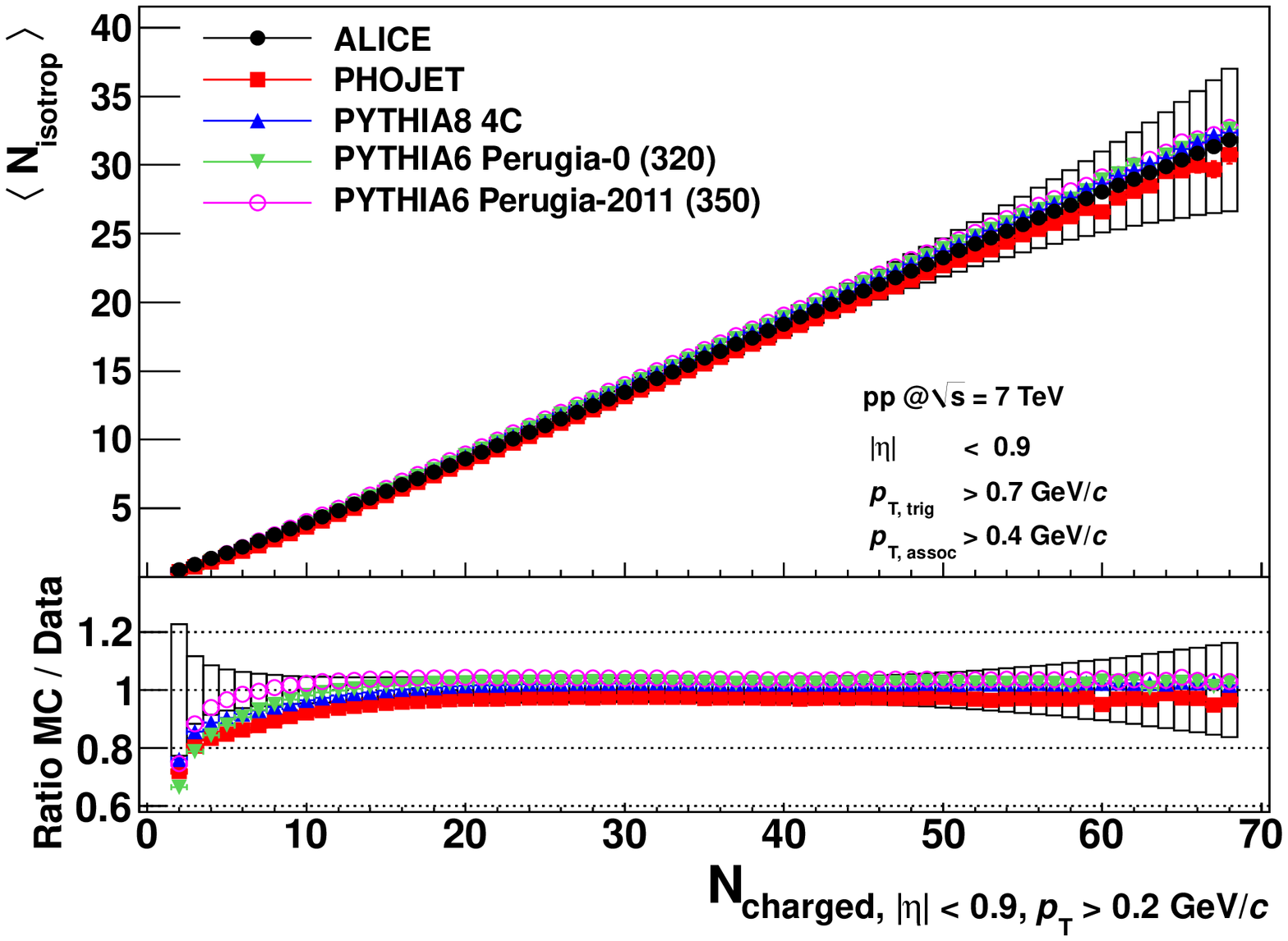} 
\end{minipage}
\begin{minipage}{0.49\textwidth}
\includegraphics[width=\textwidth]{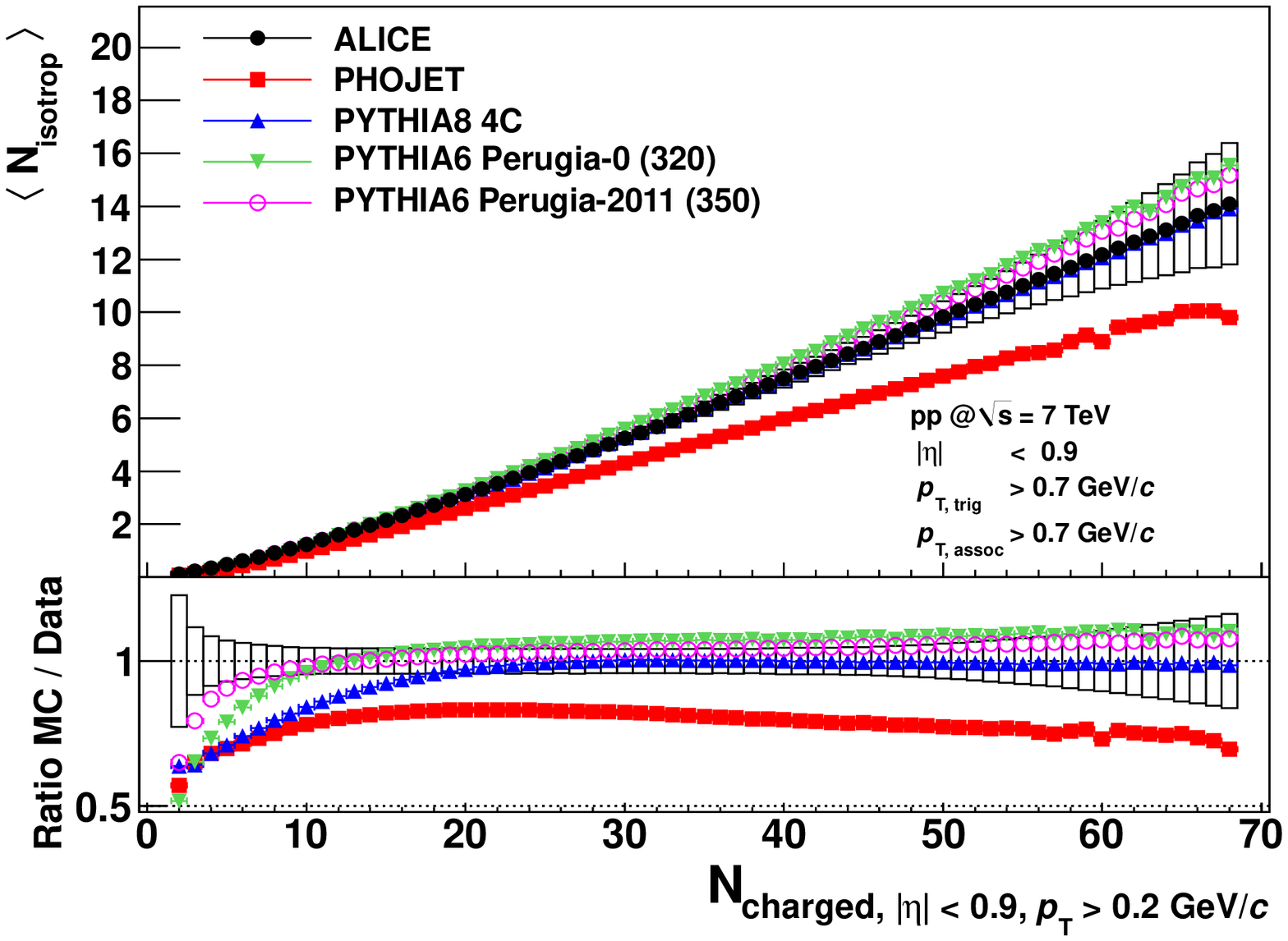} 
\end{minipage}
\vspace{-0.1cm}
\begin{minipage}{0.49\textwidth}
\includegraphics[width=\textwidth]{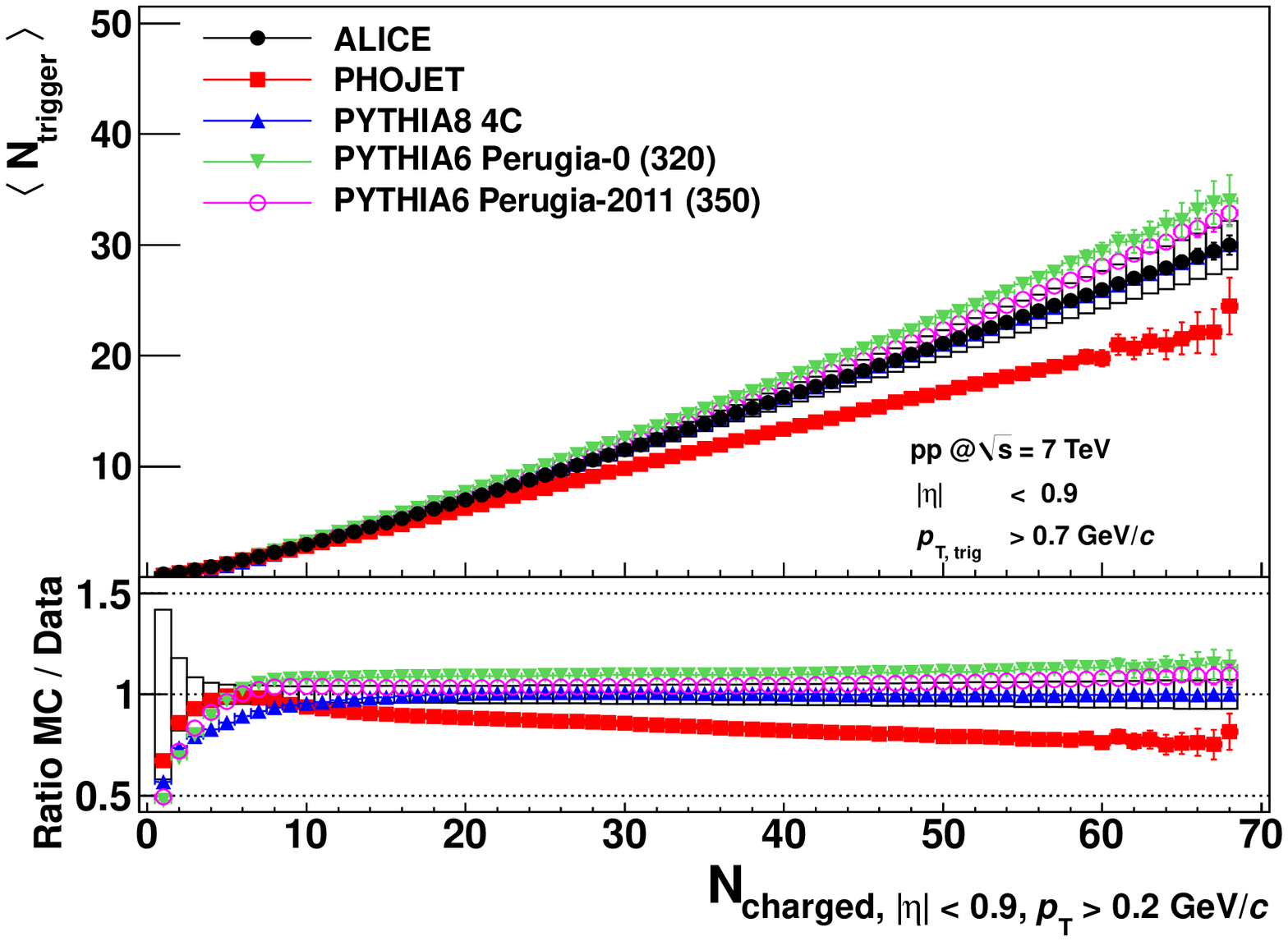} 
\end{minipage}
\begin{minipage}{0.49\textwidth}
\includegraphics[width=\textwidth]{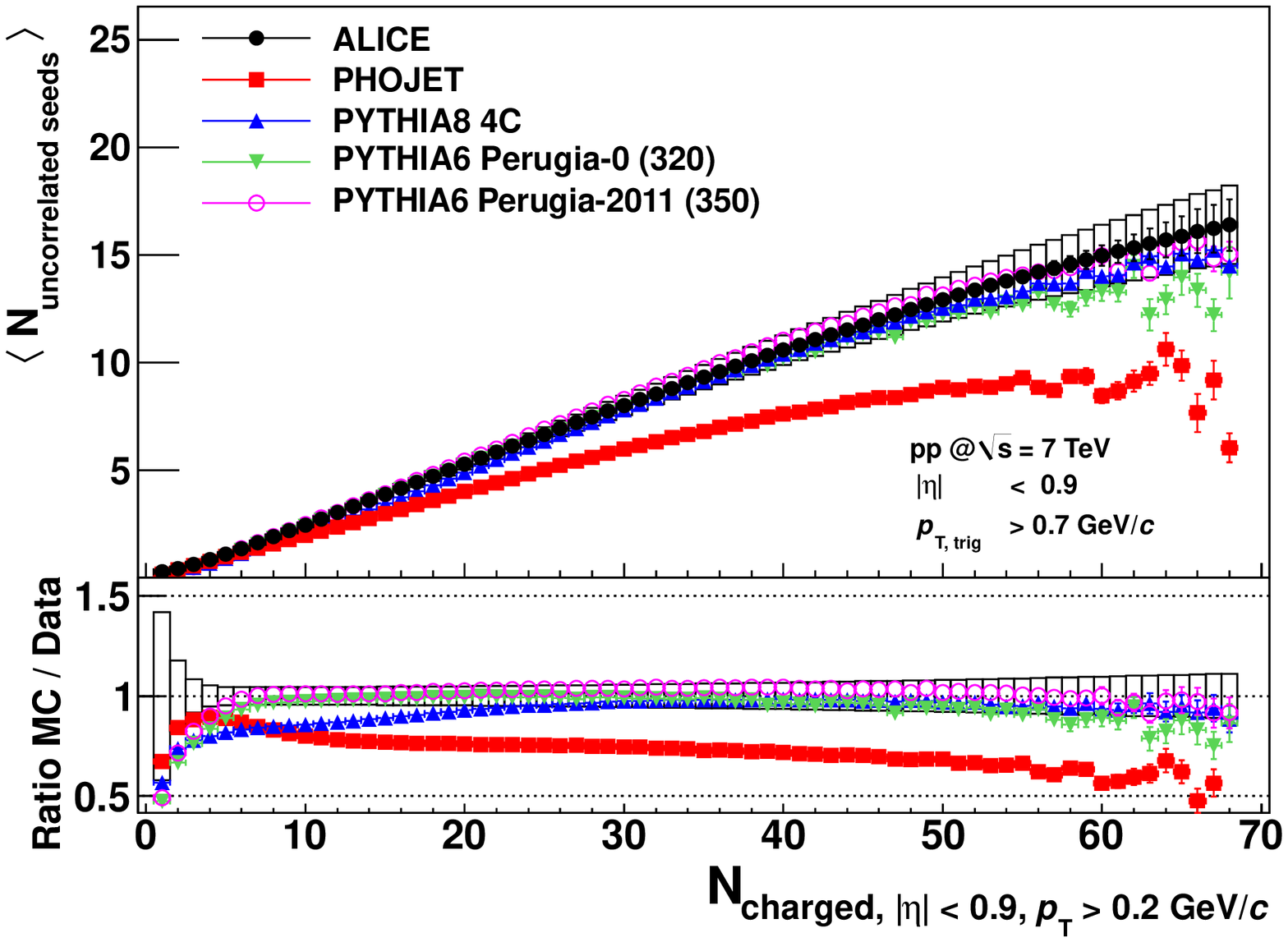} 
\end{minipage}
\vspace{0.2cm}
\caption{Per-trigger near-side pair yield (top row), per-trigger away-side pair yield (second row), per-trigger pair yield in the combinatorial background (third row), average number of trigger particles and average number of uncorrelated seeds (bottom row) measured at $\sqrts=7$\,TeV.} 
\label{fig:observables7}
\end{figure}

First, the analysis results for the highest analysed collision energy $\sqrts=7$\,TeV are presented. Next, we discuss the
collision energies $\sqrts=2.76$\,TeV and 0.9\,TeV.
\paragraph{Near-side}
The per-trigger near-side pair-yield which provides information  on the fragmentation of partons is presented in the top 
left panel of figure~\ref{fig:observables7} for $\pTtrig>0.7$\,GeV/$c$ and $\pTassoc>0.4$\,GeV/$c$. 
The measured near-side pair yield grows as a function of the charged particle multiplicity indicating a fragmentation
 bias as characteristic for a MPI distribution with a narrow plateau (tune Perugia-0, see section \ref{sec:methodMPI}).  
This general trend is reproduced by the MC generators. As expected PYTHIA6 tune Perugia-2011 and PYTHIA8 tune 4C, which already 
include LHC data, are closest to the data. 
For $\Nch > 20$ (Perugia-2011) and $\Nch > 30$ (4C) the agreement is within the systematic errors, 
while in this region, all other models  overestimate the data by up to 50\,\%. For all MCs, the agreement becomes worse moving to lower multiplicities. 
Here, Perugia-2011 also overestimates the data by up to 30\,\%. The largest deviations 
(up to 120\,\%) are found in the comparison with PHOJET.

For the higher $\pTassoc$-cut ($> 0.7 \, {\rm GeV}$) the agreement is 
with the exception of PYTHIA6 tune Perugia-2011 and PYTHIA8 at high $N_{\rm ch}$ worse (figure~\ref{fig:observables7} (top right)). In particular, for low multiplicities the deviation is between  
40\,\% and 150\,\%.
\paragraph{Away-side}
The per-trigger away-side pair yield which provides information about the fragments produced back-to-back within the detector acceptance
is presented for $\pTtrig>0.7$\,GeV/$c$ and $\pTassoc>0.4$\,GeV/$c$ in the left panel of the second row of figure~\ref{fig:observables7}. 
As with the near-side yield, the measured away-side pair yield grows as a function of the charged particle multiplicity.
Above $\Nch = 10$, the growth is significantly stronger on the away-side. Surprisingly, tune Perugia-0 now agrees 
with the data within uncertainties over the whole multiplicity range, whereas Perugia-2011 and PYTHIA8, which have the best agreement for the near-side yield, significantly underestimates the away-side yield. 
The deviations of PHOJET
is similar to the ones observed for the near-side.
When increasing the $\pTassoc$-threshold to 0.7\,GeV/$c$ (right panel of the second row of figure~\ref{fig:observables7}), 
also PYTHIA6 tune Perugia-0 overestimates the away-side pair yield by about 30\,\%, whereas tune Perugia-2011 and PYTHIA8
show the best agreement at high $\Nch$.

\paragraph{Combinatorial background}
The per-trigger pair yield in the constant combinatorial background of the correlation grows linearly as a function of the charged 
particle multiplicity as shown in the third row of figure~\ref{fig:observables7}. 
The data are well described by all models within the systematic uncertainties for all charged particle multiplicities for $\pTassoc>0.4$\,GeV/$c$ (left panel).
When increasing the $\pTassoc$-threshold to 0.7\,GeV/$c$ (right panel), PHOJET underestimates the combinatorial 
background by approximately 20\,\%.

\paragraph{Trigger particles per event}
The average number of trigger particles with $\pTtrig>0.7$\,GeV/$c$ as a function of the charged particle multiplicity 
is presented in the bottom-left panel of figure~\ref{fig:observables7}. 
The average number of trigger particles grows stronger than linearly as a function of the charged particle multiplicity.
This can be understood from the pair yield results.
As the multiplicity increases, both the number of semi-hard scatterings per event and the number of fragments per scattering increase, leading to a greater than linear increase in the number of particles above a given $\pT$-threshold.
This observation is also consistent with the observed increase of the mean transverse momentum with multiplicity~\cite{Aamodt:2010my}.
The PYTHIA6 tunes slightly overestimate the ALICE results while PHOJET underestimates the data.
The agreement with PYTHIA8 is excellent for $\Nch > 15$.
\paragraph{Number of uncorrelated seeds}
The average number of uncorrelated seeds (c.f.\ equation~\ref{eq:uncorr}) is presented in the bottom right panel of figure~\ref{fig:observables7}.
At low multiplicities, the number of uncorrelated seeds grows almost linearly. At high multiplicities, the growth decreases.
All models reproduce the qualitative development of the number of correlated seeds as a function of the charged particle multiplicity.
While the data are significantly underestimated by PHOJET, PYTHIA6 and PYTHIA8 reproduce the results reasonably well. 
\\

\begin{figure}[p]
\centering
\vspace{-0.3cm}
\begin{minipage}{0.49\textwidth}
\includegraphics[width=\textwidth]{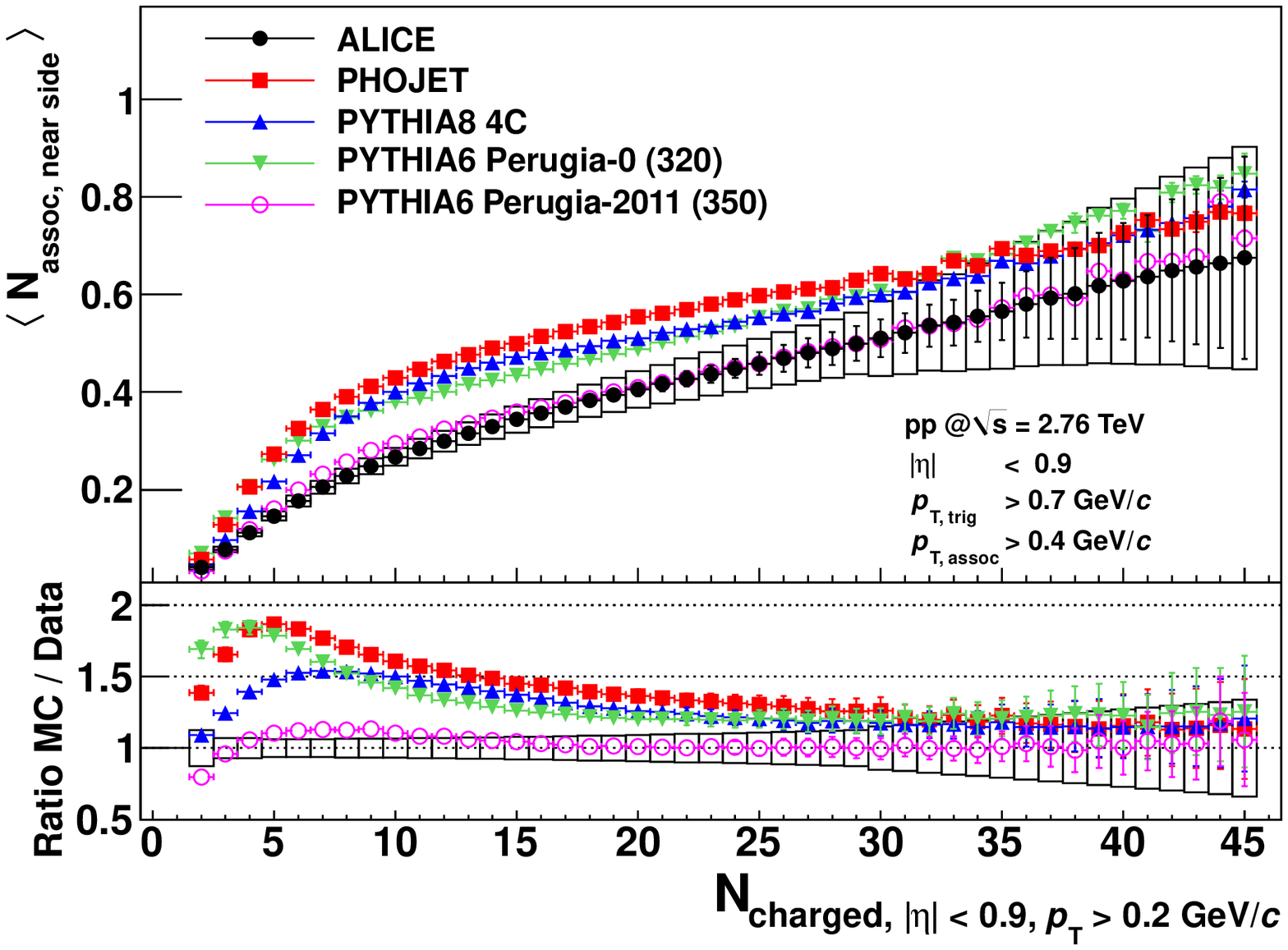} 
\end{minipage}
\begin{minipage}{0.49\textwidth}
\includegraphics[width=\textwidth]{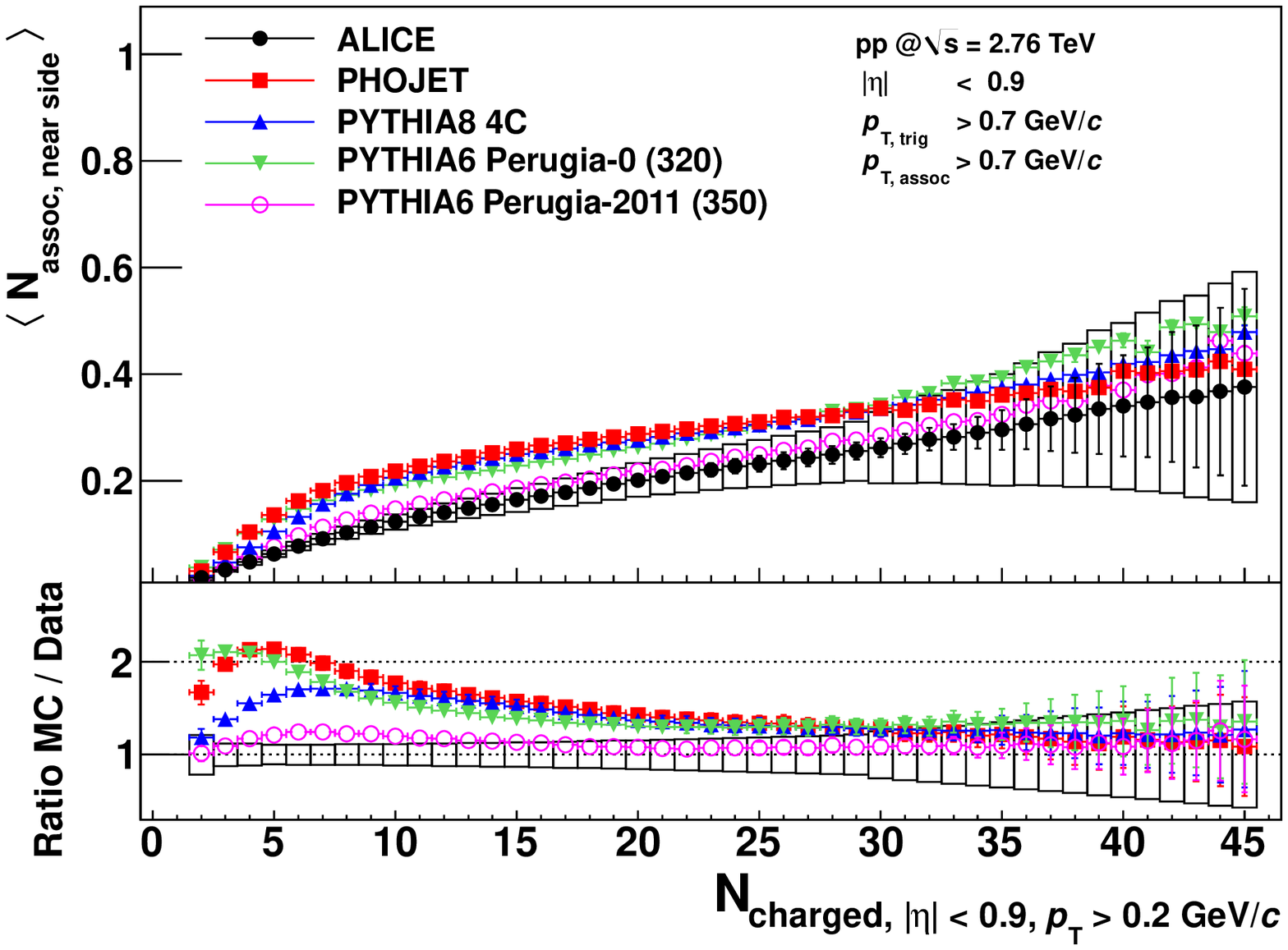} 
\end{minipage}
\vspace{-0.1cm}
\begin{minipage}{0.49\textwidth}
\includegraphics[width=\textwidth]{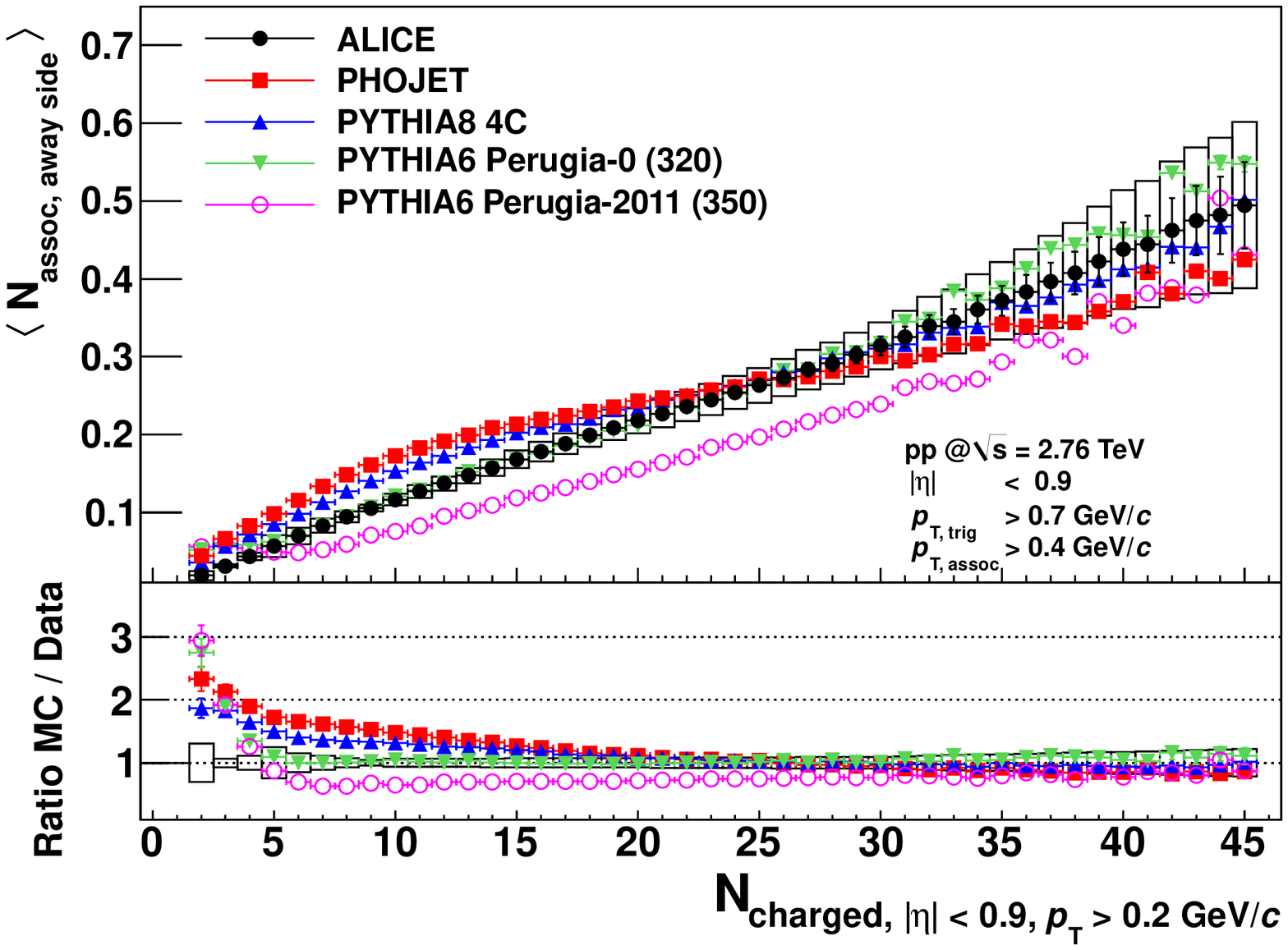} 
\end{minipage}
\begin{minipage}{0.49\textwidth}
\includegraphics[width=\textwidth]{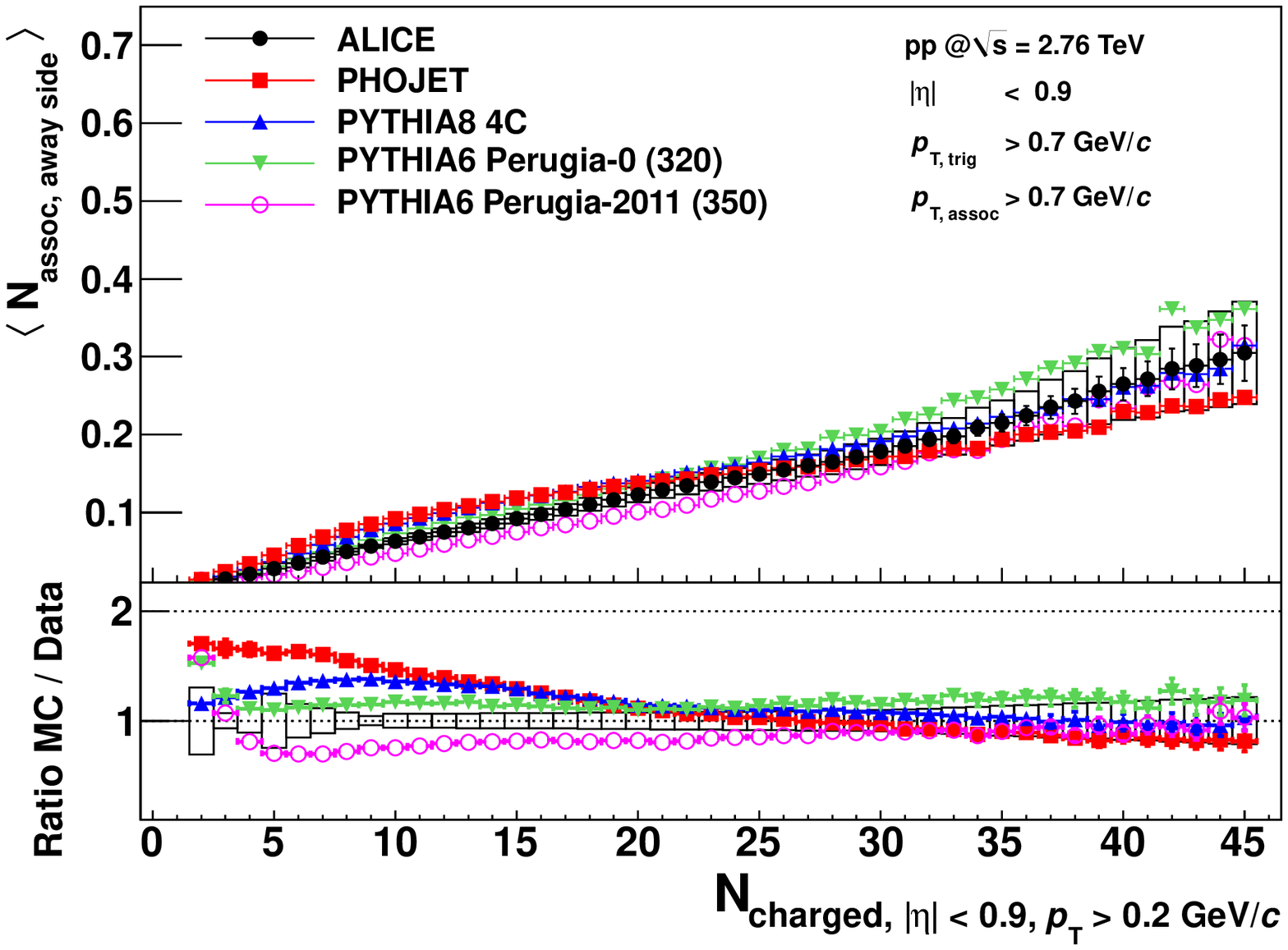} 
\end{minipage}
\vspace{-0.1cm}
\begin{minipage}{0.49\textwidth}
\includegraphics[width=\textwidth]{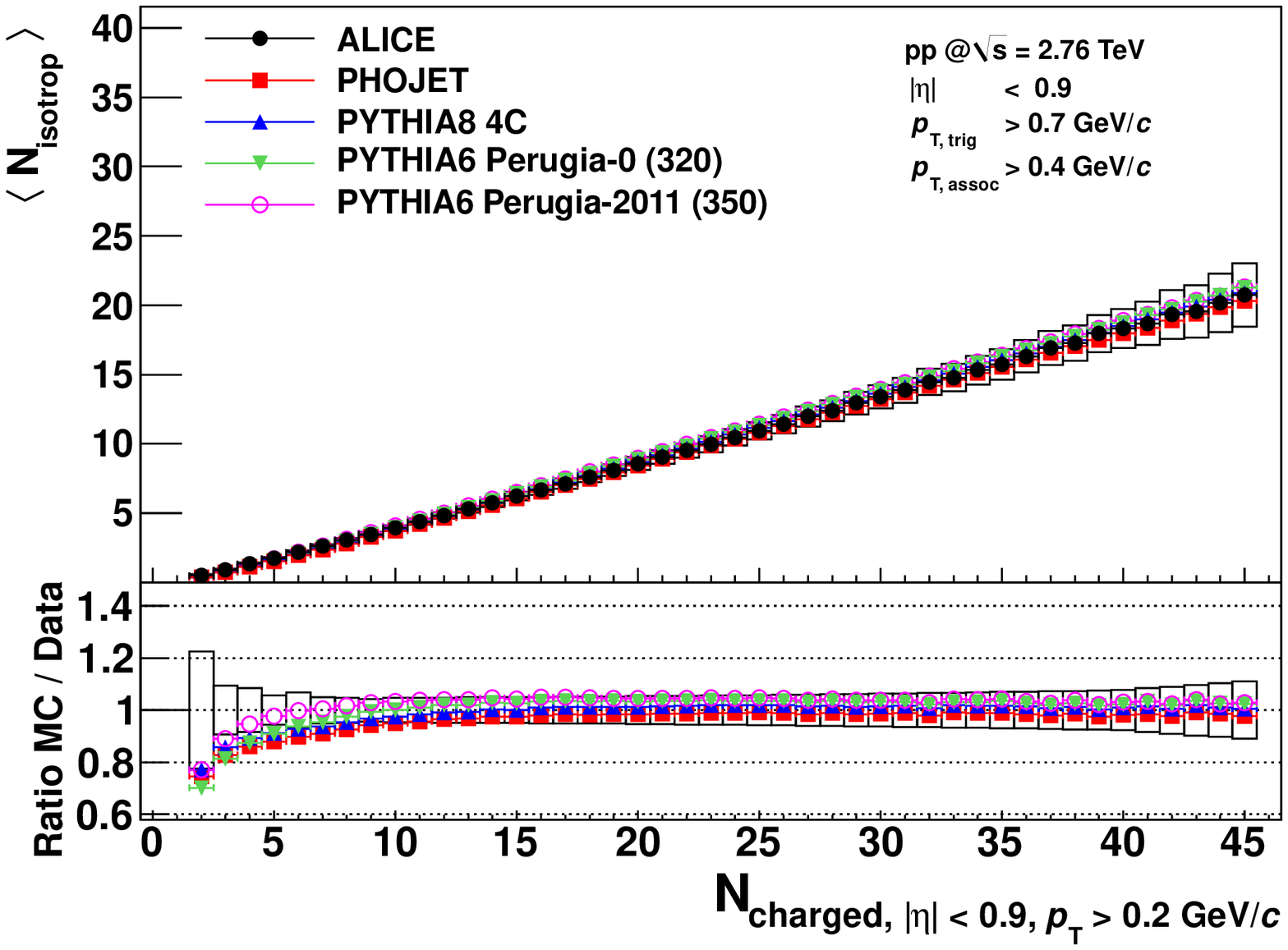} 
\end{minipage}
\begin{minipage}{0.49\textwidth}
\includegraphics[width=\textwidth]{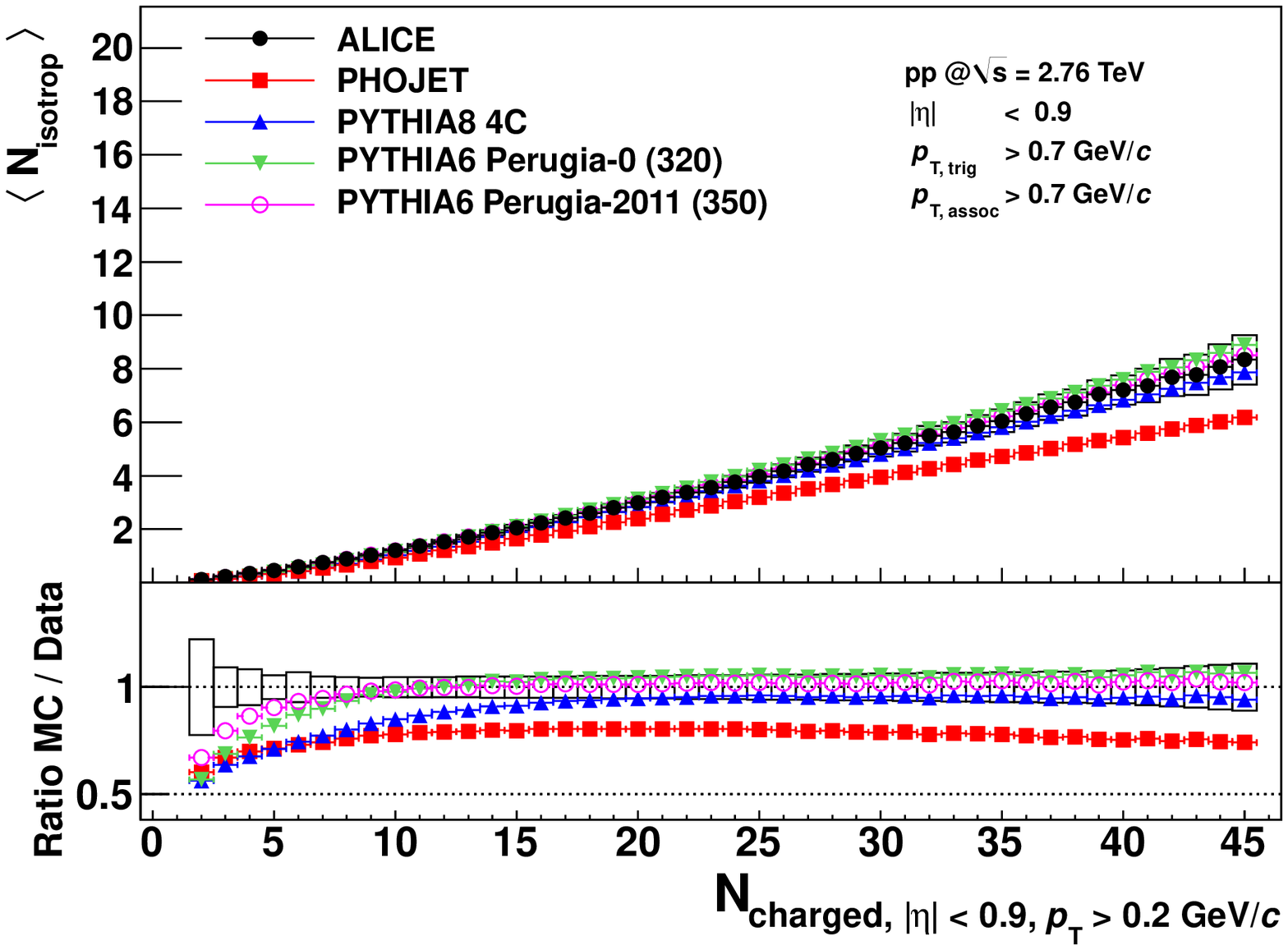} 
\end{minipage}
\vspace{-0.1cm}
\begin{minipage}{0.49\textwidth}
\includegraphics[width=\textwidth]{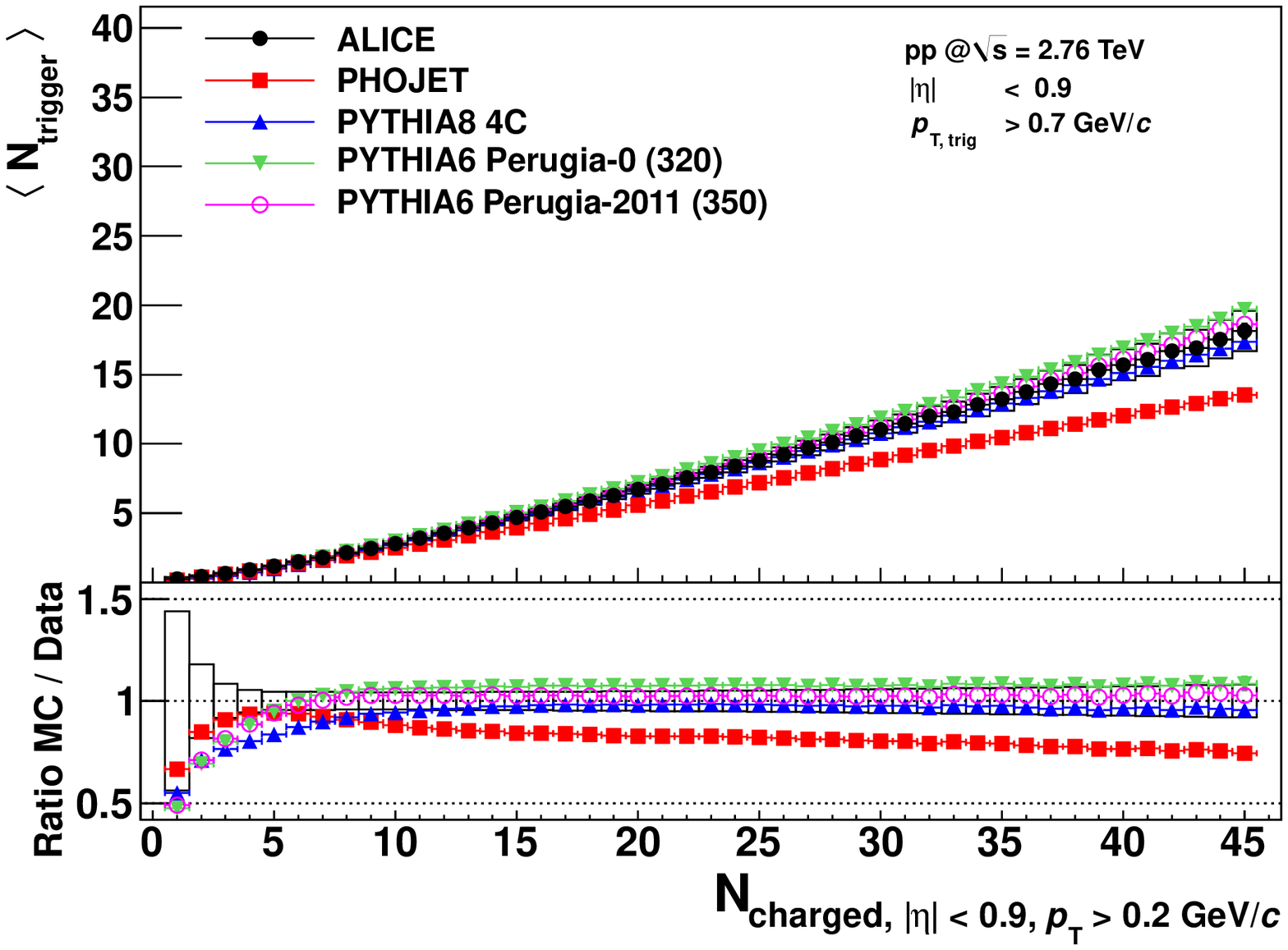} 
\end{minipage}
\begin{minipage}{0.49\textwidth}
\includegraphics[width=\textwidth]{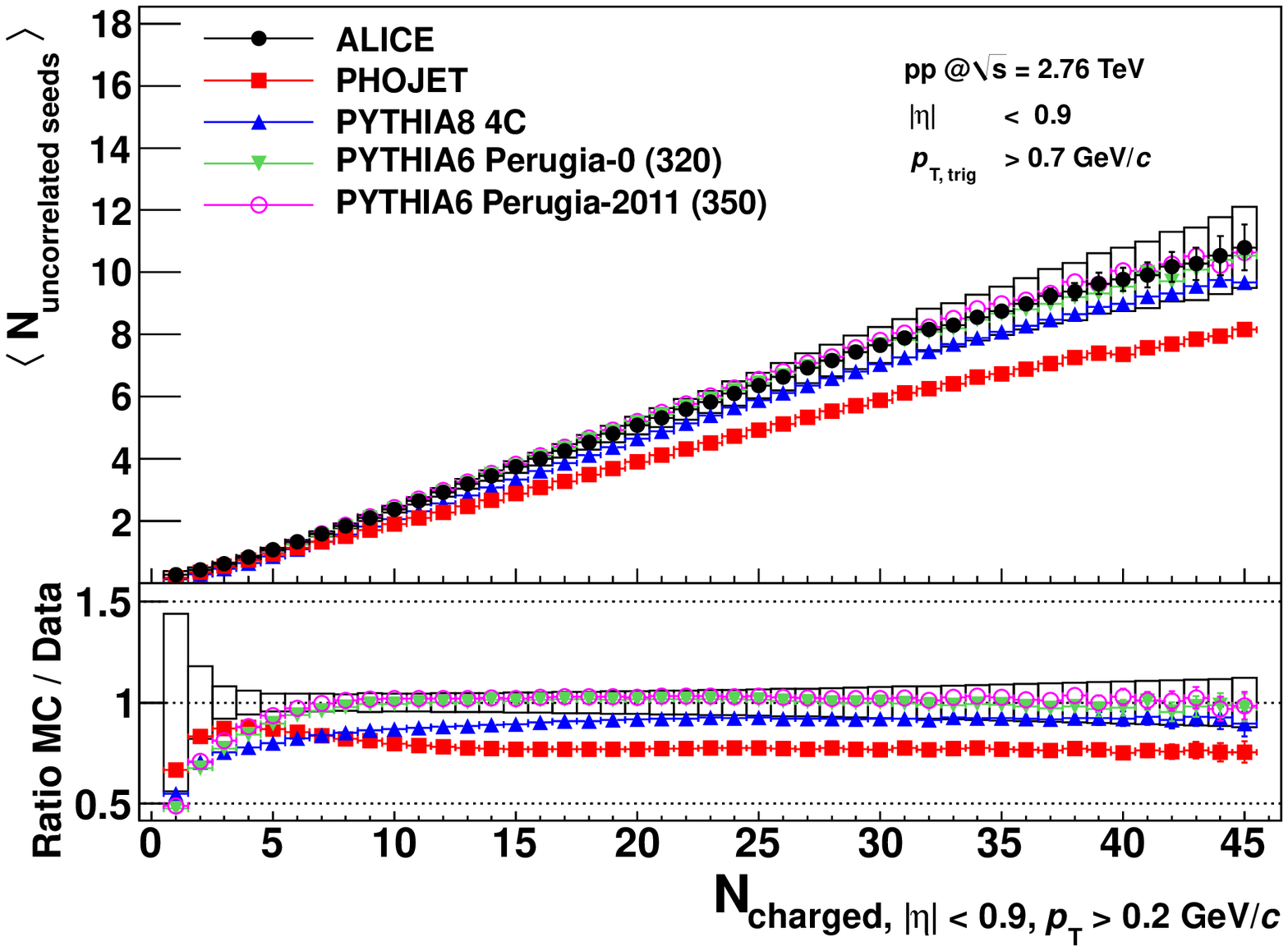} 
\end{minipage}
\vspace{0.2cm}
\caption{Per-trigger near-side pair yield (top row), per-trigger away-side pair yield (second row), per-trigger pair yield in the combinatorial background (third row), average number of trigger particles and average number of uncorrelated seeds (bottom row) measured at $\sqrts=2.76$\,TeV.} 
\label{fig:observables2}
\end{figure}

\begin{figure}[p]
\centering
\vspace{-0.3cm}
\begin{minipage}{0.49\textwidth}
\includegraphics[width=\textwidth]{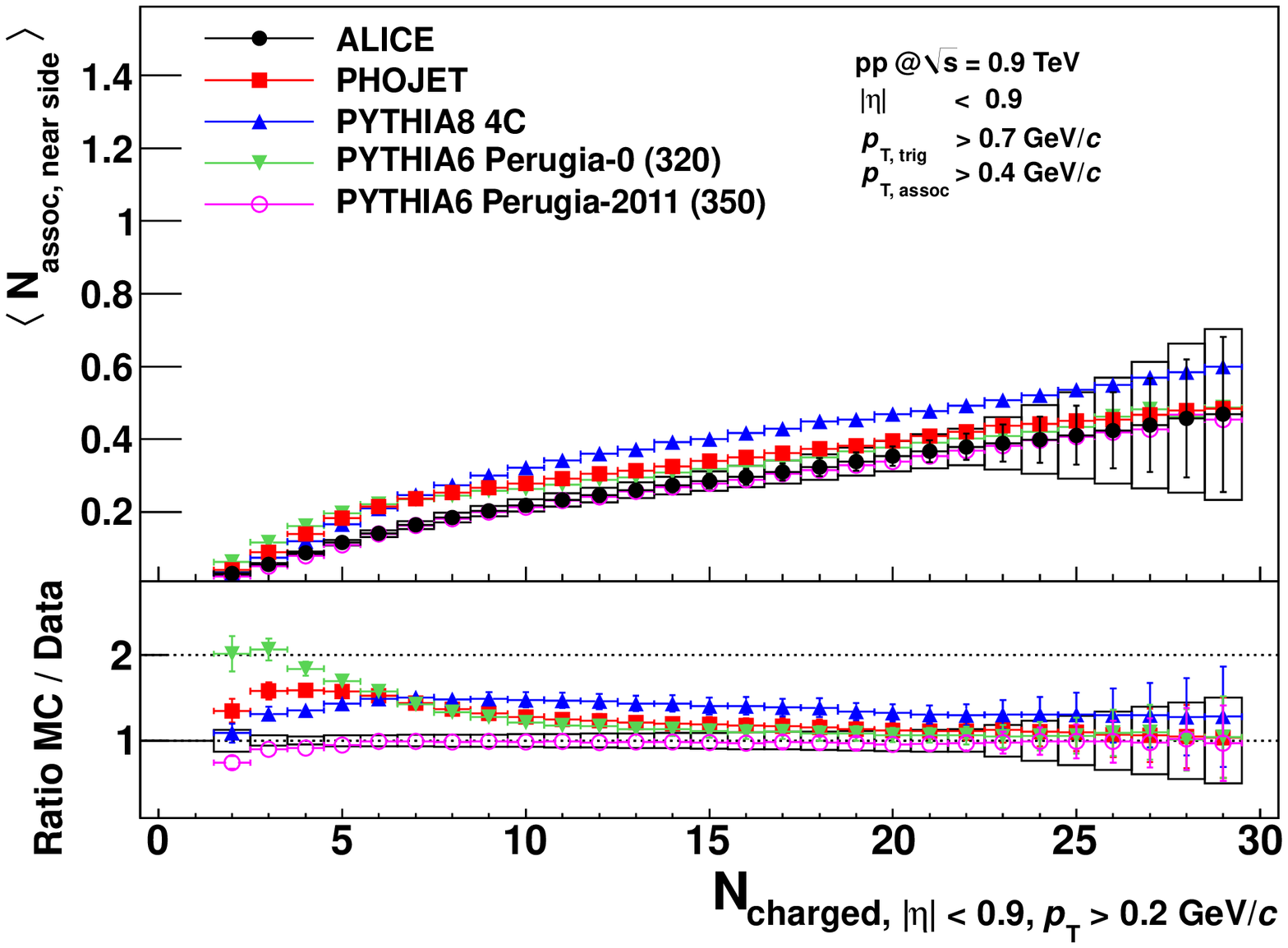} 
\end{minipage}
\begin{minipage}{0.49\textwidth}
\includegraphics[width=\textwidth]{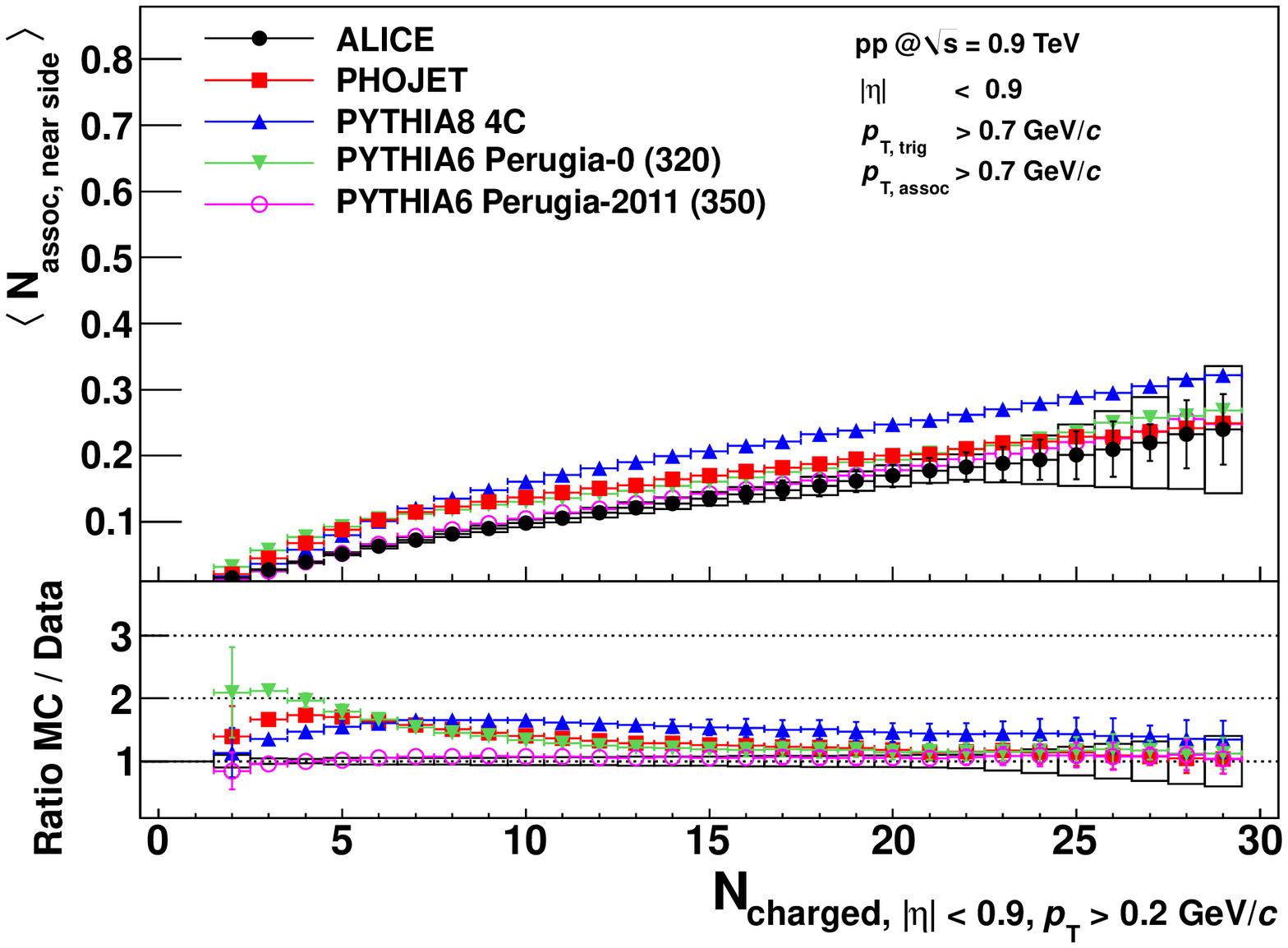} 
\end{minipage}
\begin{minipage}{0.49\textwidth}
\includegraphics[width=\textwidth]{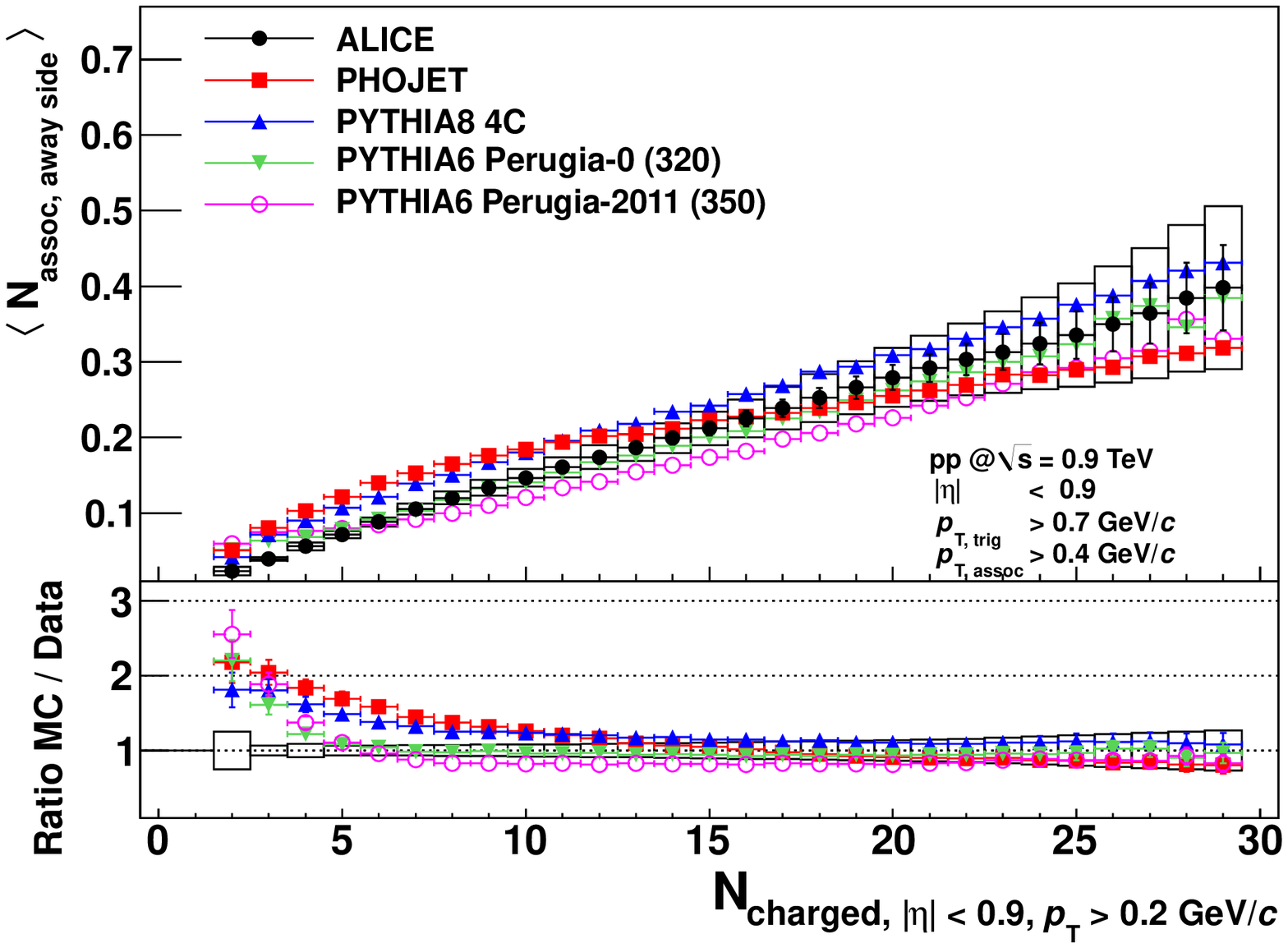} 
\end{minipage}
\begin{minipage}{0.49\textwidth}
\includegraphics[width=\textwidth]{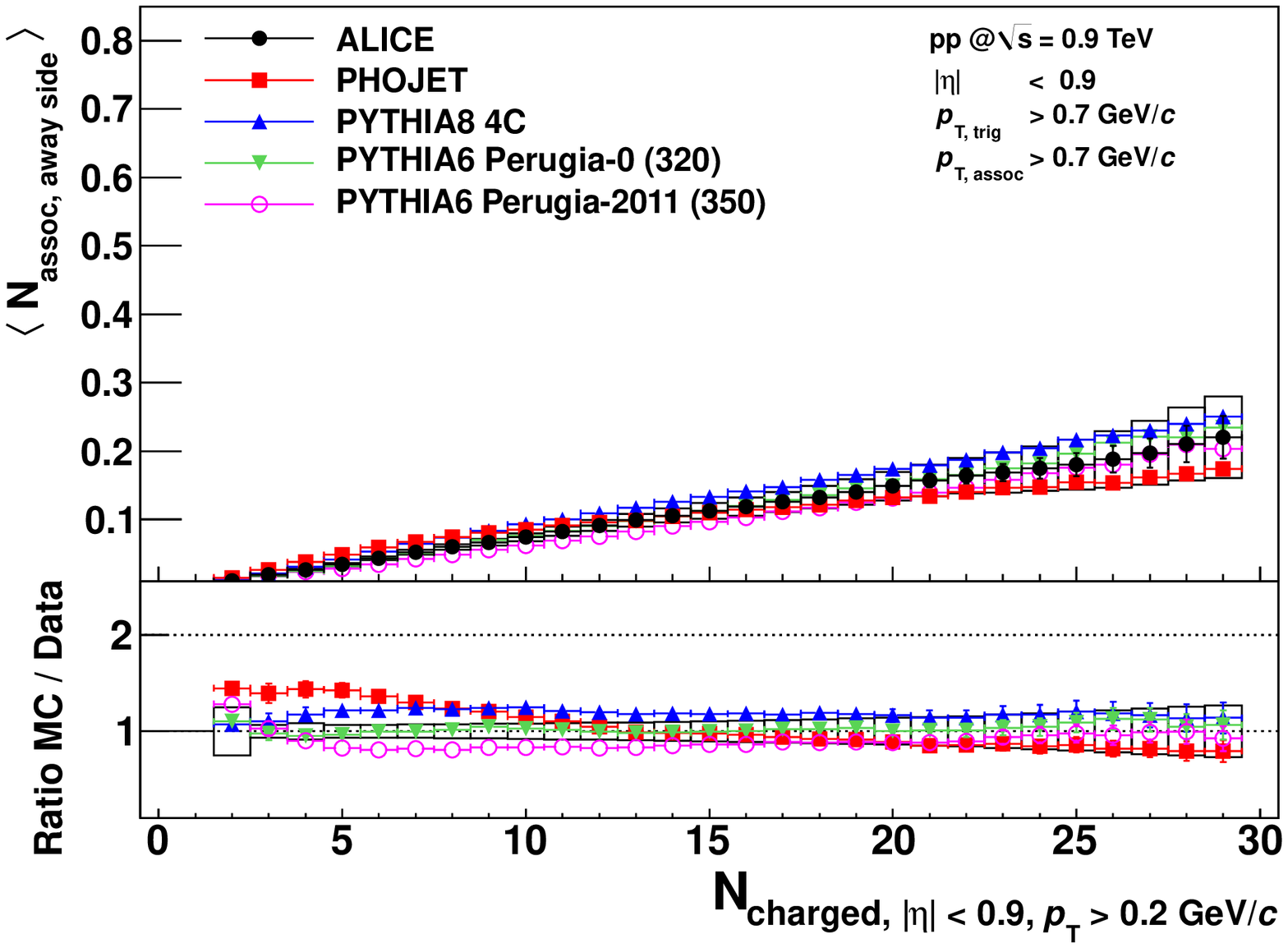} 
\end{minipage}
\vspace{-0.1cm}
\begin{minipage}{0.49\textwidth}
\includegraphics[width=\textwidth]{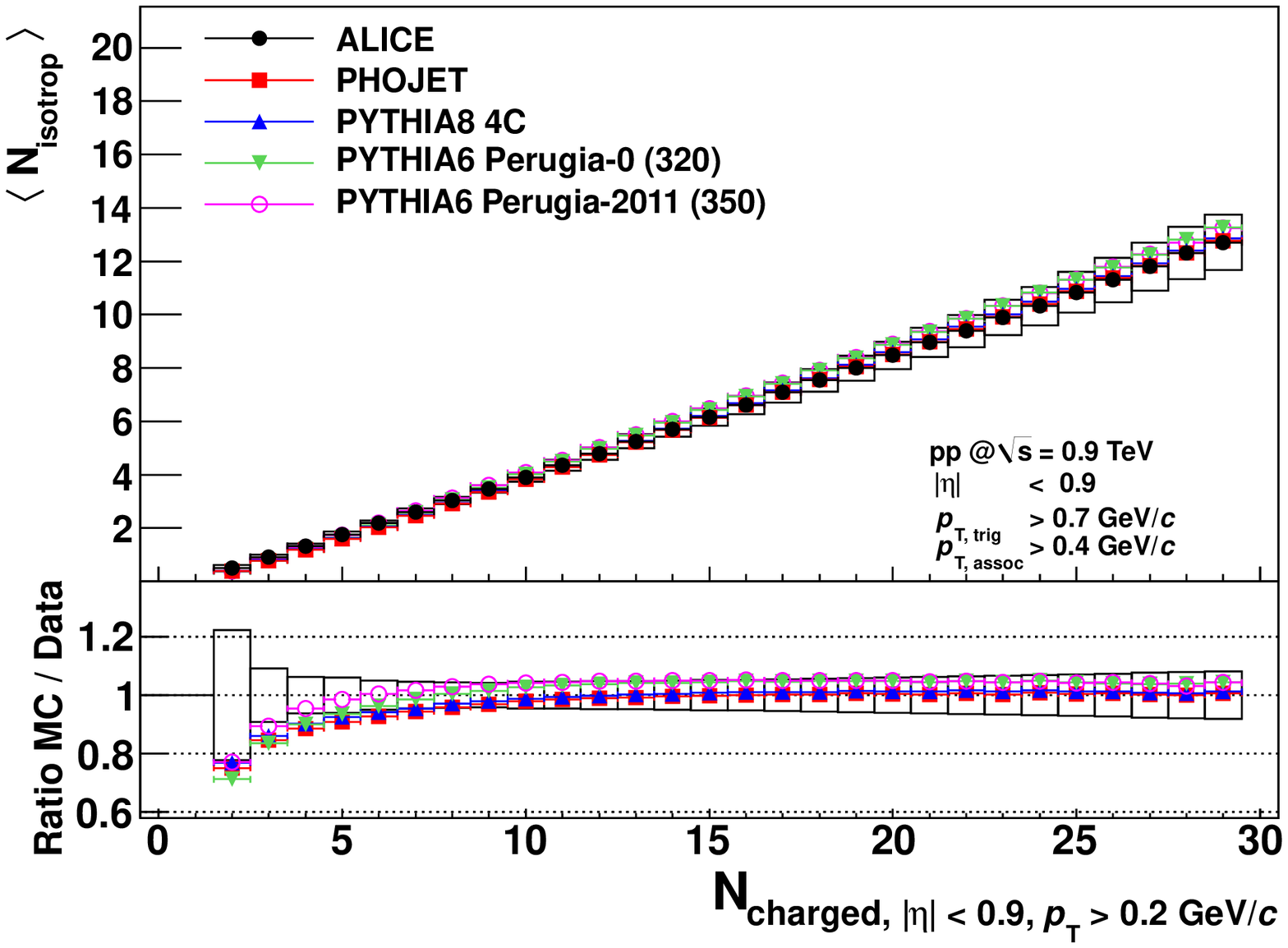} 
\end{minipage}
\begin{minipage}{0.49\textwidth}
\includegraphics[width=\textwidth]{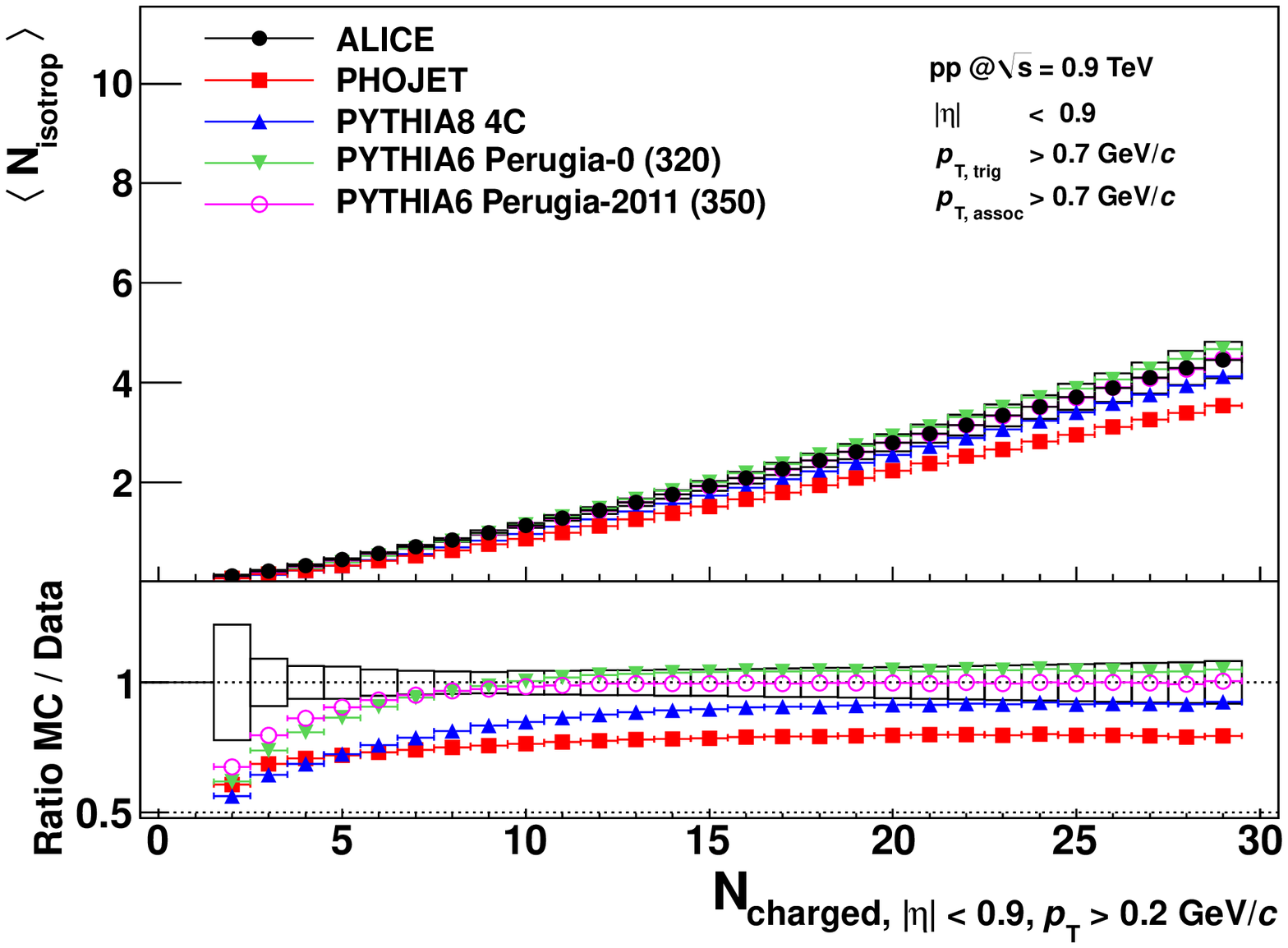} 
\end{minipage}
\begin{minipage}{0.49\textwidth}
\includegraphics[width=\textwidth]{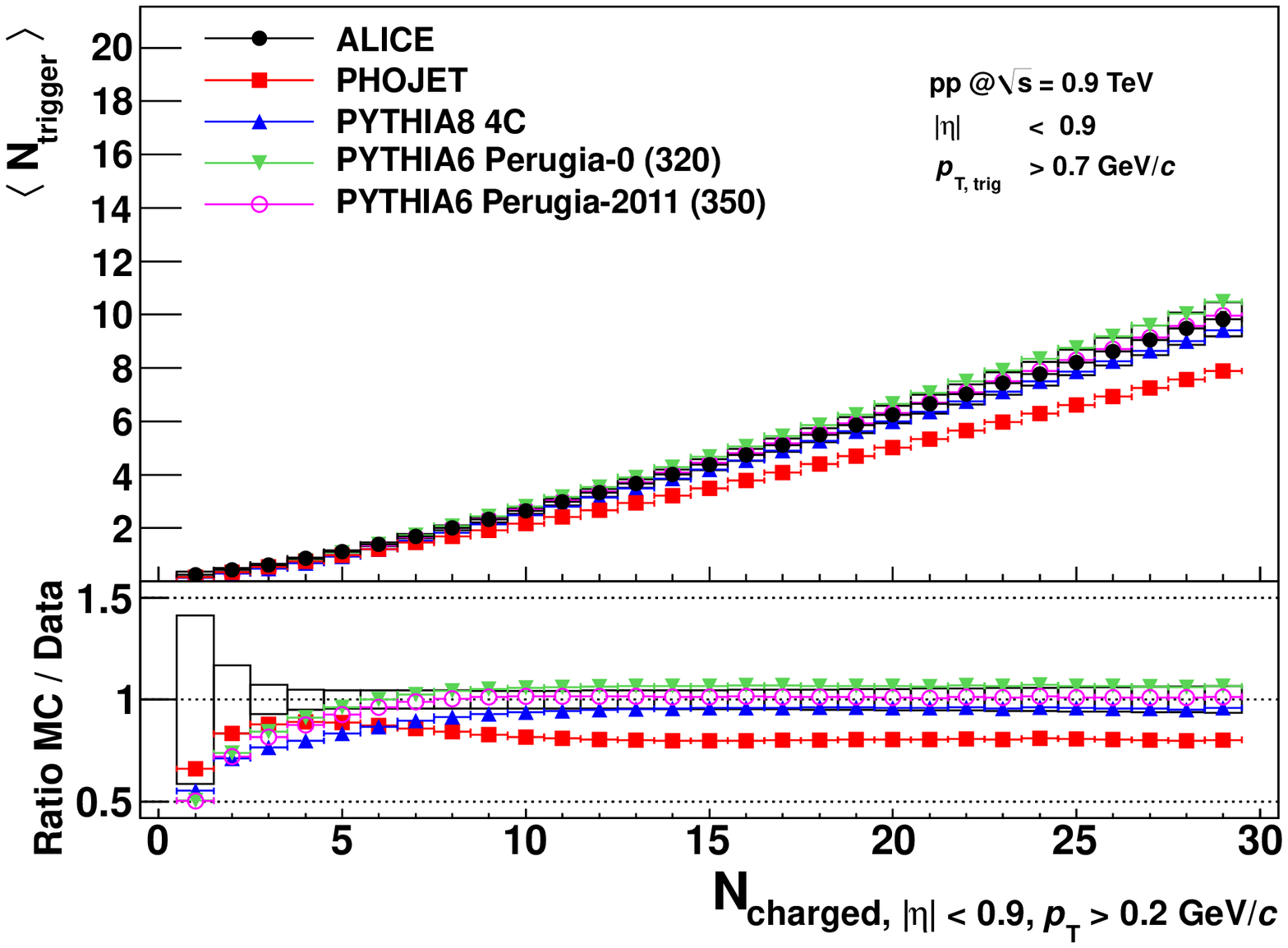} 
\end{minipage}
\begin{minipage}{0.49\textwidth}
\includegraphics[width=\textwidth]{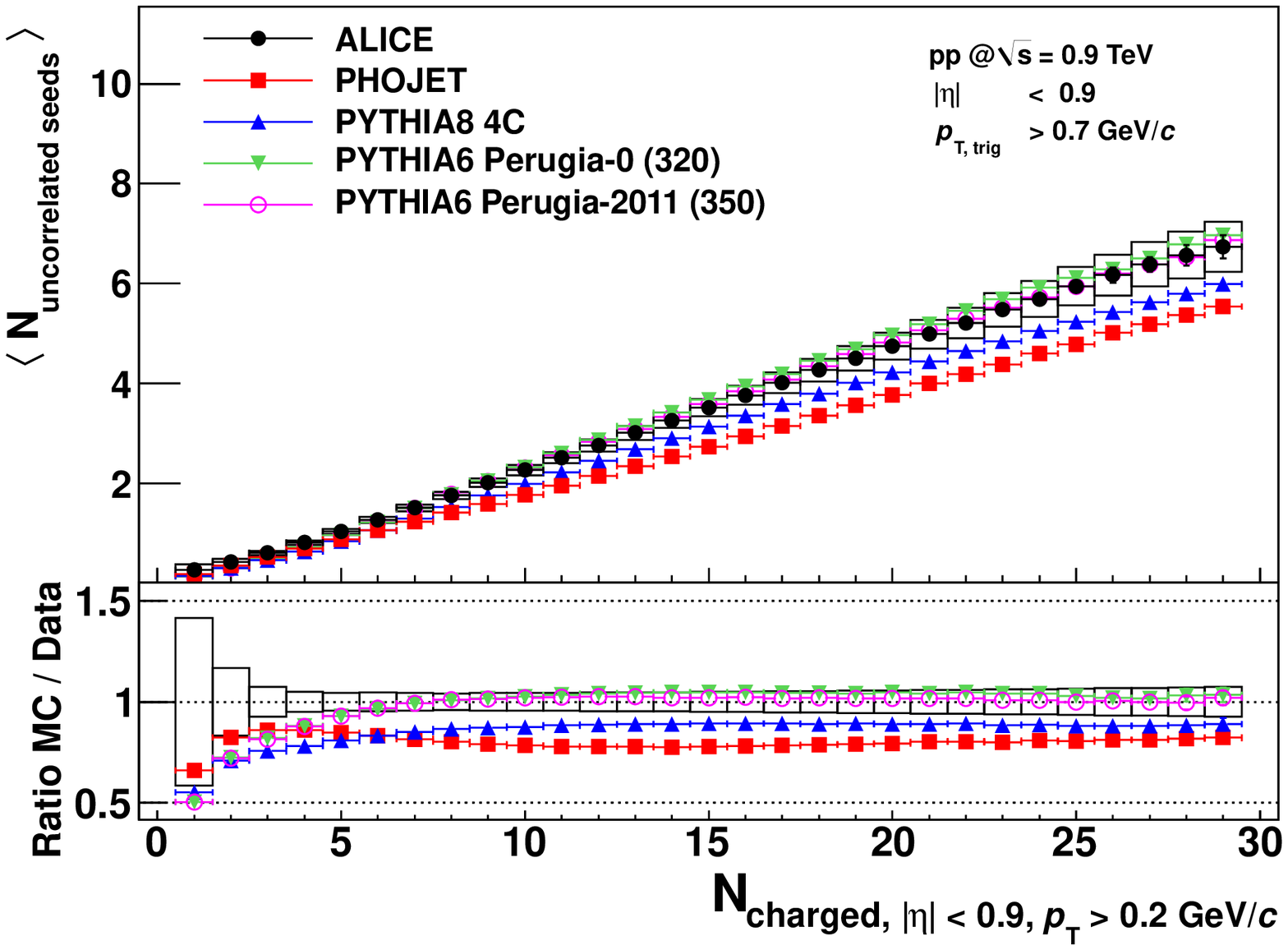} 
\end{minipage}
\vspace{0.2cm}
\caption{Per-trigger near-side pair yield (top row), per-trigger away-side pair yield (second row), per-trigger pair yield in the combinatorial background (third row), average number of trigger particles and average number of uncorrelated seeds (bottom row) measured at $\sqrts=0.9$\,TeV.} 
\label{fig:observables9}
\end{figure}

\subsection{Centre-of-mass energy dependence}
\begin{figure}[p]
\centering
\vspace{-0.3cm}
\begin{minipage}{0.49\textwidth}
\includegraphics[width=\textwidth]{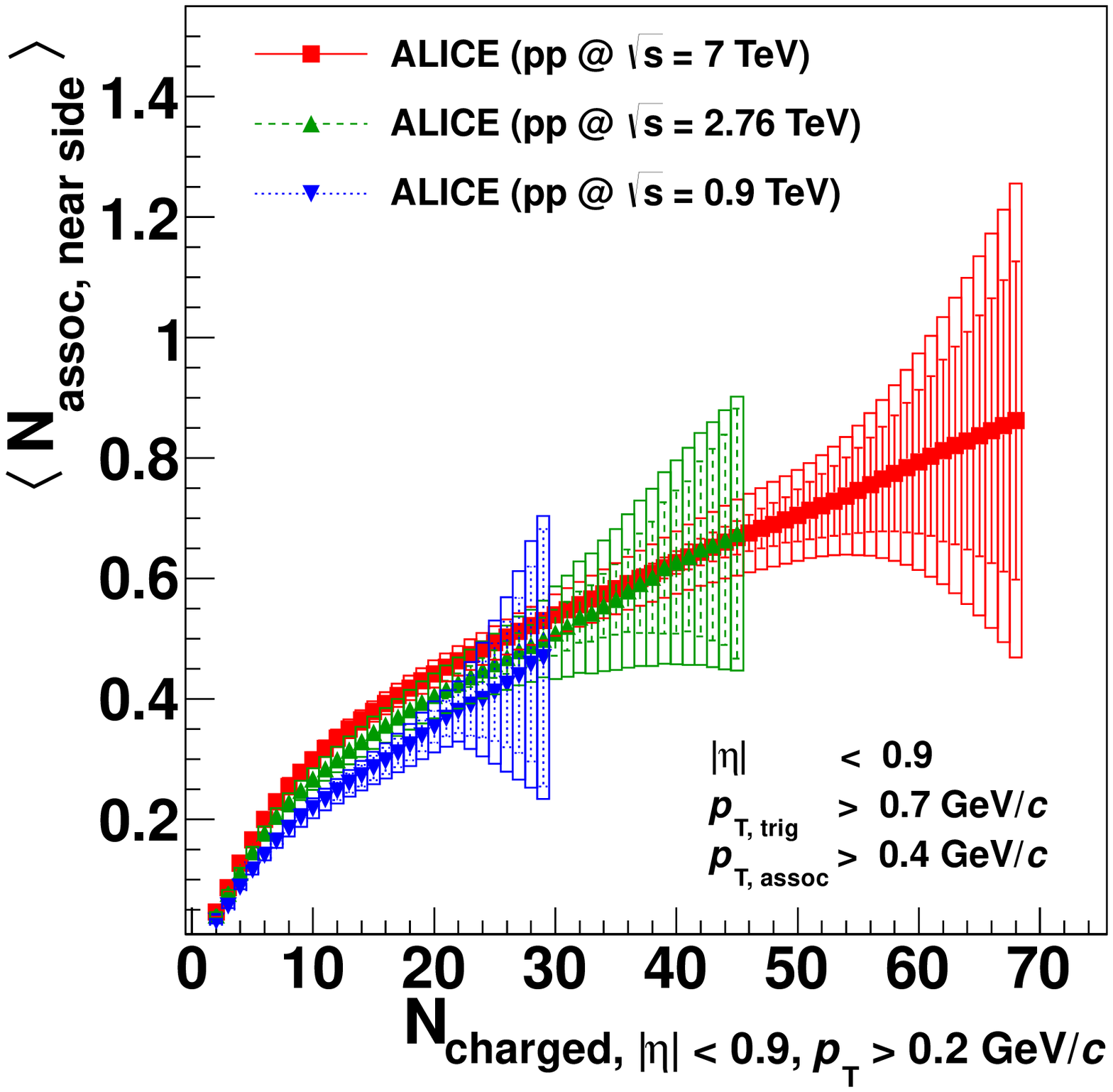} 
\end{minipage}
\begin{minipage}{0.49\textwidth}
\includegraphics[width=\textwidth]{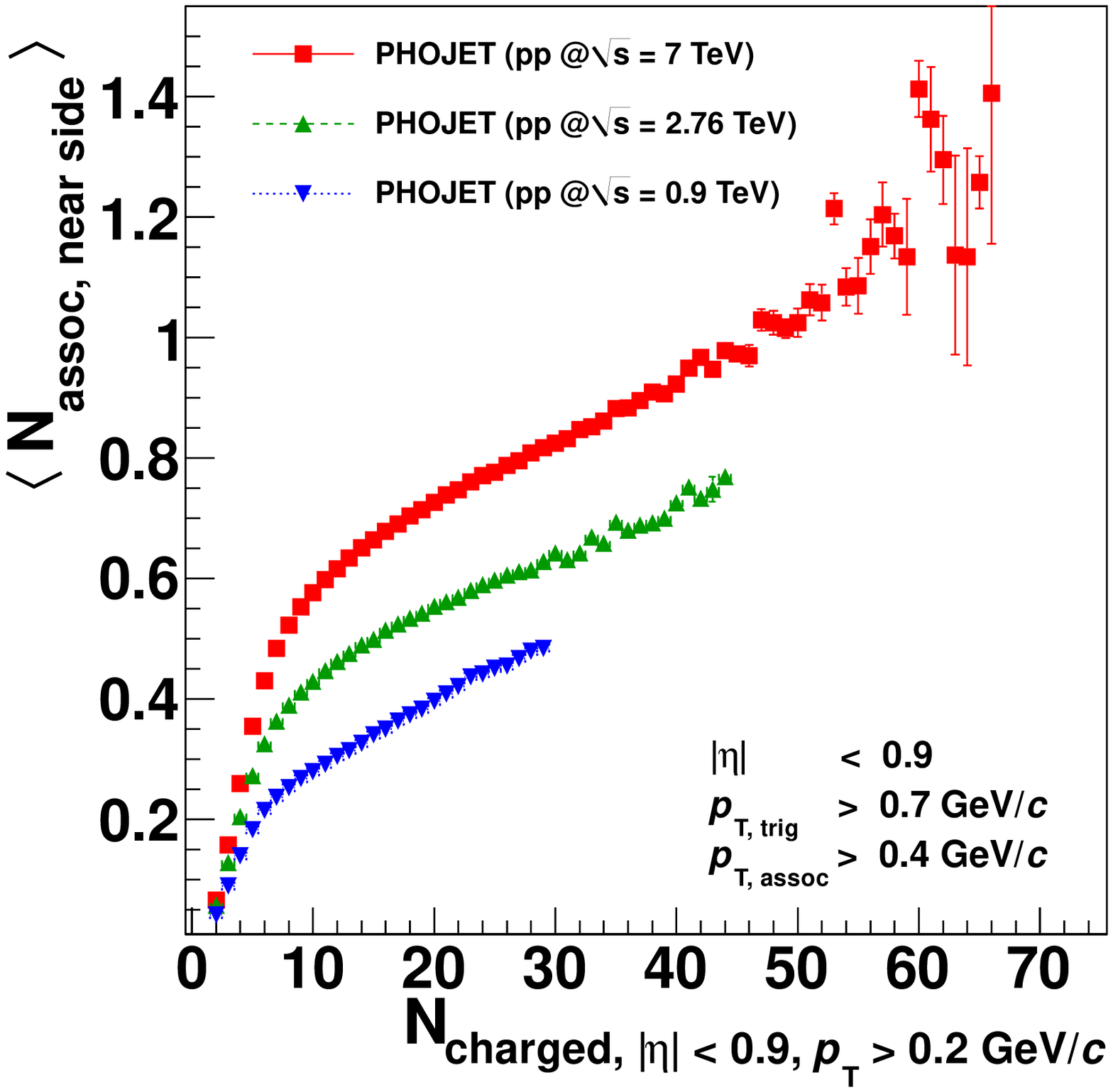} 
\end{minipage}
\begin{minipage}{0.49\textwidth}
\includegraphics[width=\textwidth]{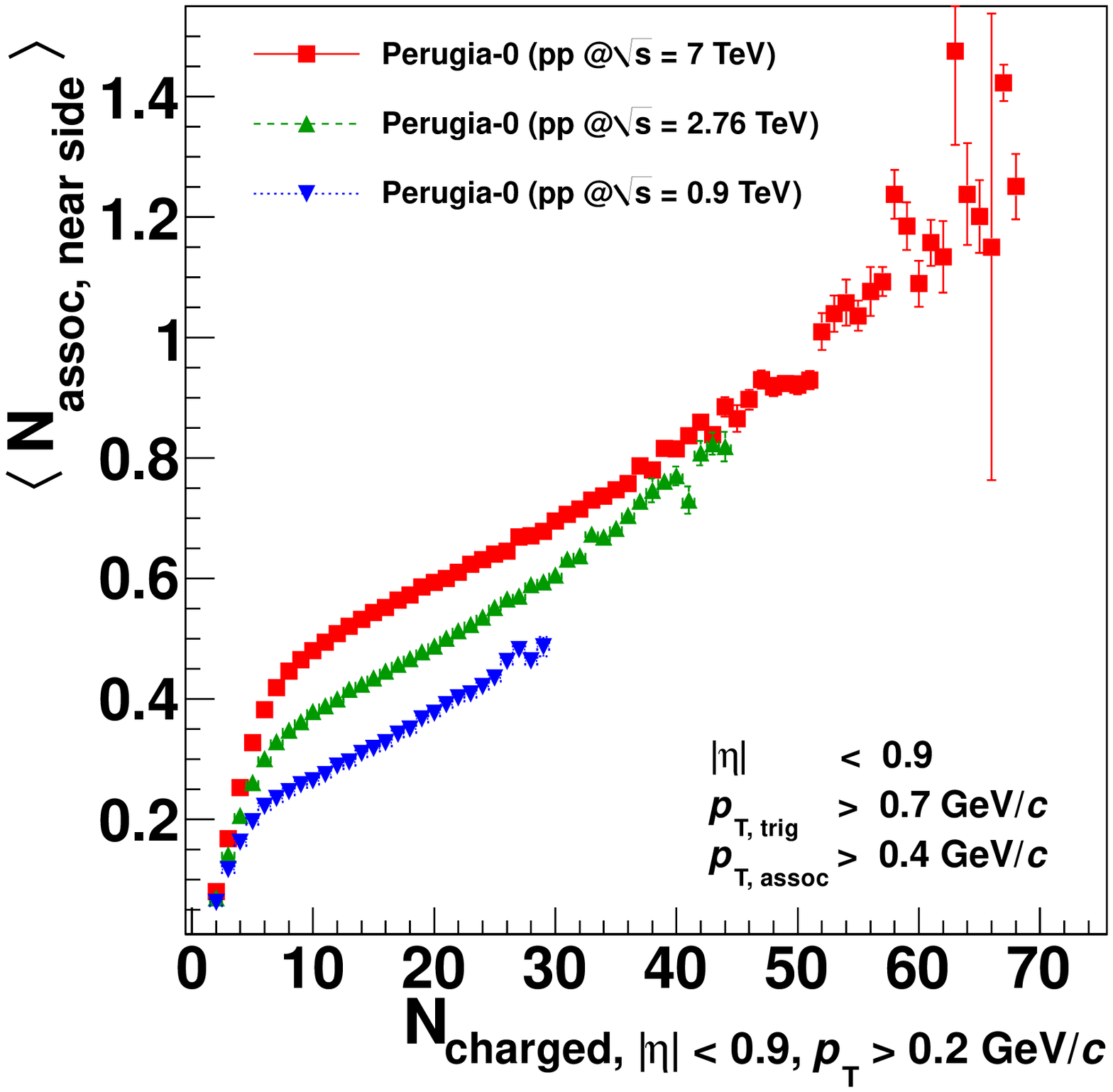} 
\end{minipage}
\vspace{-0.1cm}
\begin{minipage}{0.49\textwidth}
\includegraphics[width=\textwidth]{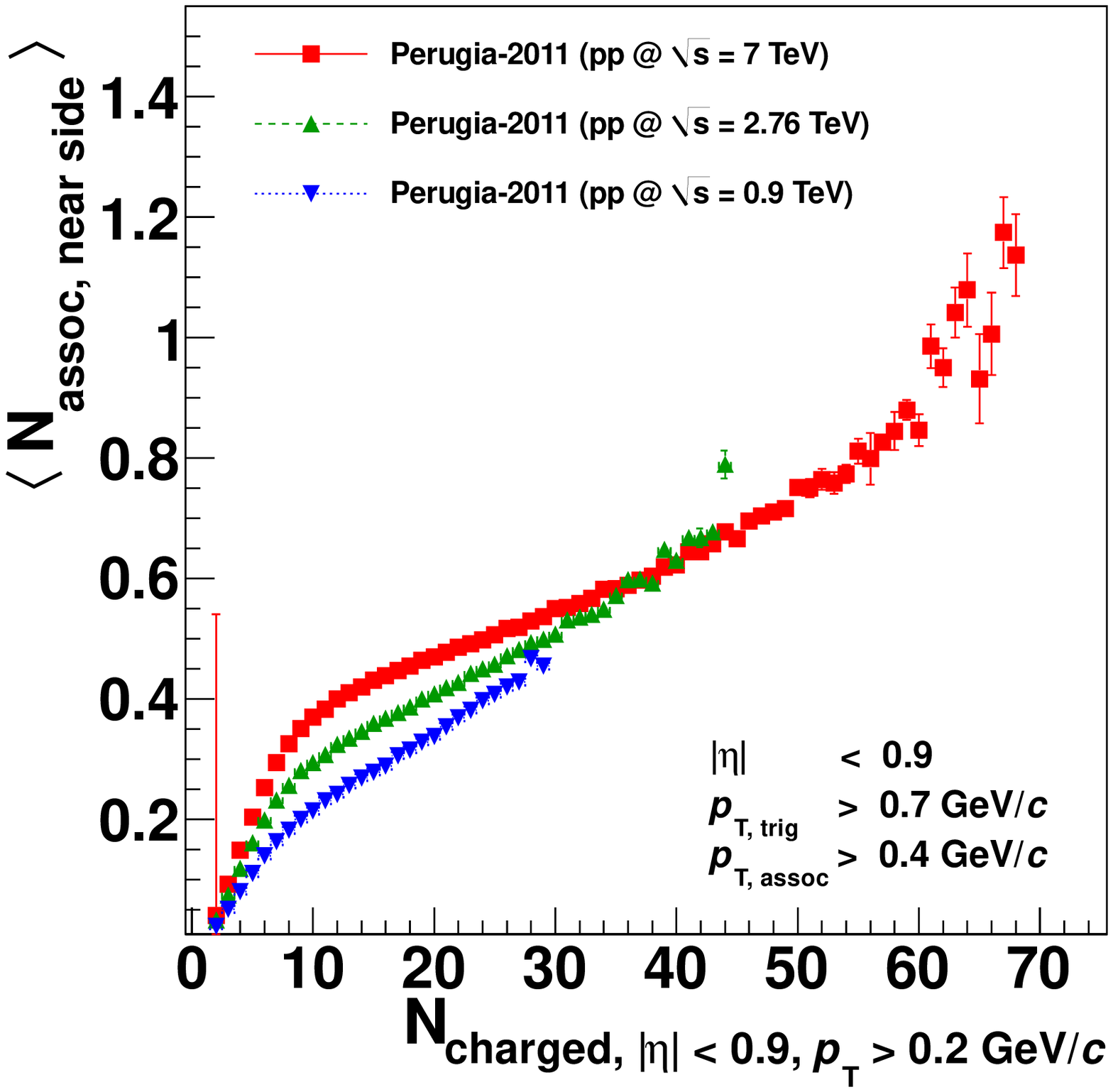} 
\end{minipage}
\begin{minipage}{0.49\textwidth}
\includegraphics[width=\textwidth]{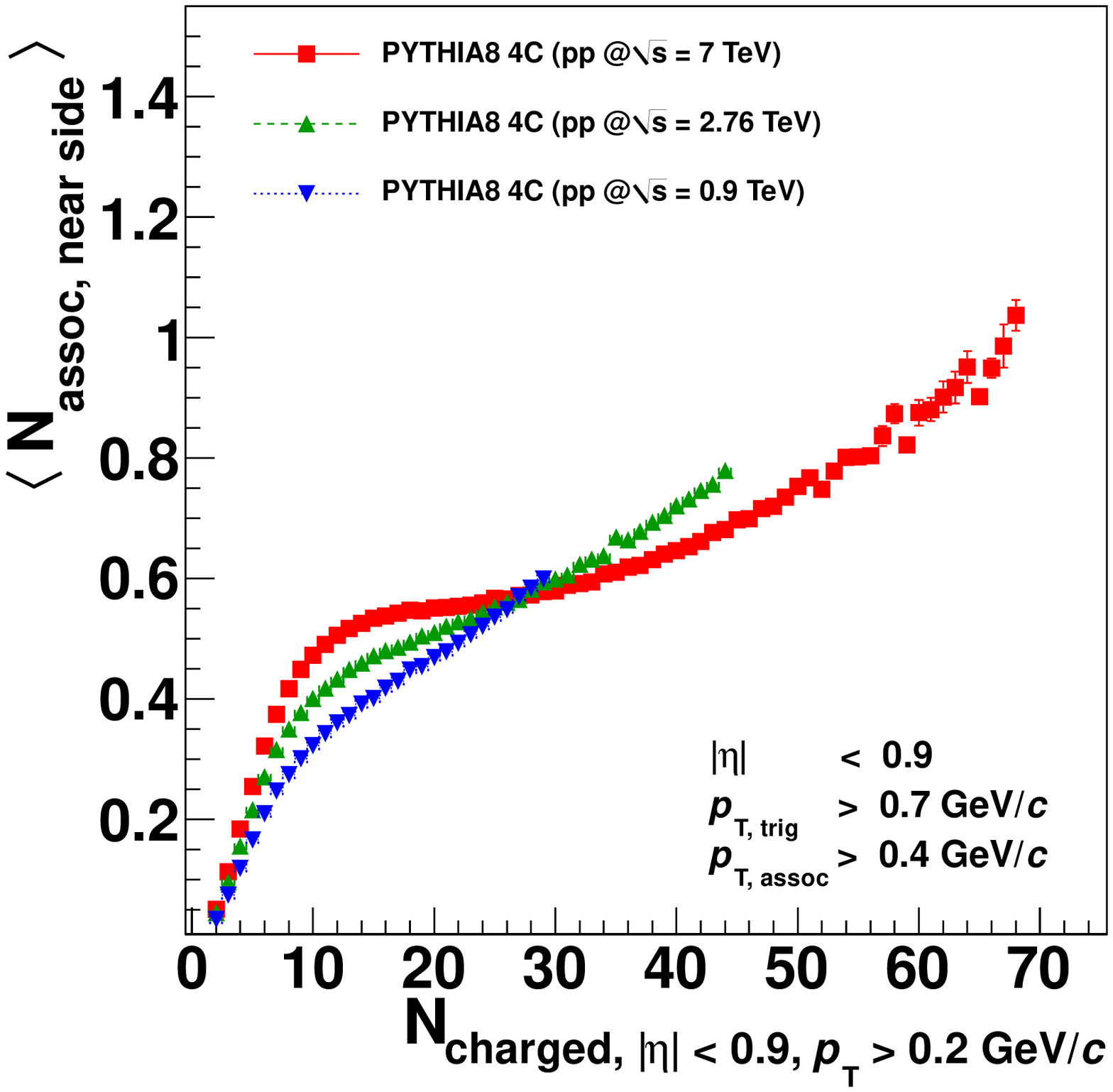} 
\end{minipage}
\caption{Per-trigger near-side pair yield as a function of the charged particle multiplicity measured for $\sqrts=0.9$, 2.76, and 7\,TeV.} 
\label{fig:nearsideCMS}
\end{figure}

\begin{figure}[p]
\centering
\vspace{-0.3cm}
\begin{minipage}{0.49\textwidth}
\includegraphics[width=\textwidth]{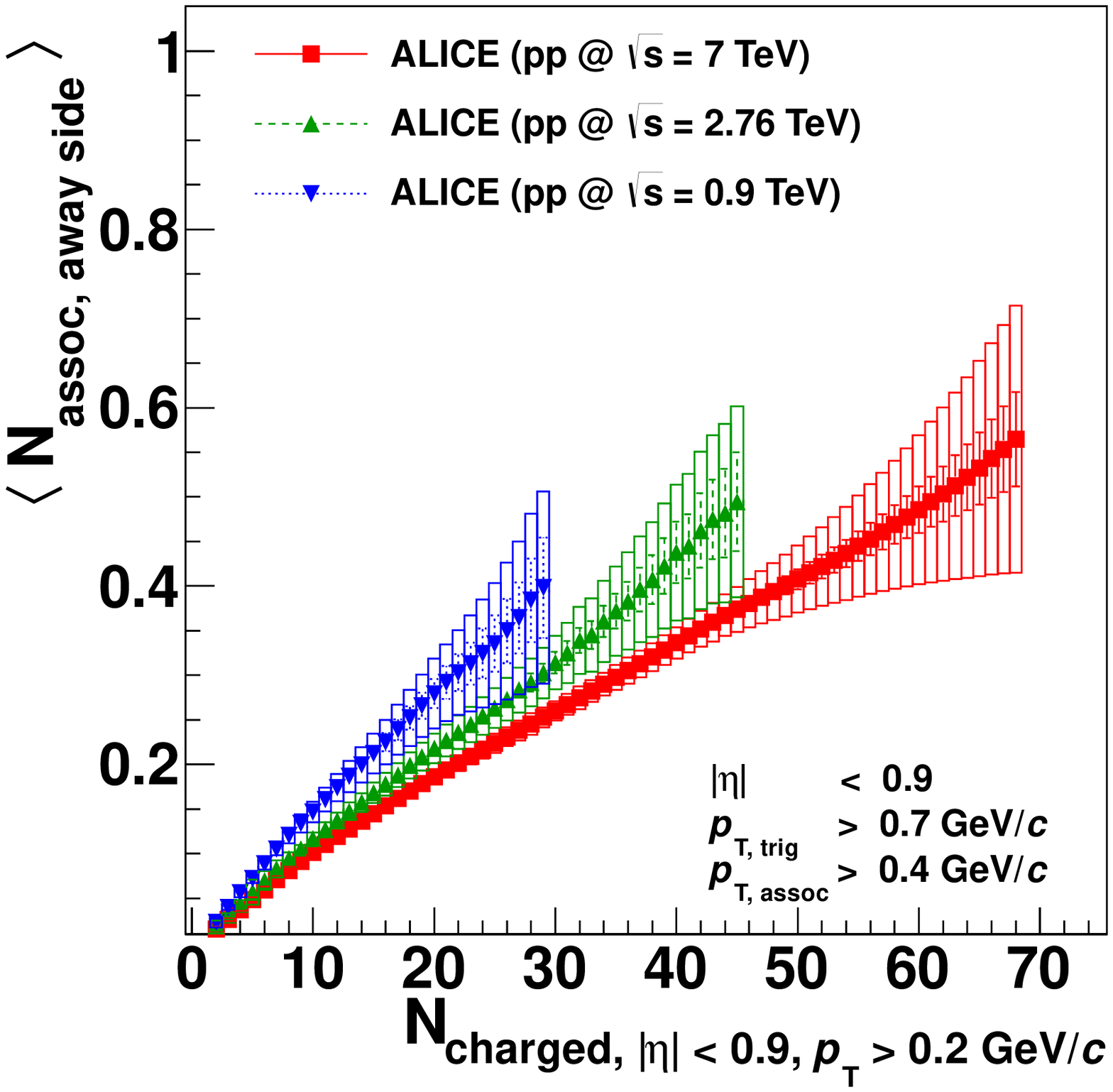} 
\end{minipage}
\begin{minipage}{0.49\textwidth}
\includegraphics[width=\textwidth]{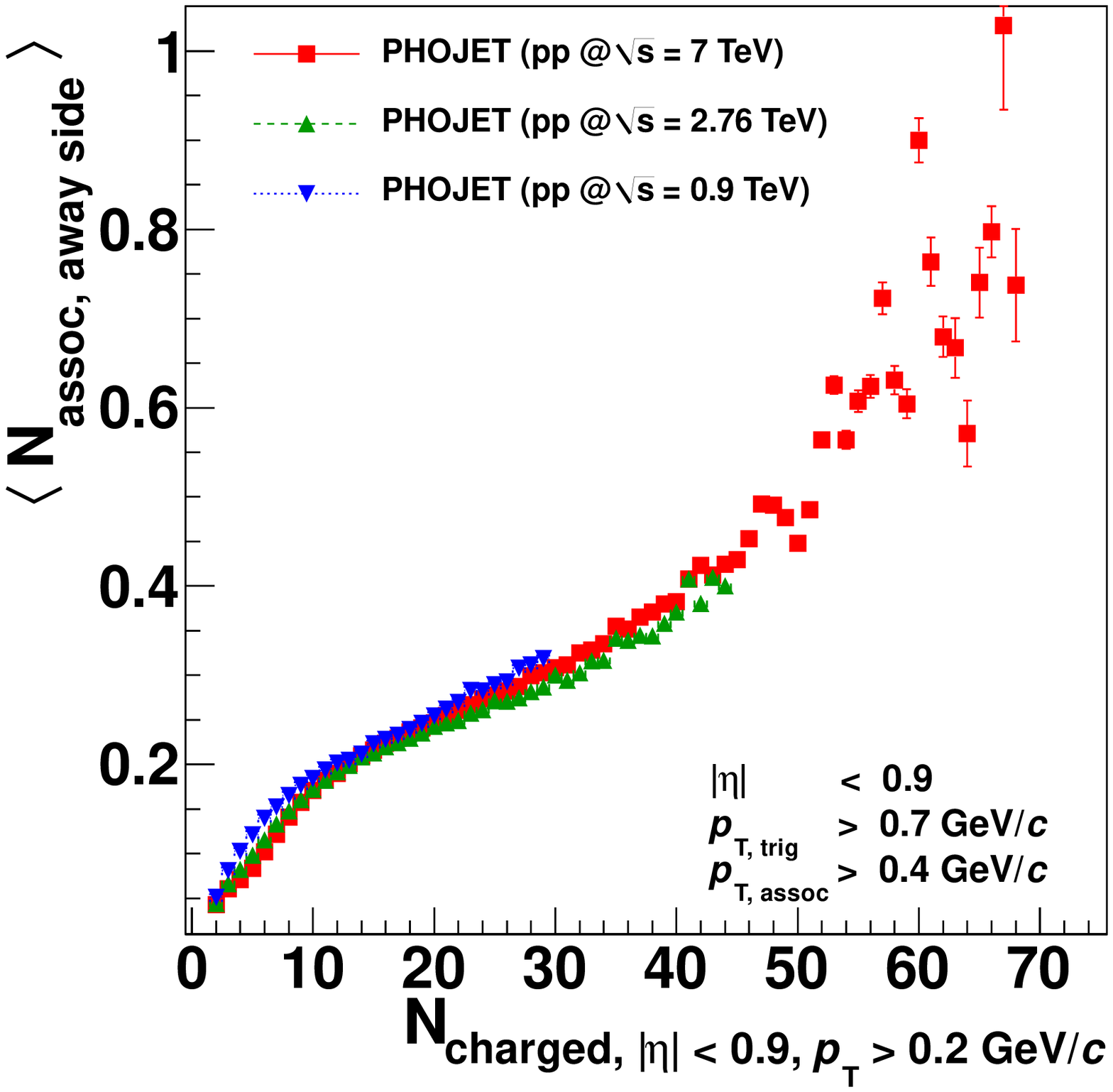} 
\end{minipage}
\vspace{-0.1cm}
\begin{minipage}{0.49\textwidth}
\includegraphics[width=\textwidth]{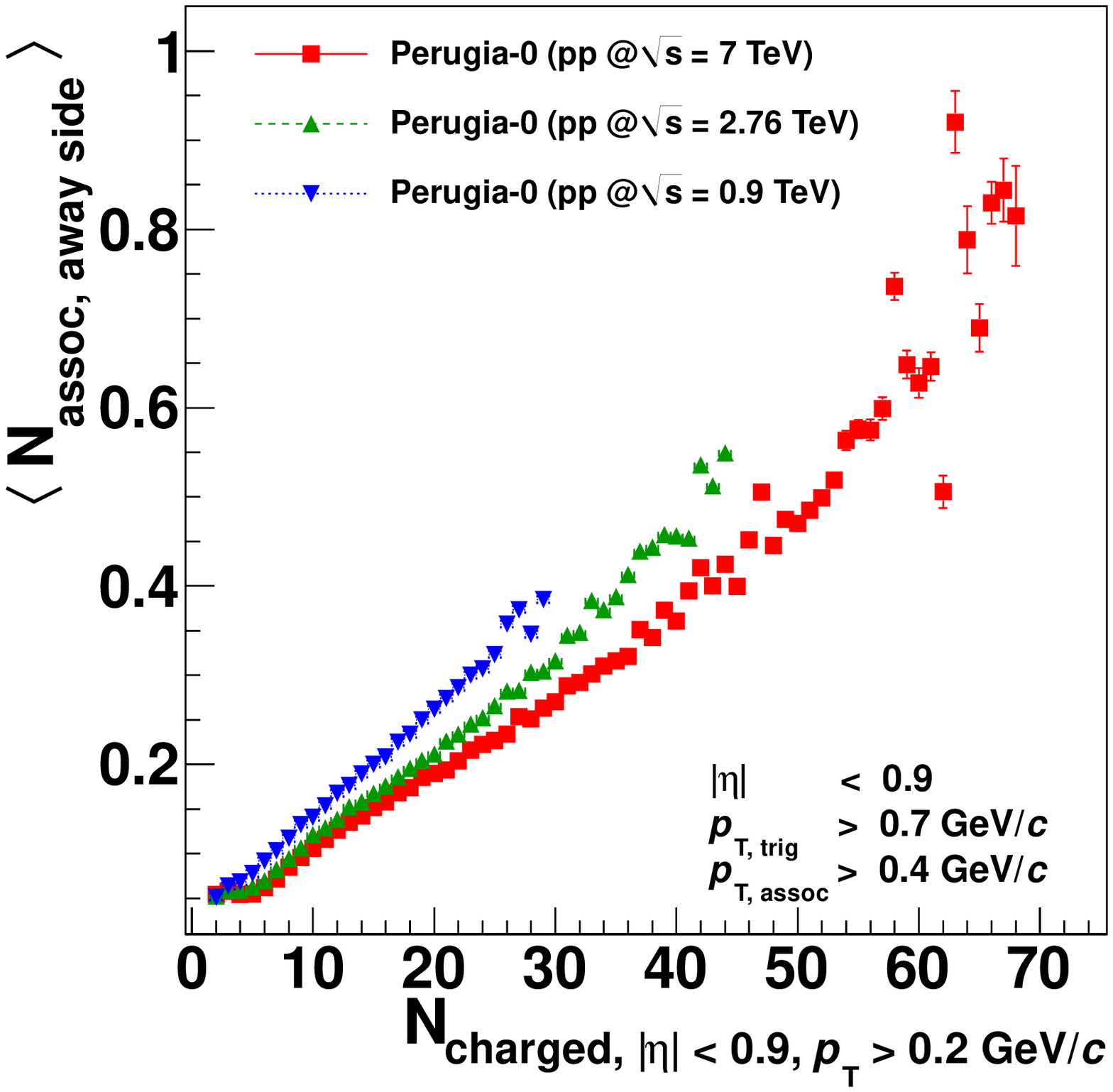} 
\end{minipage}
\begin{minipage}{0.49\textwidth}
\includegraphics[width=\textwidth]{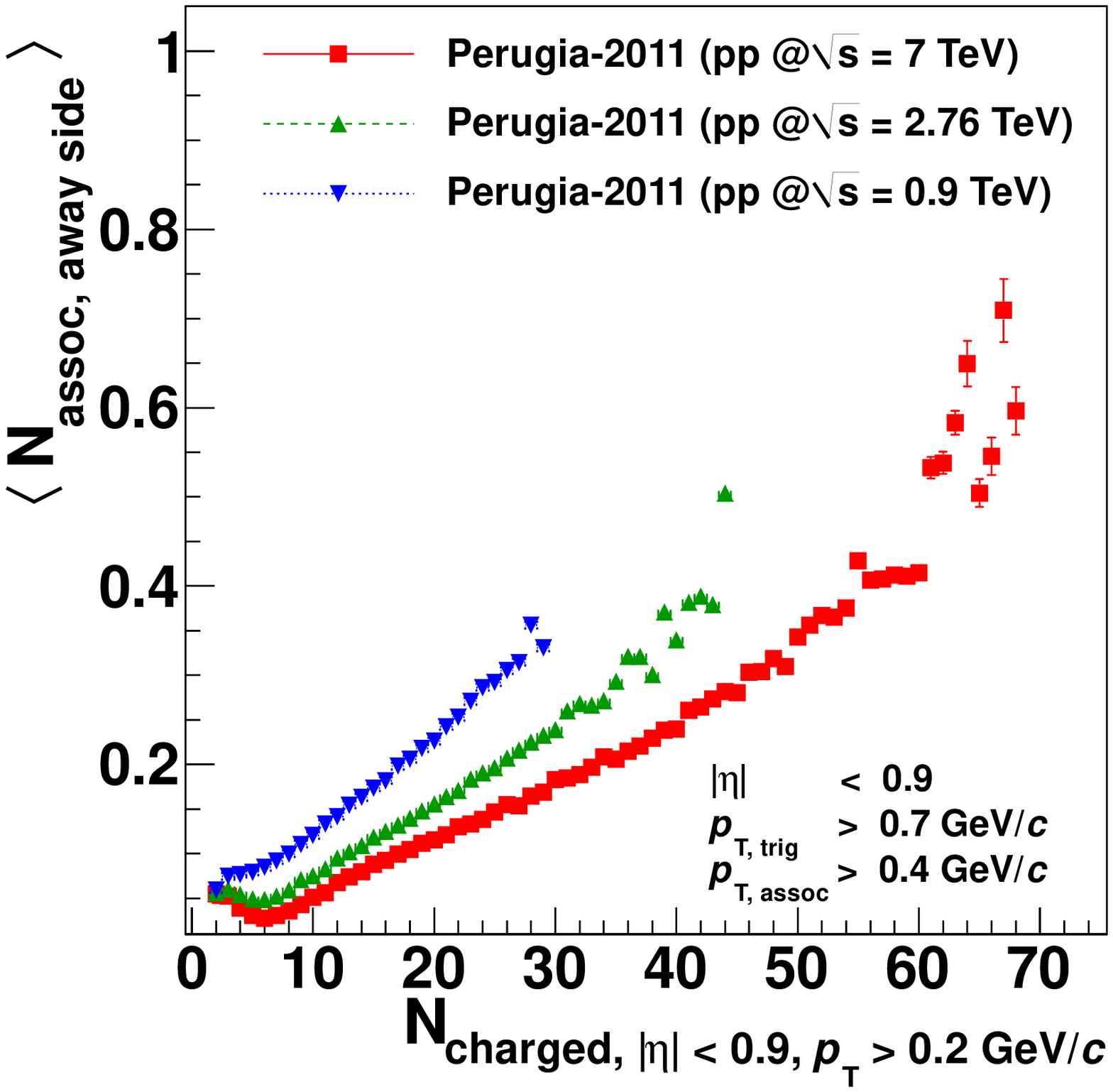} 
\end{minipage}
\begin{minipage}{0.49\textwidth}
\includegraphics[width=\textwidth]{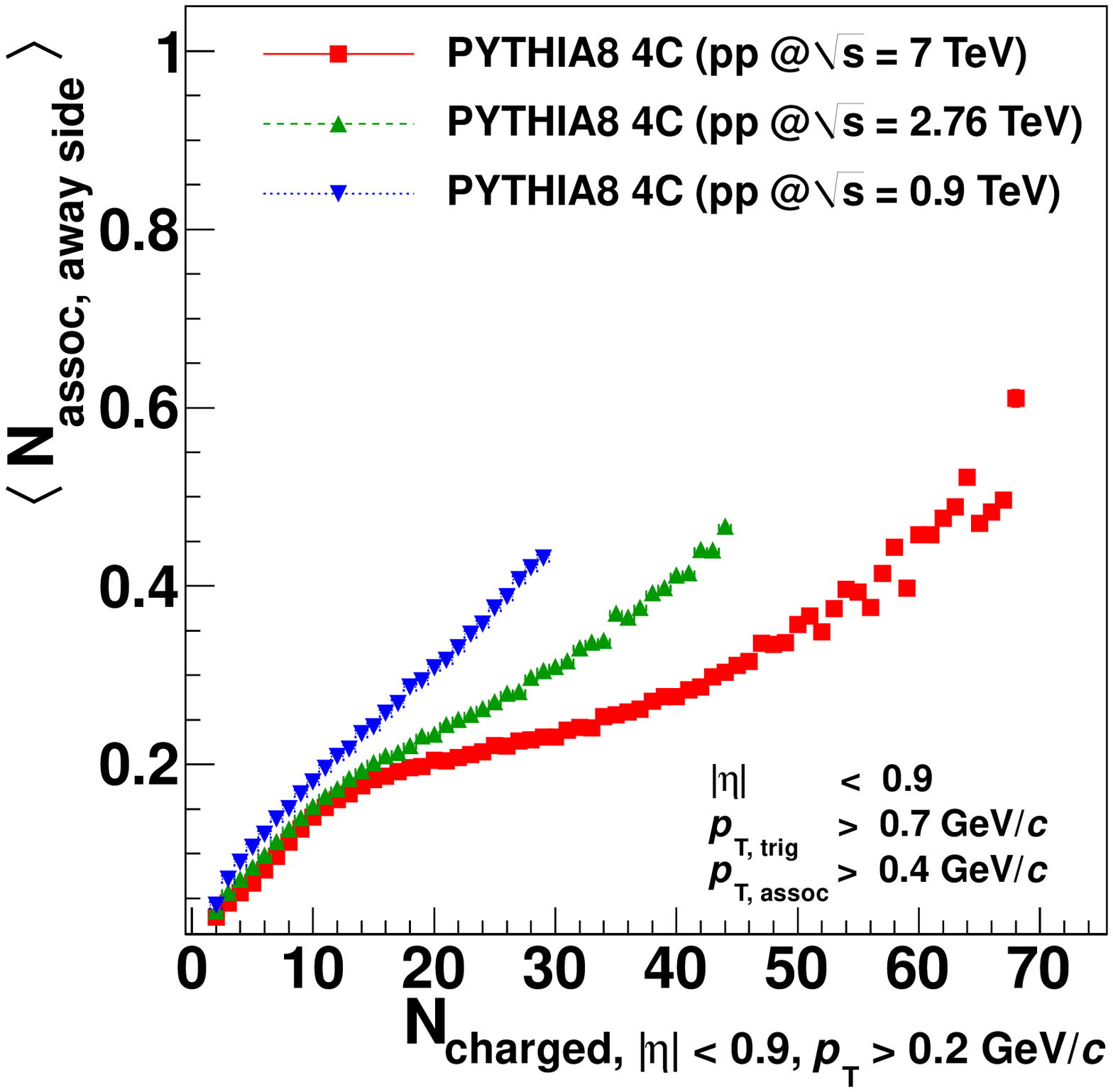} 
\end{minipage}
\caption{Per-trigger away-side pair yield as a function of the charged particle multiplicity measured for $\sqrts=0.9$, 2.76, and 7\,TeV.} 
\label{fig:awaysideCMS}
\end{figure}

\begin{figure}[p]
\centering
\vspace{-0.3cm}
\begin{minipage}{0.49\textwidth}
\includegraphics[width=\textwidth]{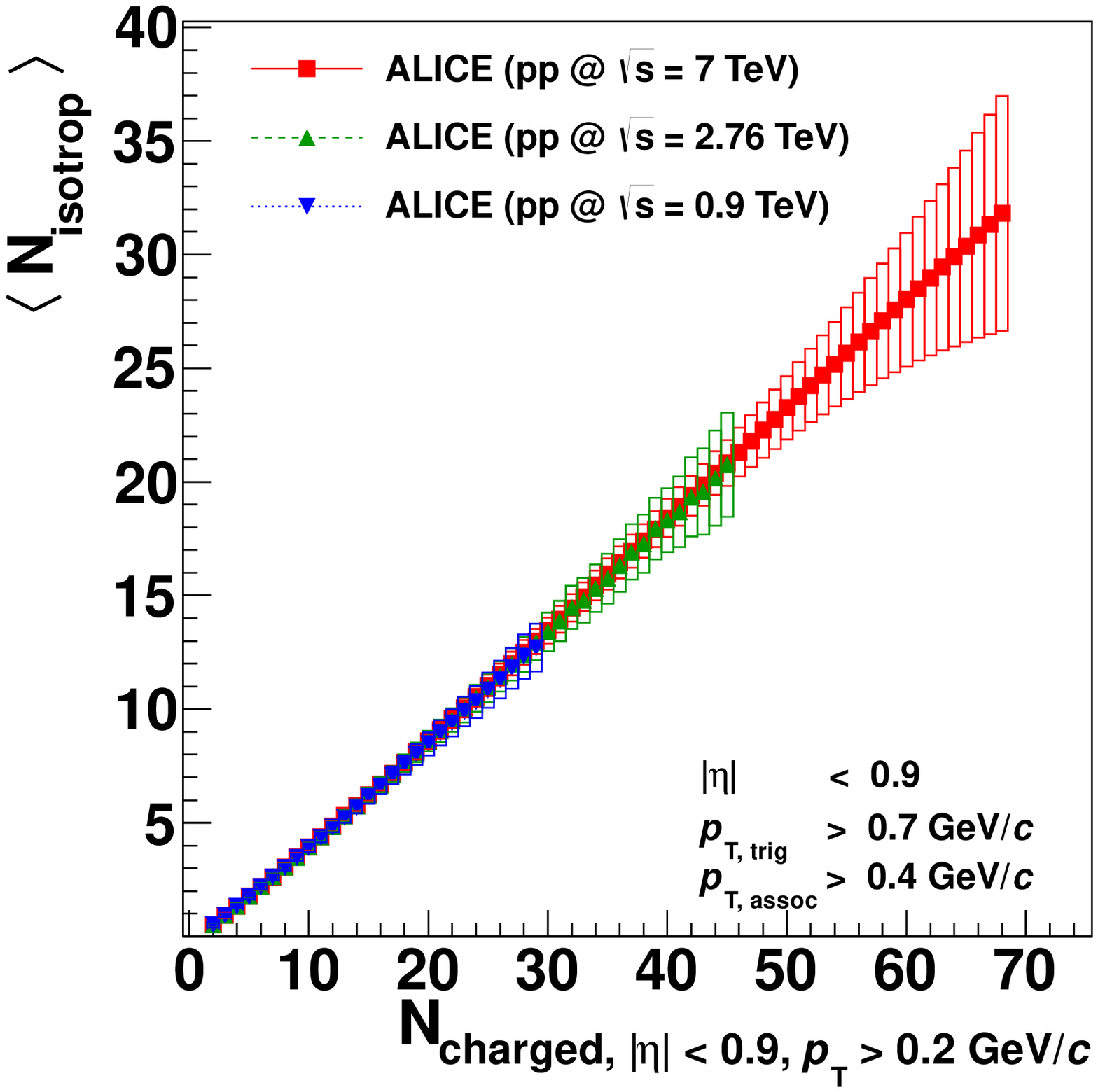} 
\end{minipage}
\begin{minipage}{0.49\textwidth}
\includegraphics[width=\textwidth]{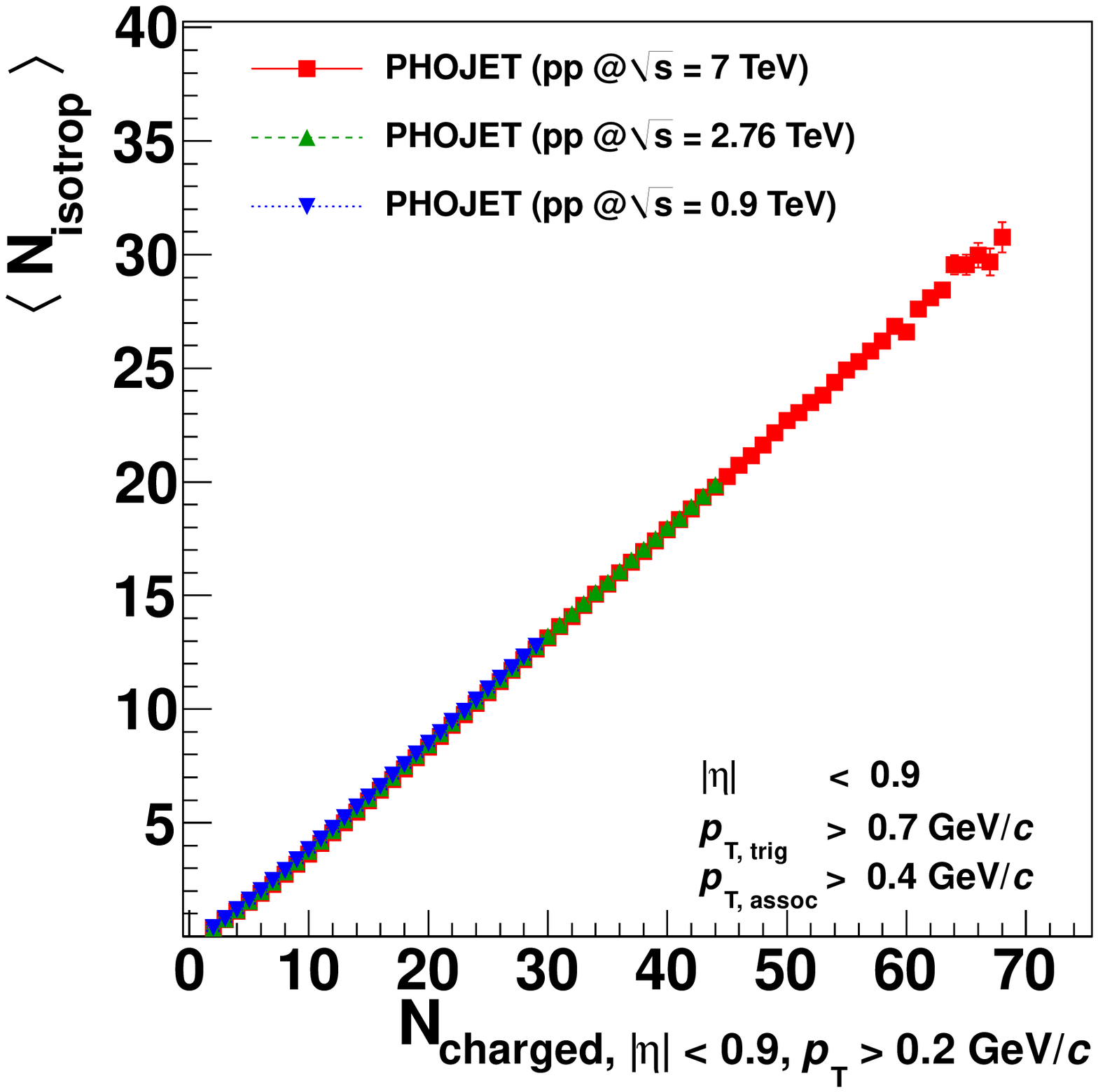} 
\end{minipage}
\vspace{-0.1cm}
\begin{minipage}{0.49\textwidth}
\includegraphics[width=\textwidth]{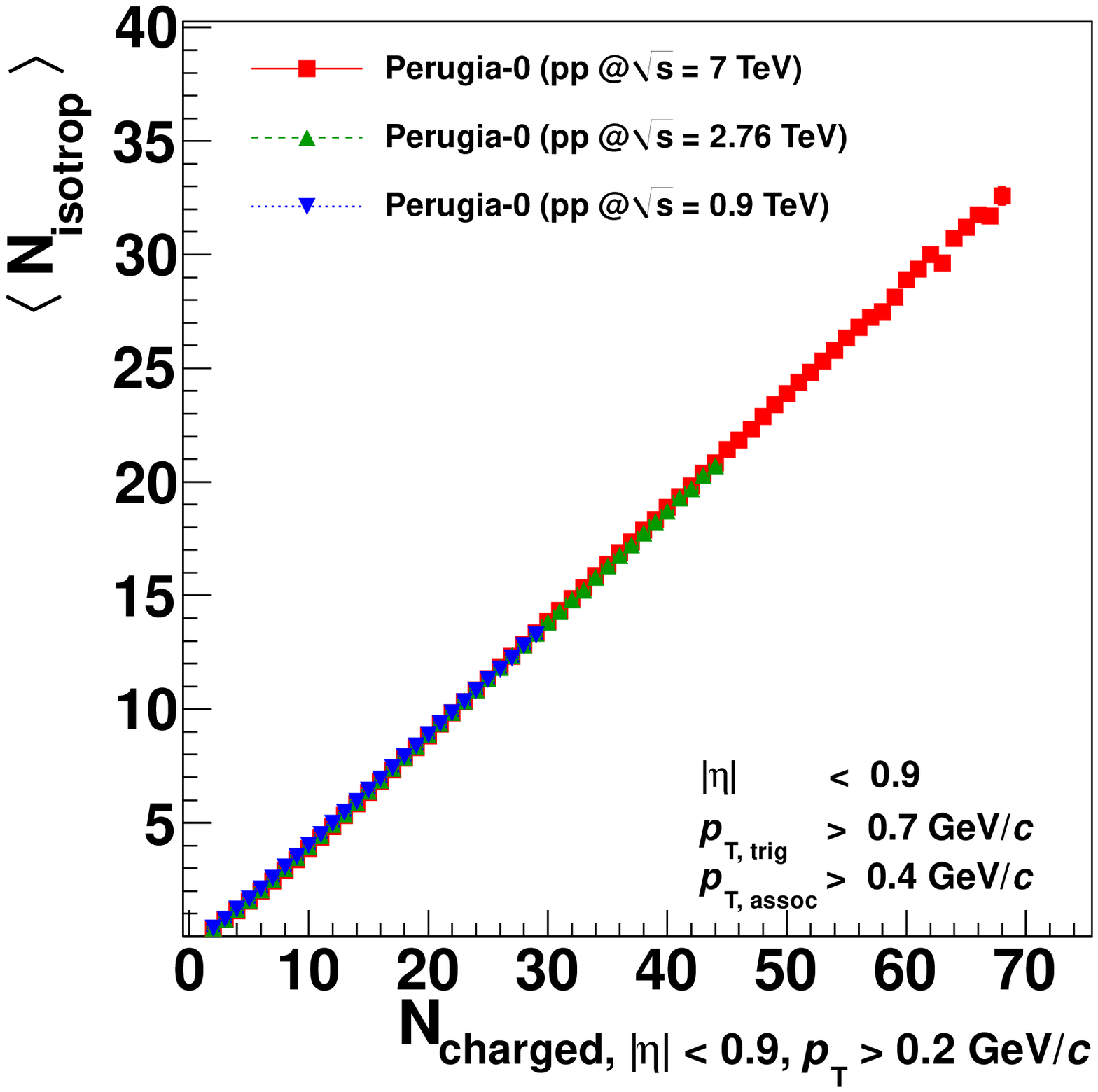} 
\end{minipage}
\begin{minipage}{0.49\textwidth}
\includegraphics[width=\textwidth]{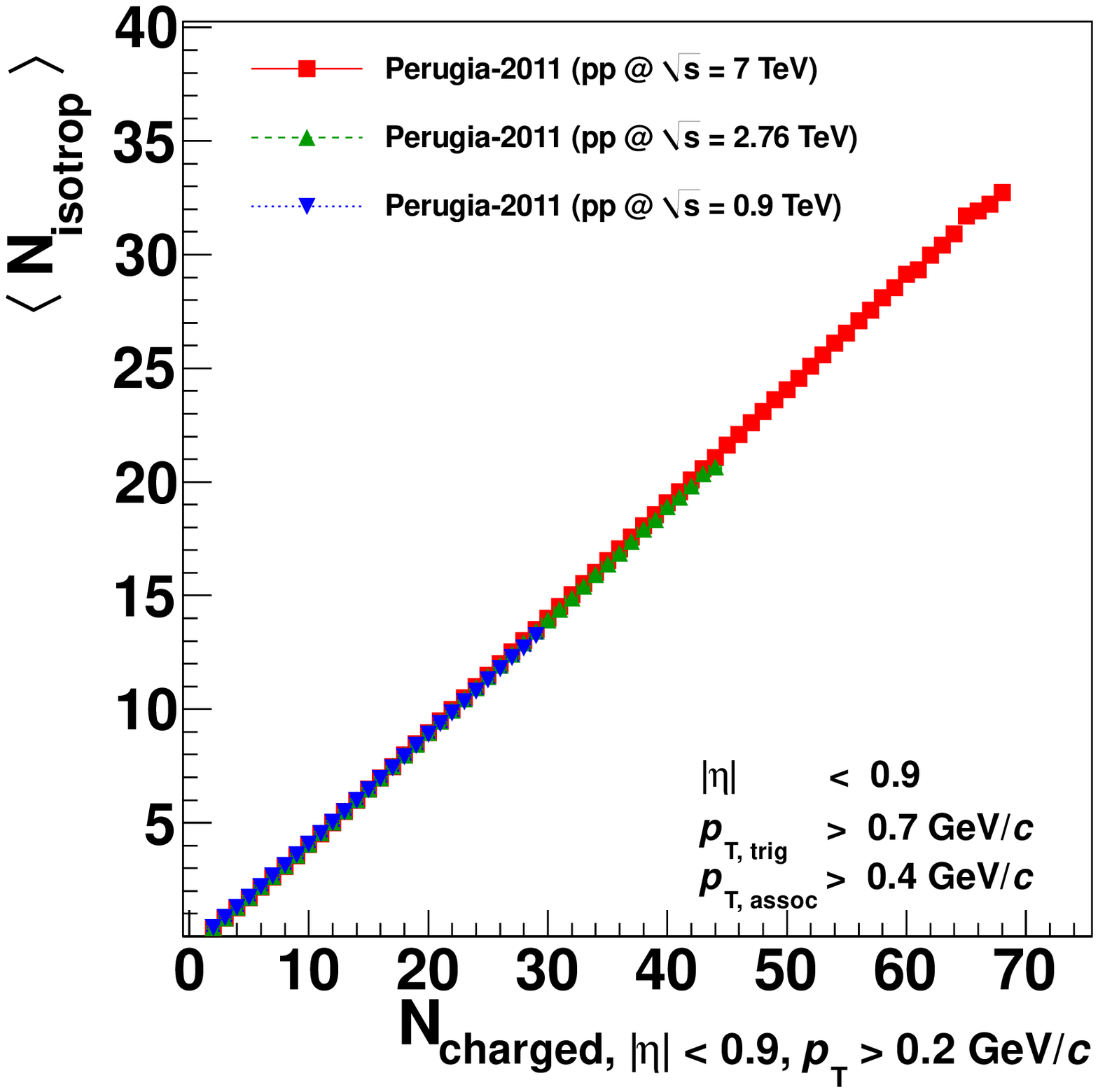} 
\end{minipage}
\begin{minipage}{0.49\textwidth}
\includegraphics[width=\textwidth]{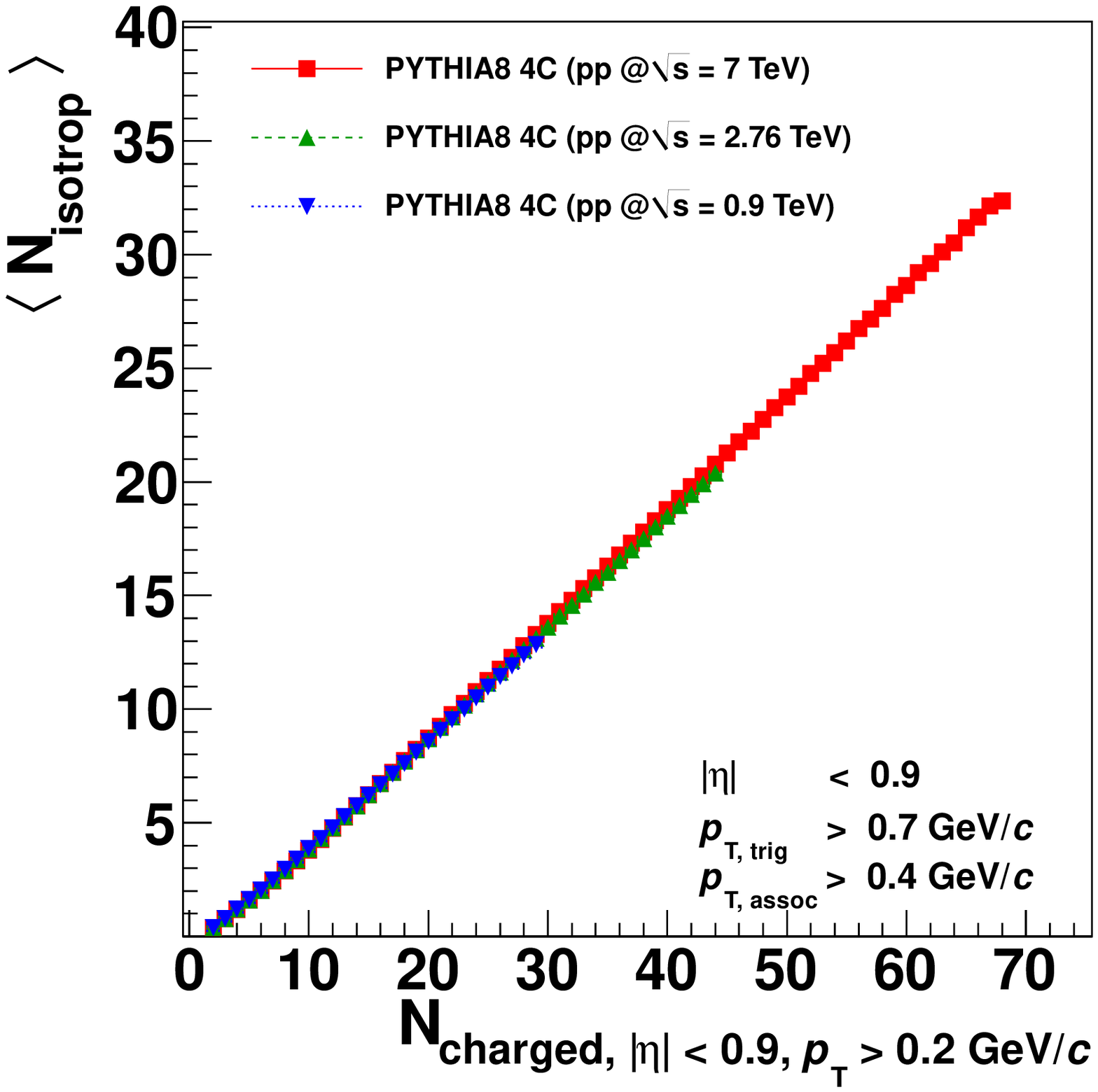} 
\end{minipage}
\caption{Per-trigger pair yield in the combinatorial background as a function of the charged particle multiplicity measured for $\sqrts=0.9$, 2.76, and 7\,TeV.} 
\label{fig:nisotropCMS}
\end{figure}

\begin{figure}[p]
\centering
\vspace{-0.3cm}
\begin{minipage}{0.49\textwidth}
\includegraphics[width=\textwidth]{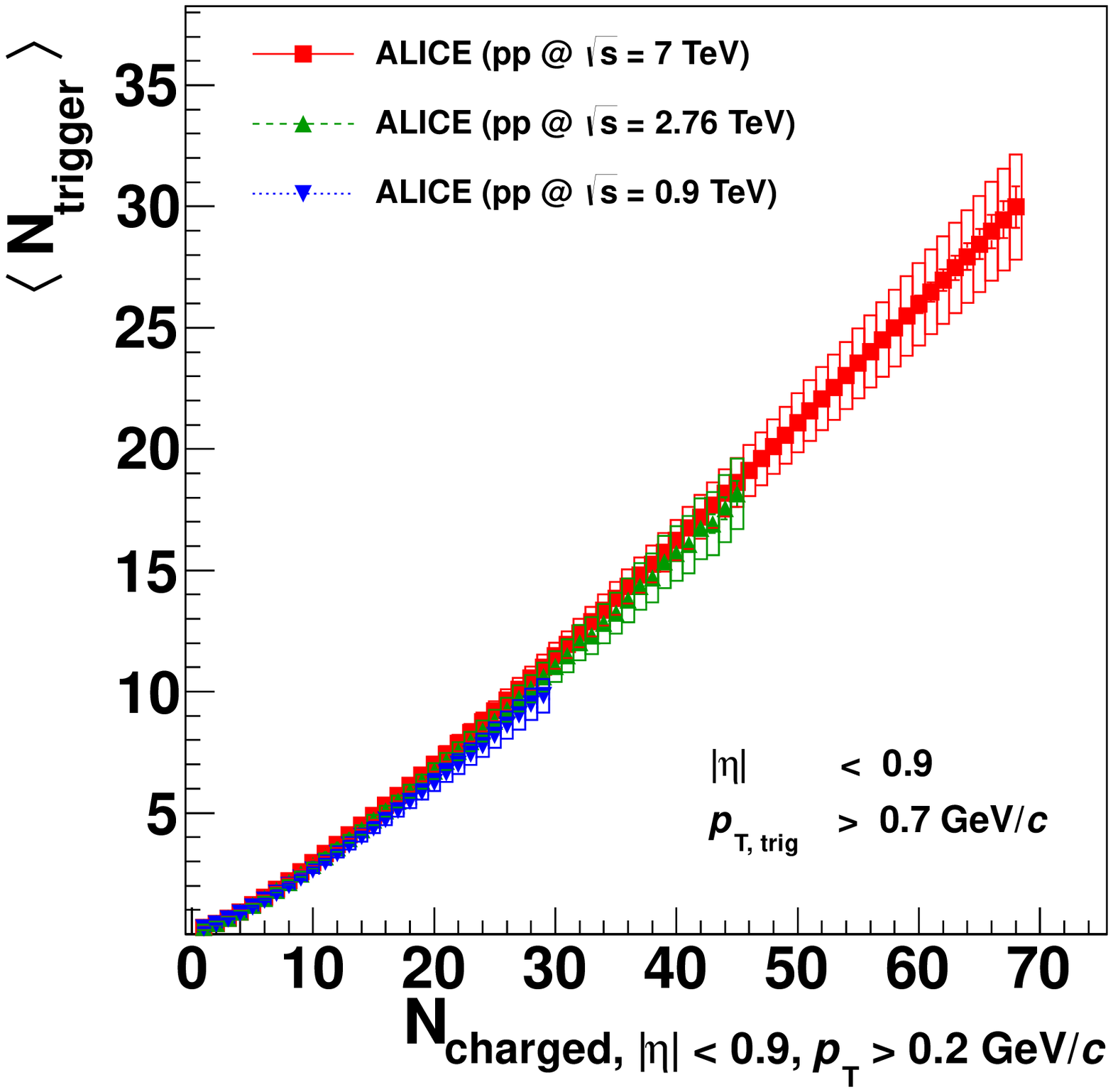} 
\end{minipage}
\begin{minipage}{0.49\textwidth}
\includegraphics[width=\textwidth]{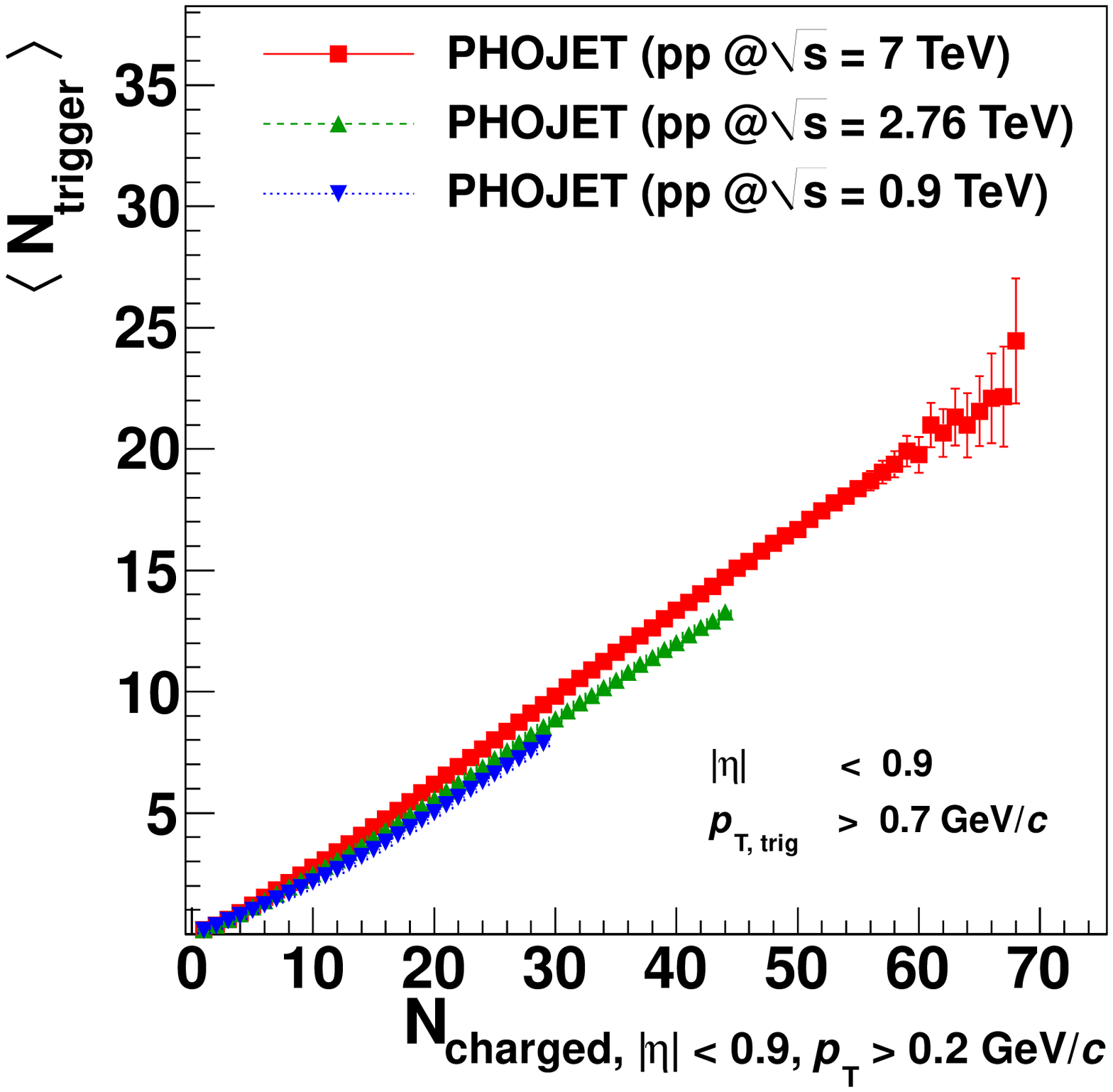} 
\end{minipage}
\vspace{-0.1cm}
\begin{minipage}{0.49\textwidth}
\includegraphics[width=\textwidth]{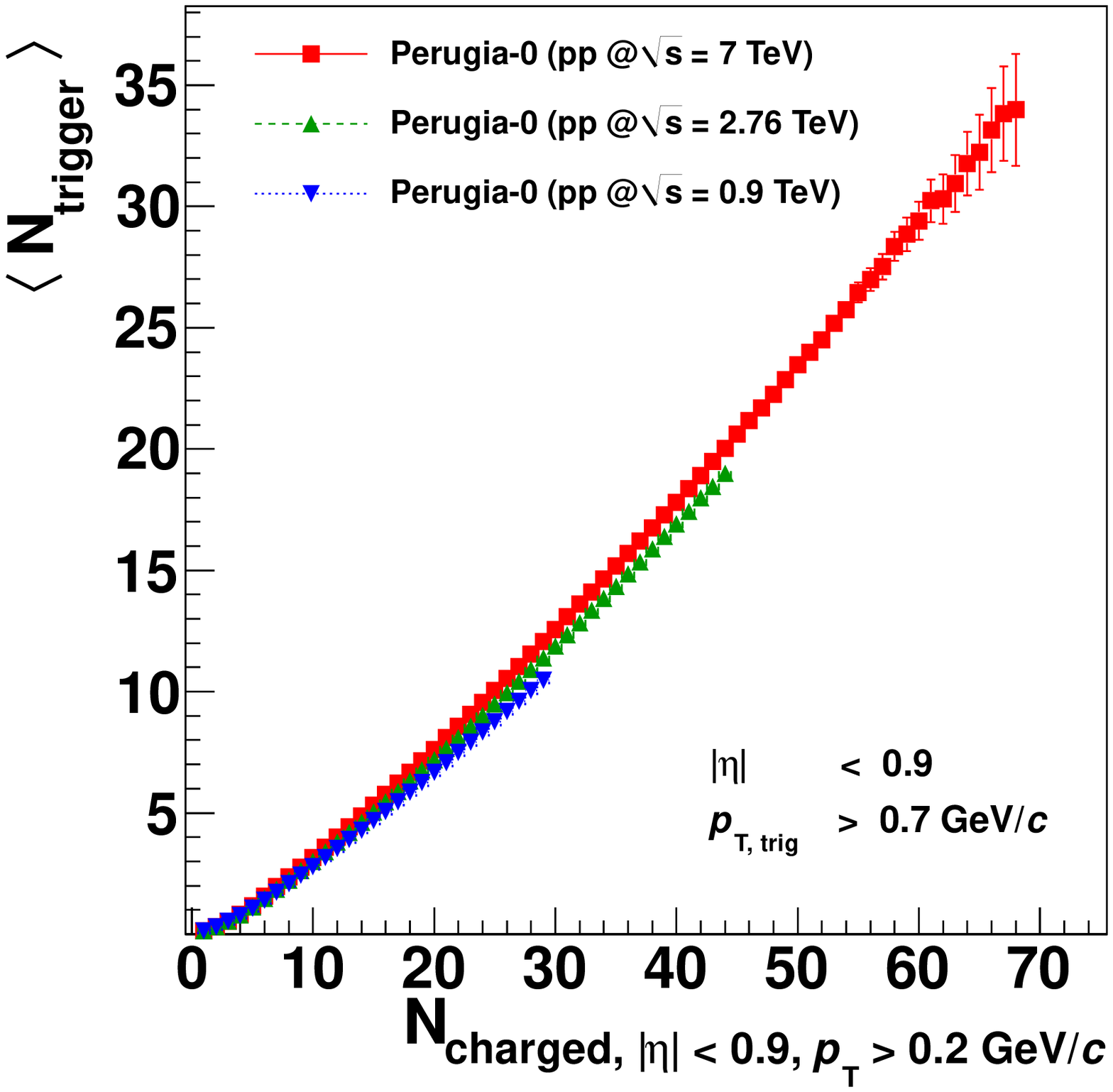} 
\end{minipage}
\begin{minipage}{0.49\textwidth}
\includegraphics[width=\textwidth]{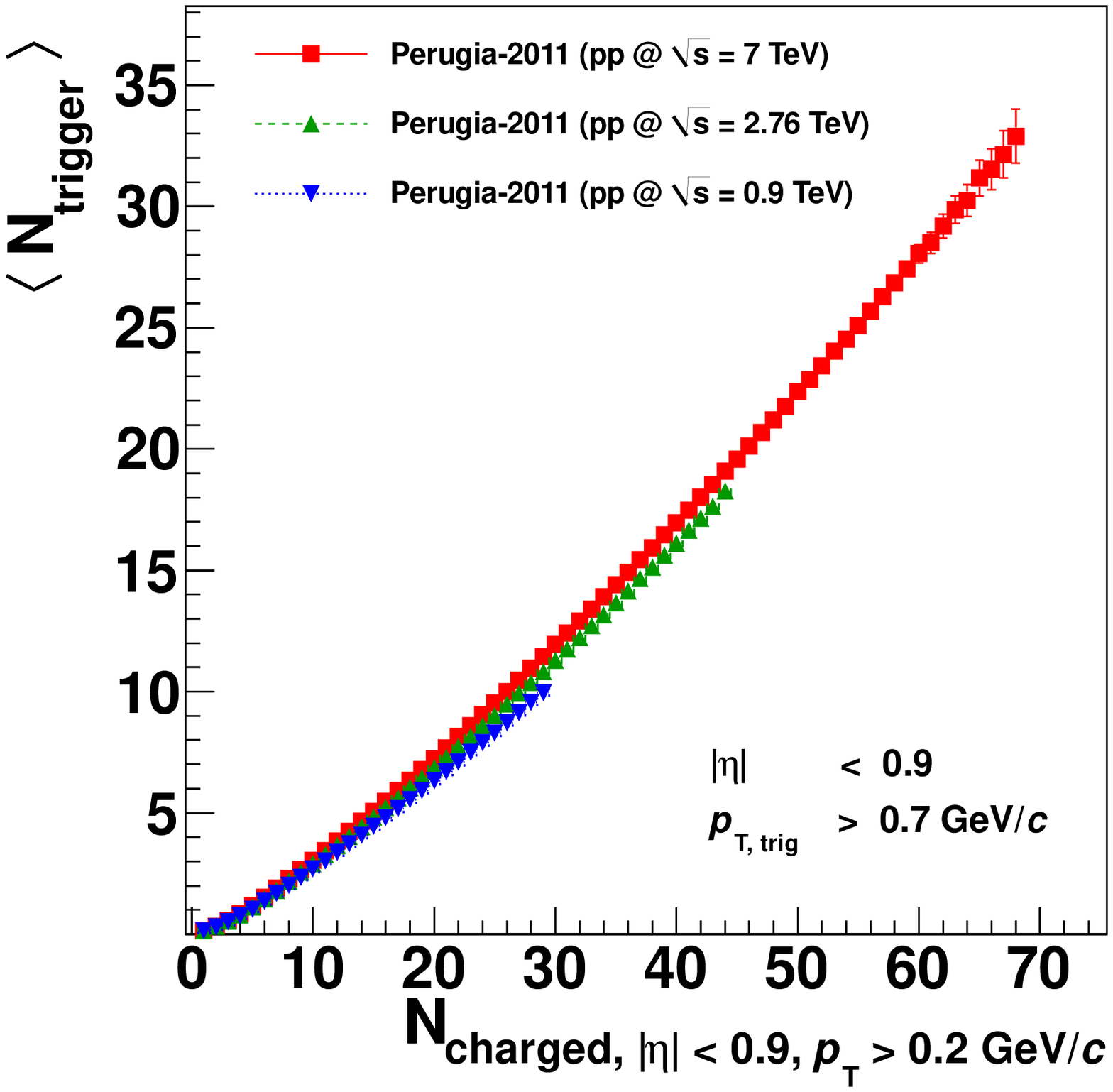} 
\end{minipage}
\begin{minipage}{0.49\textwidth}
\includegraphics[width=\textwidth]{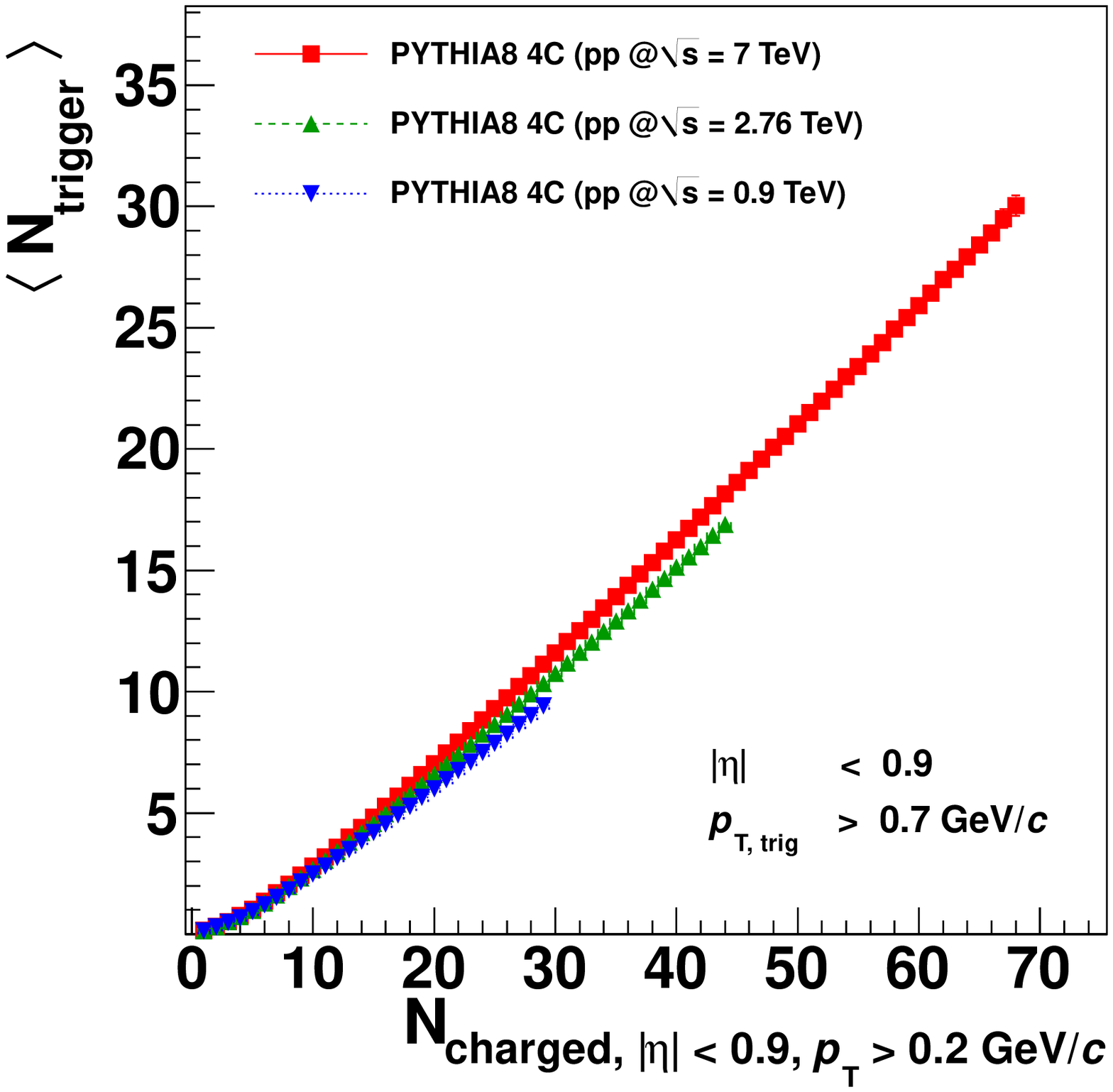} 
\end{minipage}
\caption{Average number of trigger particles per event as a function of the charged particle multiplicity measured for $\sqrts=0.9$, 2.76, and 7\,TeV.} 
\label{fig:ntriggerCMS}
\end{figure}

\begin{figure}[p]
\centering
\vspace{-0.3cm}
\begin{minipage}{0.49\textwidth}
\includegraphics[width=\textwidth]{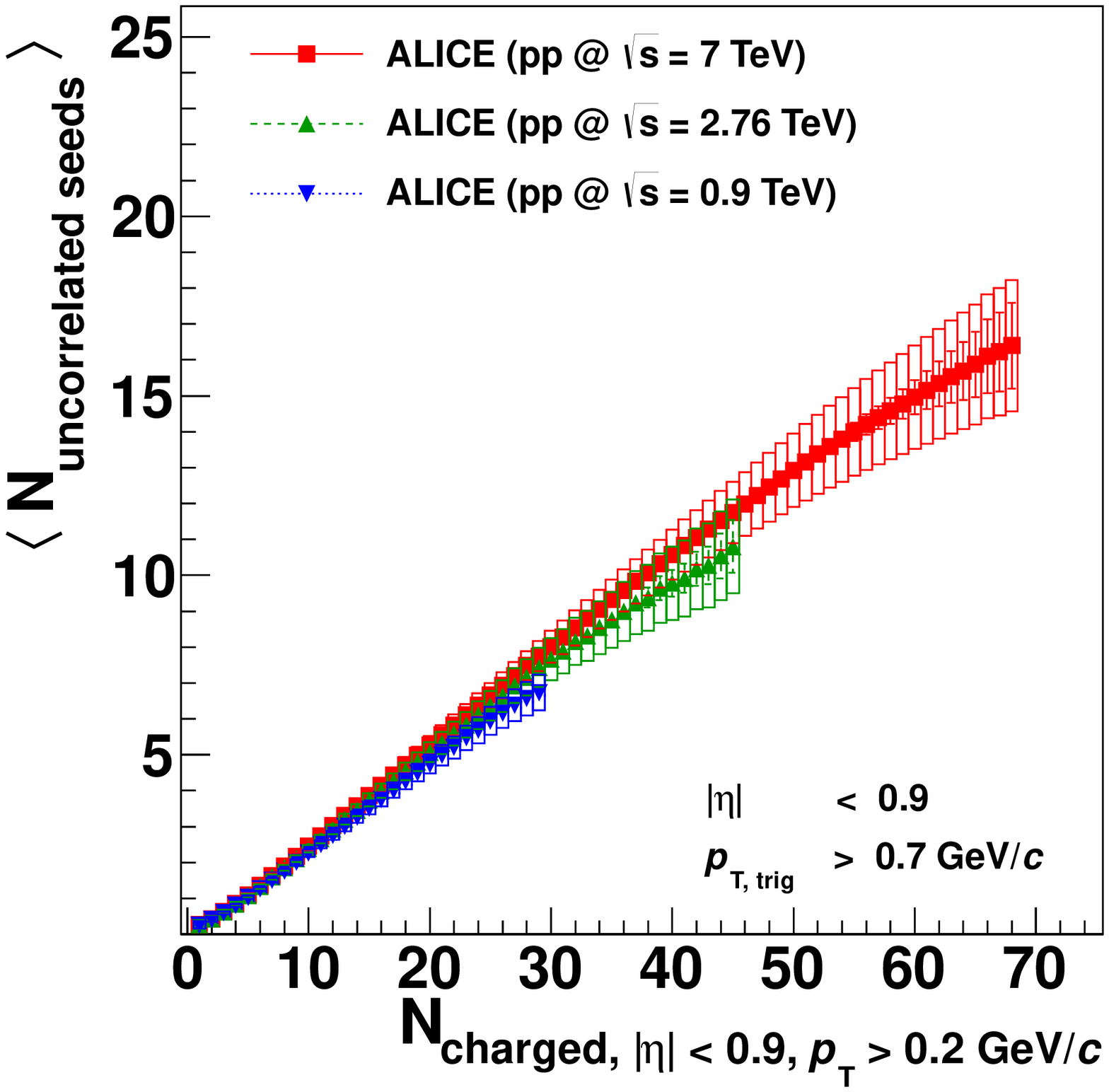} 
\end{minipage}
\begin{minipage}{0.49\textwidth}
\includegraphics[width=\textwidth]{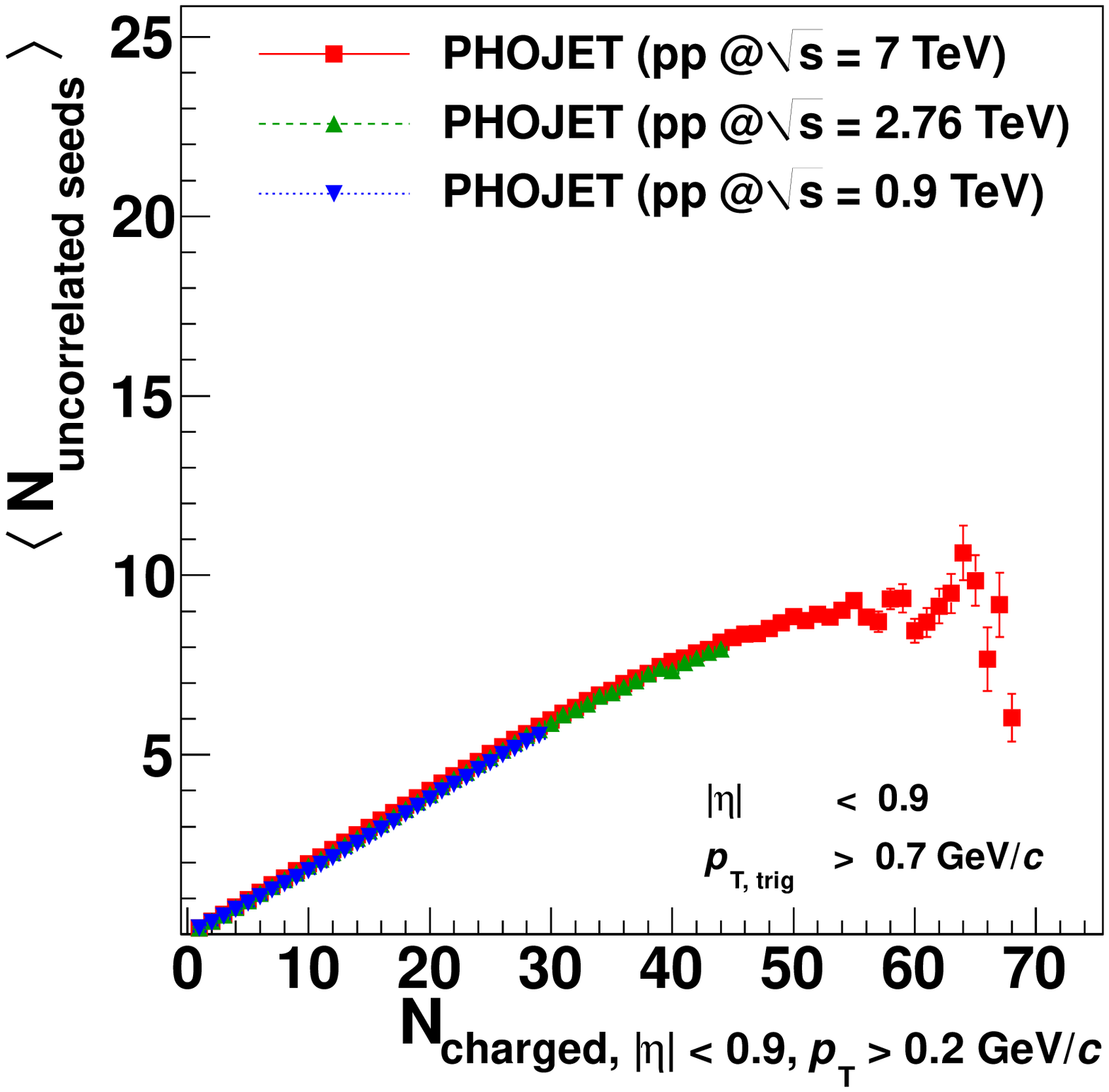} 
\end{minipage}
\vspace{-0.1cm}
\begin{minipage}{0.49\textwidth}
\includegraphics[width=\textwidth]{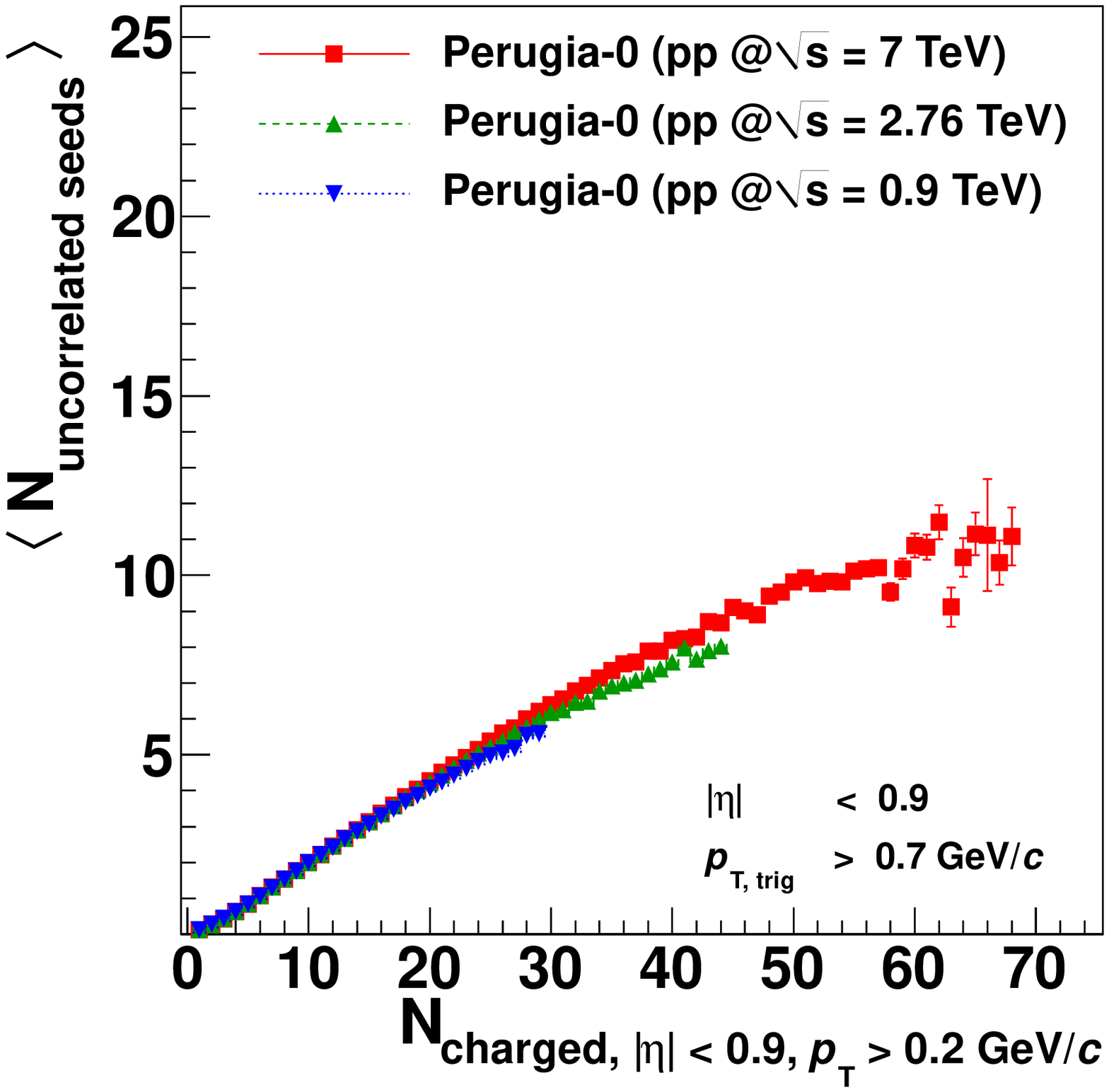} 
\end{minipage}
\begin{minipage}{0.49\textwidth}
\includegraphics[width=\textwidth]{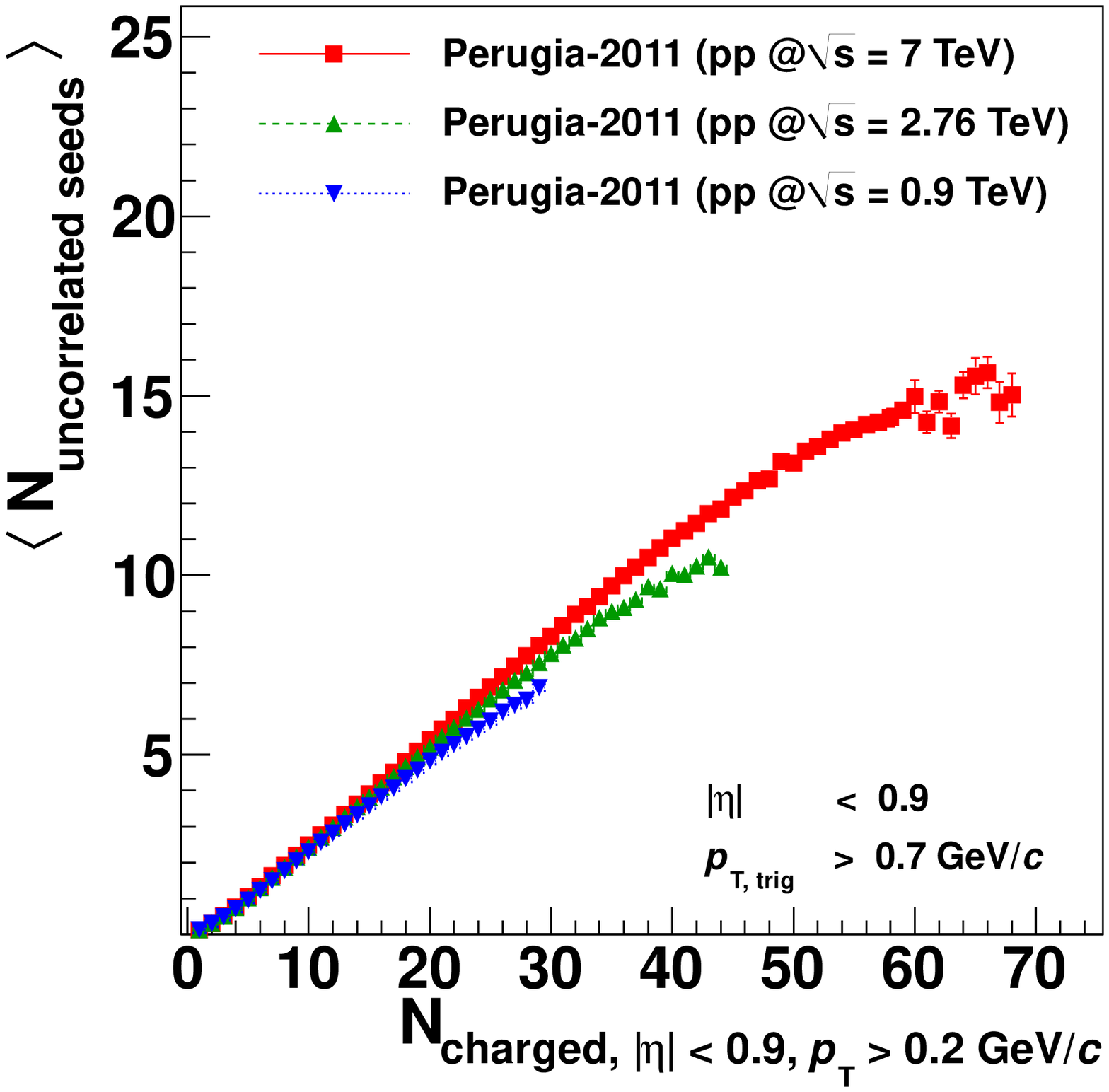} 
\end{minipage}
\begin{minipage}{0.49\textwidth}
\includegraphics[width=\textwidth]{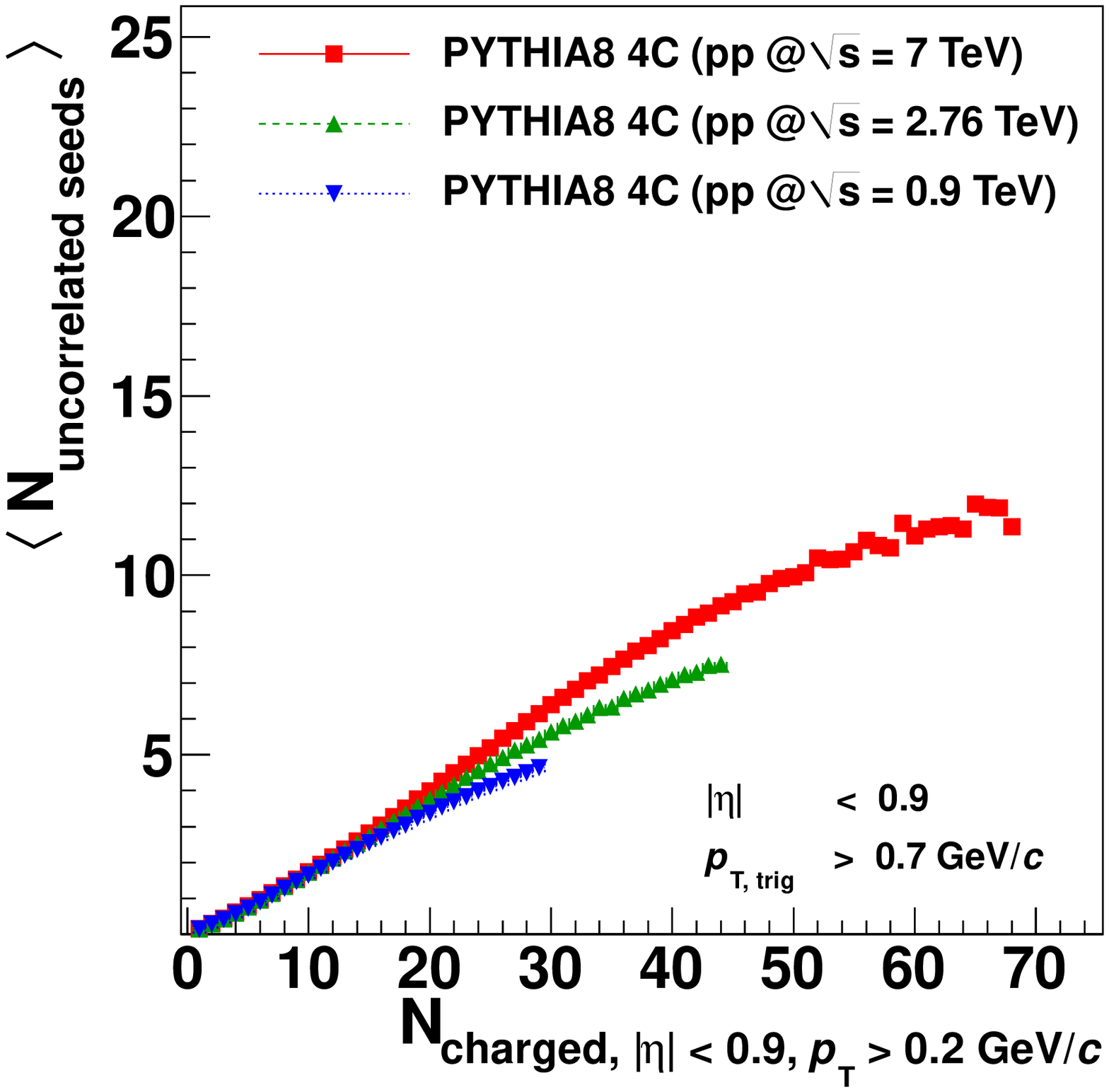} 
\end{minipage}
\caption{Average number of uncorrelated seeds per event as a function of the charged particle multiplicity measured for $\sqrts=0.9$, 2.76, and 7\,TeV.} 
\label{fig:njetCMS}
\end{figure}

Figures~\ref{fig:observables2} and \ref{fig:observables9} show the observables discussed above measured at the two
lower centre-of-mass energies $\sqrts=2.76$  and 0.9\,TeV. 
On average, the agreement between the model calculations and the ALICE results improves with decreasing collision
energy. However, qualitatively the behaviour of the different models is similar. Tune Perugia-2011 agrees best with 
the measured near-side yield and under-predicts the away-side yield, for which Perugia-0 has the best agreement.
PHOJET generally shows the worst agreement. However, the agreement between PHOJET and the ALICE results in terms of the near- and away-side
yields is good for $\sqrt{s} = 900 \, {\rm GeV}$ at high multiplicity, whereas PYTHIA8 has the largest disagreement 
in this region.\\
To allow for a more direct comparison of the trends as a function of centre-of-mass energy, figures \ref{fig:nearsideCMS}-\ref{fig:njetCMS} show in the same plots the multiplicity dependence for the three
energies for data (top left) and for the various MC generators. We note that the colors now indicate the different beam energies. 
In data, the near-side pair yield in a fixed charged particle multiplicity bin (figure~\ref{fig:nearsideCMS})
grows as a function of $\sqrt{s}$.
While all event generators reproduce this increase qualitatively, PHOJET shows a significantly stronger energy dependence 
than the data and the PYTHIA results.
The away-side pair yield in a fixed charged particle multiplicity bin measured by ALICE decreases 
as a function of the centre-of-mass energy as shown in figure \ref{fig:awaysideCMS}.
This decrease is explained by the limited $\eta$-acceptance. Due to the longitudinal momentum distribution
of partons in the colliding protons, the scattered partons have a wide relative $\Delta \eta$ distribution
that increases with increasing $\sqrt{s}$.
While all PYTHIA tunes reproduce the away-side yield decrease, PHOJET does not show a clear energy 
dependence of the yield in the studied centre-of-mass energy range.\\
The combinatorial background in a fixed charged particle multiplicity bin does not show any centre-of-mass energy dependence (figure~\ref{fig:nisotropCMS}). 
This behaviour is well reproduced by all Monte Carlo generators.
The average number of trigger particle shown in figure \ref{fig:ntriggerCMS} grows slowly as a function of the centre-of-mass energy.
The average number of uncorrelated seeds (figure \ref{fig:njetCMS}) also grows slowly as a function of the centre-of-mass energy. 
This increase is smallest for PHOJET.
The qualitative centre-of-mass energy dependence of the average number of trigger particle and the average number of uncorrelated 
seeds is well reproduced by the Monte Carlo generators.

\FloatBarrier

\begin{figure}[ht]
\begin{center}
\includegraphics[width=0.55\textwidth]{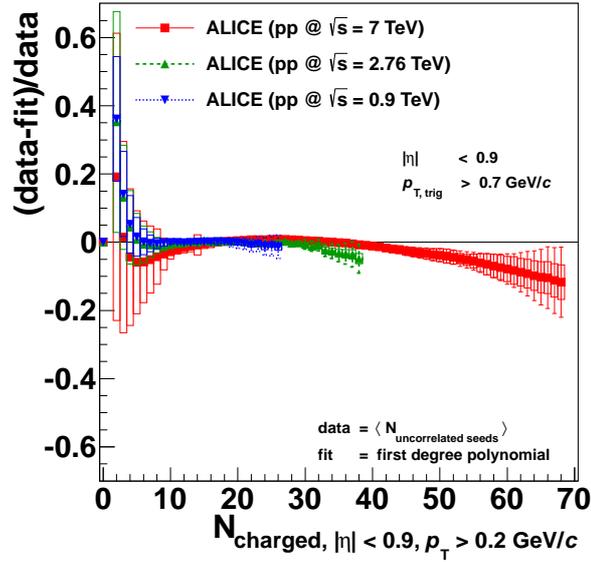}
\caption{Residual between the number of uncorrelated seeds and linear fit functions.}
\label{fig:MPI}
\end{center}
\end{figure}

\subsection{Multiple parton interactions}
Interpreted in the context of the PYTHIA model, the number of uncorrelated seeds (c.f.\ equation~\ref{eq:uncorr}) provides information about the number of semi-hard parton--parton interactions per event as discussed in section \ref{sec:method}. 
In the top left panel of figure \ref{fig:njetCMS}, the average number of uncorrelated seeds as a function of the charged particle multiplicity 
is presented for the centre-of-mass energies $\sqrts=0.9$, 2.76, and 7\,TeV. 
Figure~\ref{fig:MPI} shows the residuals between the data points and linear fit functions ((data-fit)/data). 
It can be observed that the charged particle multiplicity increases approximately linearly with the number of uncorrelated seeds. 
However, it deviates from the linear dependence at large charged particle multiplicities. Here, 
the rise of the number of uncorrelated seeds levels off. 
This observation is consistent with the assumption that at highest multiplicities a further increase of the number of 
multiple parton interactions becomes extremely improbable. In this scenario, high charged particle multiplicities 
can only be reached by selecting events with many high-multiplicity jets.

\FloatBarrier
\section{Conclusions}\label{sec:conclusion}
We have studied the pair-yields per trigger in  two-particle azimuthal correlations between charged trigger and associated particles in pp collisions at $\sqrts=0.9$, 2.76, and 7\,TeV. 
The correlations have been measured for charged particles recorded with the ALICE central barrel detectors ITS and TPC covering the full azimuth and a pseudorapidity range of $\arrowvert\eta\arrowvert<0.9$.
The analysis has been performed as a function of the charged particle multiplicity and for the transverse momentum thresholds for trigger particles of $\pTtrig>0.7\,$GeV/$c$ and for associated particles of $\pTassoc>0.4$ and 0.7\,GeV/$c$. 

The azimuthal correlations have been decomposed into the pair yield in the combinatorial background, the pair yield in the near-side peak ($\dphi \approx 0$), and the pair yield in the away-side peak ($\dphi \approx \pi$). 
Furthermore, the average number of trigger particles per event have been measured. 
While the per-trigger near-side and away-side pair yield provide information about fragmentation properties of low-$\pT$ partons, the average number of trigger particles includes information from both the number of sources of particle production and the fragmentation. 
In order to increase the sensitivity  to the number of sources of particle production, we have defined an observable, number of uncorrelated seeds, in which the impact of the fragmentation is reduced. 
Using PYTHIA simulations on generator level, we have shown that the number of uncorrelated seeds is proportional to the number of semi-hard parton--parton interactions in pp collision. However, the factor of proportionality depends on the tune and, hence, no absolute number of interactions 
can be derived from this procedure.

The per-trigger near- and away-side pair-yields as a function of the charged particle multiplicity increase with 
multiplicity.  This increase can be explained by the fact that the correlations and the multiplicity are measured in the 
same pseudo-rapidity region and that the probability distribution of the number of multi-parton interactions is steeply 
falling. Under these conditions, high multiplicities are reached through a high number of multi-parton interactions and
a higher than average number of fragments per parton.  This is also consistent with our observation that the number of trigger particles above a
$\pT$ threshold ($0.7 \, {\rm GeV}$ considered here) increases stronger than linearly with multiplicity. The symmetric
bin $\pTtrig , \pTassoc >0.7\,$GeV/$c$ has been used to reduce the multiplicative effect of fragmentation and to determine the number of uncorrelated trigger particles. The latter increases linearly with multiplicity up to the highest 
multiplicities where it starts to level off. This effect is observed for all centre-of-mass energies. Interpreted within the 
PYTHIA model of multi-parton interactions this is evidence for a limitation of the number of MPIs above a certain threshold. Independent of its physical interpretation the observed systematics are important for any study performed as a function of multiplicity.

We have compared our results to the event generators PYTHIA6, PYTHIA8, and PHOJET.
While the constant, combinatorial background of the correlation is described fairly well by all models, the models have difficulties to describe the per-trigger pair-yields in the near-side peak and the away-side peaks. 
The PYTHIA tunes reproduce the centre-of-mass dependence of the near and the away-side pair yield. PHOJET overestimates the increase of the near-side yield with the centre-of-mass energy, while it does not show any centre-of-mass dependence of the away-side yield.
The development of the number of uncorrelated seeds with charged particle multiplicity is described well by all models.
These findings are expected to provide important input for future Monte Carlo tunes and will help to constrain the models used in these generators.

\newenvironment{acknowledgement}{\relax}{\relax}
\begin{acknowledgement}
\section*{Acknowledgements}
The ALICE collaboration would like to thank all its engineers and technicians for their invaluable contributions to the construction of the experiment and the CERN accelerator teams for the outstanding performance of the LHC complex.
\\
The ALICE collaboration acknowledges the following funding agencies for their support in building and
running the ALICE detector:
 \\
State Committee of Science,  World Federation of Scientists (WFS)
and Swiss Fonds Kidagan, Armenia,
 \\
Conselho Nacional de Desenvolvimento Cient\'{\i}fico e Tecnol\'{o}gico (CNPq), Financiadora de Estudos e Projetos (FINEP),
Funda\c{c}\~{a}o de Amparo \`{a} Pesquisa do Estado de S\~{a}o Paulo (FAPESP);
 \\
National Natural Science Foundation of China (NSFC), the Chinese Ministry of Education (CMOE)
and the Ministry of Science and Technology of China (MSTC);
 \\
Ministry of Education and Youth of the Czech Republic;
 \\
Danish Natural Science Research Council, the Carlsberg Foundation and the Danish National Research Foundation;
 \\
The European Research Council under the European Community's Seventh Framework Programme;
 \\
Helsinki Institute of Physics and the Academy of Finland;
 \\
French CNRS-IN2P3, the `Region Pays de Loire', `Region Alsace', `Region Auvergne' and CEA, France;
 \\
German BMBF and the Helmholtz Association;
\\
General Secretariat for Research and Technology, Ministry of
Development, Greece;
\\
Hungarian OTKA and National Office for Research and Technology (NKTH);
 \\
Department of Atomic Energy and Department of Science and Technology of the Government of India;
 \\
Istituto Nazionale di Fisica Nucleare (INFN) and Centro Fermi -
Museo Storico della Fisica e Centro Studi e Ricerche "Enrico
Fermi", Italy;
 \\
MEXT Grant-in-Aid for Specially Promoted Research, Ja\-pan;
 \\
Joint Institute for Nuclear Research, Dubna;
 \\
National Research Foundation of Korea (NRF);
 \\
CONACYT, DGAPA, M\'{e}xico, ALFA-EC and the EPLANET Program
(European Particle Physics Latin American Network)
 \\
Stichting voor Fundamenteel Onderzoek der Materie (FOM) and the Nederlandse Organisatie voor Wetenschappelijk Onderzoek (NWO), Netherlands;
 \\
Research Council of Norway (NFR);
 \\
Polish Ministry of Science and Higher Education;
 \\
National Authority for Scientific Research - NASR (Autoritatea Na\c{t}ional\u{a} pentru Cercetare \c{S}tiin\c{t}ific\u{a} - ANCS);
 \\
Ministry of Education and Science of Russian Federation, Russian
Academy of Sciences, Russian Federal Agency of Atomic Energy,
Russian Federal Agency for Science and Innovations and The Russian
Foundation for Basic Research;
 \\
Ministry of Education of Slovakia;
 \\
Department of Science and Technology, South Africa;
 \\
CIEMAT, EELA, Ministerio de Econom\'{i}a y Competitividad (MINECO) of Spain, Xunta de Galicia (Conseller\'{\i}a de Educaci\'{o}n),
CEA\-DEN, Cubaenerg\'{\i}a, Cuba, and IAEA (International Atomic Energy Agency);
 \\
Swedish Research Council (VR) and Knut $\&$ Alice Wallenberg
Foundation (KAW);
 \\
Ukraine Ministry of Education and Science;
 \\
United Kingdom Science and Technology Facilities Council (STFC);
 \\
The United States Department of Energy, the United States National
Science Foundation, the State of Texas, and the State of Ohio.
\end{acknowledgement}


\newpage
\appendix
\section{The ALICE Collaboration}
\label{app:collab}



\begingroup
\small
\begin{flushleft}
B.~Abelev\Irefn{org1234}\And
J.~Adam\Irefn{org1274}\And
D.~Adamov\'{a}\Irefn{org1283}\And
A.M.~Adare\Irefn{org1260}\And
M.M.~Aggarwal\Irefn{org1157}\And
G.~Aglieri~Rinella\Irefn{org1192}\And
M.~Agnello\Irefn{org1313}\textsuperscript{,}\Irefn{org1017688}\And
A.G.~Agocs\Irefn{org1143}\And
A.~Agostinelli\Irefn{org1132}\And
Z.~Ahammed\Irefn{org1225}
% ...
\renewcommand\labelenumi{\textsuperscript{\theenumi}~}

\section*{Affiliation notes}
\renewcommand\theenumi{\roman{enumi}}
\begin{Authlist}
\item \Adef{0}Deceased
\item \Adef{idp4197632}{Also at: M.V.Lomonosov Moscow State University, D.V.Skobeltsyn Institute of Nuclear Physics, Moscow, Russia}
\item \Adef{idp4490016}{Also at: University of Belgrade, Faculty of Physics and "Vin\v{c}a" Institute of Nuclear Sciences, Belgrade, Serbia}
\item \Adef{idp5510336}{Also at: Institute of Theoretical Physics, University of Wroclaw, Wroclaw, Poland}
\end{Authlist}

\section*{Collaboration Institutes}
\renewcommand\theenumi{\arabic{enumi}~}
\begin{Authlist}

\item \Idef{org36632}Academy of Scientific Research and Technology (ASRT), Cairo, Egypt
\item \Idef{org1332}A. I. Alikhanyan National Science Laboratory (Yerevan Physics Institute) Foundation, Yerevan, Armenia
\item \Idef{org1279}Benem\'{e}rita Universidad Aut\'{o}noma de Puebla, Puebla, Mexico
\item \Idef{org1220}Bogolyubov Institute for Theoretical Physics, Kiev, Ukraine
\item \Idef{org20959}Bose Institute, Department of Physics and Centre for Astroparticle Physics and Space Science (CAPSS), Kolkata, India
\item \Idef{org1262}Budker Institute for Nuclear Physics, Novosibirsk, Russia
\item \Idef{org1292}California Polytechnic State University, San Luis Obispo, California, United States
\item \Idef{org1329}Central China Normal University, Wuhan, China
\item \Idef{org14939}Centre de Calcul de l'IN2P3, Villeurbanne, France
\item \Idef{org1197}Centro de Aplicaciones Tecnol\'{o}gicas y Desarrollo Nuclear (CEADEN), Havana, Cuba
\item \Idef{org1242}Centro de Investigaciones Energ\'{e}ticas Medioambientales y Tecnol\'{o}gicas (CIEMAT), Madrid, Spain
\item \Idef{org1244}Centro de Investigaci\'{o}n y de Estudios Avanzados (CINVESTAV), Mexico City and M\'{e}rida, Mexico
\item \Idef{org1335}Centro Fermi - Museo Storico della Fisica e Centro Studi e Ricerche ``Enrico Fermi'', Rome, Italy
\item \Idef{org17347}Chicago State University, Chicago, United States
\item \Idef{org1288}Commissariat \`{a} l'Energie Atomique, IRFU, Saclay, France
\item \Idef{org15782}COMSATS Institute of Information Technology (CIIT), Islamabad, Pakistan
\item \Idef{org1294}Departamento de F\'{\i}sica de Part\'{\i}culas and IGFAE, Universidad de Santiago de Compostela, Santiago de Compostela, Spain
\item \Idef{org1106}Department of Physics Aligarh Muslim University, Aligarh, India
\item \Idef{org1121}Department of Physics and Technology, University of Bergen, Bergen, Norway
\item \Idef{org1162}Department of Physics, Ohio State University, Columbus, Ohio, United States
\item \Idef{org1300}Department of Physics, Sejong University, Seoul, South Korea
\item \Idef{org1268}Department of Physics, University of Oslo, Oslo, Norway
\item \Idef{org1315}Dipartimento di Fisica dell'Universit\`{a} and Sezione INFN, Trieste, Italy
\item \Idef{org1145}Dipartimento di Fisica dell'Universit\`{a} and Sezione INFN, Cagliari, Italy
\item \Idef{org1312}Dipartimento di Fisica dell'Universit\`{a} and Sezione INFN, Turin, Italy
\item \Idef{org1285}Dipartimento di Fisica dell'Universit\`{a} `La Sapienza' and Sezione INFN, Rome, Italy
\item \Idef{org1154}Dipartimento di Fisica e Astronomia dell'Universit\`{a} and Sezione INFN, Catania, Italy
\item \Idef{org1132}Dipartimento di Fisica e Astronomia dell'Universit\`{a} and Sezione INFN, Bologna, Italy
\item \Idef{org1270}Dipartimento di Fisica e Astronomia dell'Universit\`{a} and Sezione INFN, Padova, Italy
\item \Idef{org1290}Dipartimento di Fisica `E.R.~Caianiello' dell'Universit\`{a} and Gruppo Collegato INFN, Salerno, Italy
\item \Idef{org1103}Dipartimento di Scienze e Innovazione Tecnologica dell'Universit\`{a} del Piemonte Orientale and Gruppo Collegato INFN, Alessandria, Italy
\item \Idef{org1114}Dipartimento Interateneo di Fisica `M.~Merlin' and Sezione INFN, Bari, Italy
\item \Idef{org1237}Division of Experimental High Energy Physics, University of Lund, Lund, Sweden
\item \Idef{org1192}European Organization for Nuclear Research (CERN), Geneva, Switzerland
\item \Idef{org1227}Fachhochschule K\"{o}ln, K\"{o}ln, Germany
\item \Idef{org1122}Faculty of Engineering, Bergen University College, Bergen, Norway
\item \Idef{org1136}Faculty of Mathematics, Physics and Informatics, Comenius University, Bratislava, Slovakia
\item \Idef{org1274}Faculty of Nuclear Sciences and Physical Engineering, Czech Technical University in Prague, Prague, Czech Republic
\item \Idef{org1229}Faculty of Science, P.J.~\v{S}af\'{a}rik University, Ko\v{s}ice, Slovakia
\item \Idef{org1184}Frankfurt Institute for Advanced Studies, Johann Wolfgang Goethe-Universit\"{a}t Frankfurt, Frankfurt, Germany
\item \Idef{org1215}Gangneung-Wonju National University, Gangneung, South Korea
\item \Idef{org20958}Gauhati University, Department of Physics, Guwahati, India
\item \Idef{org1212}Helsinki Institute of Physics (HIP) and University of Jyv\"{a}skyl\"{a}, Jyv\"{a}skyl\"{a}, Finland
\item \Idef{org1203}Hiroshima University, Hiroshima, Japan
\item \Idef{org1254}Indian Institute of Technology Bombay (IIT), Mumbai, India
\item \Idef{org36378}Indian Institute of Technology Indore, Indore, India (IITI)
\item \Idef{org1266}Institut de Physique Nucl\'{e}aire d'Orsay (IPNO), Universit\'{e} Paris-Sud, CNRS-IN2P3, Orsay, France
\item \Idef{org1277}Institute for High Energy Physics, Protvino, Russia
\item \Idef{org1249}Institute for Nuclear Research, Academy of Sciences, Moscow, Russia
\item \Idef{org1320}Nikhef, National Institute for Subatomic Physics and Institute for Subatomic Physics of Utrecht University, Utrecht, Netherlands
\item \Idef{org1250}Institute for Theoretical and Experimental Physics, Moscow, Russia
\item \Idef{org1230}Institute of Experimental Physics, Slovak Academy of Sciences, Ko\v{s}ice, Slovakia
\item \Idef{org1127}Institute of Physics, Bhubaneswar, India
\item \Idef{org1275}Institute of Physics, Academy of Sciences of the Czech Republic, Prague, Czech Republic
\item \Idef{org1139}Institute of Space Sciences (ISS), Bucharest, Romania
\item \Idef{org27399}Institut f\"{u}r Informatik, Johann Wolfgang Goethe-Universit\"{a}t Frankfurt, Frankfurt, Germany
\item \Idef{org1185}Institut f\"{u}r Kernphysik, Johann Wolfgang Goethe-Universit\"{a}t Frankfurt, Frankfurt, Germany
\item \Idef{org1177}Institut f\"{u}r Kernphysik, Technische Universit\"{a}t Darmstadt, Darmstadt, Germany
\item \Idef{org1256}Institut f\"{u}r Kernphysik, Westf\"{a}lische Wilhelms-Universit\"{a}t M\"{u}nster, M\"{u}nster, Germany
\item \Idef{org1246}Instituto de Ciencias Nucleares, Universidad Nacional Aut\'{o}noma de M\'{e}xico, Mexico City, Mexico
\item \Idef{org1247}Instituto de F\'{\i}sica, Universidad Nacional Aut\'{o}noma de M\'{e}xico, Mexico City, Mexico
\item \Idef{org1308}Institut Pluridisciplinaire Hubert Curien (IPHC), Universit\'{e} de Strasbourg, CNRS-IN2P3, Strasbourg, France
\item \Idef{org1182}Joint Institute for Nuclear Research (JINR), Dubna, Russia
\item \Idef{org1199}Kirchhoff-Institut f\"{u}r Physik, Ruprecht-Karls-Universit\"{a}t Heidelberg, Heidelberg, Germany
\item \Idef{org20954}Korea Institute of Science and Technology Information, Daejeon, South Korea
\item \Idef{org1017642}KTO Karatay University, Konya, Turkey
\item \Idef{org1160}Laboratoire de Physique Corpusculaire (LPC), Clermont Universit\'{e}, Universit\'{e} Blaise Pascal, CNRS--IN2P3, Clermont-Ferrand, France
\item \Idef{org1194}Laboratoire de Physique Subatomique et de Cosmologie (LPSC), Universit\'{e} Joseph Fourier, CNRS-IN2P3, Institut Polytechnique de Grenoble, Grenoble, France
\item \Idef{org1187}Laboratori Nazionali di Frascati, INFN, Frascati, Italy
\item \Idef{org1232}Laboratori Nazionali di Legnaro, INFN, Legnaro, Italy
\item \Idef{org1125}Lawrence Berkeley National Laboratory, Berkeley, California, United States
\item \Idef{org1234}Lawrence Livermore National Laboratory, Livermore, California, United States
\item \Idef{org1251}Moscow Engineering Physics Institute, Moscow, Russia
\item \Idef{org1322}National Centre for Nuclear Studies, Warsaw, Poland
\item \Idef{org1140}National Institute for Physics and Nuclear Engineering, Bucharest, Romania
\item \Idef{org1017626}National Institute of Science Education and Research, Bhubaneswar, India
\item \Idef{org1165}Niels Bohr Institute, University of Copenhagen, Copenhagen, Denmark
\item \Idef{org1109}Nikhef, National Institute for Subatomic Physics, Amsterdam, Netherlands
\item \Idef{org1283}Nuclear Physics Institute, Academy of Sciences of the Czech Republic, \v{R}e\v{z} u Prahy, Czech Republic
\item \Idef{org1264}Oak Ridge National Laboratory, Oak Ridge, Tennessee, United States
\item \Idef{org1189}Petersburg Nuclear Physics Institute, Gatchina, Russia
\item \Idef{org1170}Physics Department, Creighton University, Omaha, Nebraska, United States
\item \Idef{org1157}Physics Department, Panjab University, Chandigarh, India
\item \Idef{org1112}Physics Department, University of Athens, Athens, Greece
\item \Idef{org1152}Physics Department, University of Cape Town and  iThemba LABS, National Research Foundation, Somerset West, South Africa
\item \Idef{org1209}Physics Department, University of Jammu, Jammu, India
\item \Idef{org1207}Physics Department, University of Rajasthan, Jaipur, India
\item \Idef{org1200}Physikalisches Institut, Ruprecht-Karls-Universit\"{a}t Heidelberg, Heidelberg, Germany
\item \Idef{org1017688}Politecnico di Torino, Turin, Italy
\item \Idef{org1325}Purdue University, West Lafayette, Indiana, United States
\item \Idef{org1281}Pusan National University, Pusan, South Korea
\item \Idef{org1176}Research Division and ExtreMe Matter Institute EMMI, GSI Helmholtzzentrum f\"ur Schwerionenforschung, Darmstadt, Germany
\item \Idef{org1334}Rudjer Bo\v{s}kovi\'{c} Institute, Zagreb, Croatia
\item \Idef{org1298}Russian Federal Nuclear Center (VNIIEF), Sarov, Russia
\item \Idef{org1252}Russian Research Centre Kurchatov Institute, Moscow, Russia
\item \Idef{org1224}Saha Institute of Nuclear Physics, Kolkata, India
\item \Idef{org1130}School of Physics and Astronomy, University of Birmingham, Birmingham, United Kingdom
\item \Idef{org1338}Secci\'{o}n F\'{\i}sica, Departamento de Ciencias, Pontificia Universidad Cat\'{o}lica del Per\'{u}, Lima, Peru
\item \Idef{org1155}Sezione INFN, Catania, Italy
\item \Idef{org1313}Sezione INFN, Turin, Italy
\item \Idef{org1271}Sezione INFN, Padova, Italy
\item \Idef{org1133}Sezione INFN, Bologna, Italy
\item \Idef{org1146}Sezione INFN, Cagliari, Italy
\item \Idef{org1316}Sezione INFN, Trieste, Italy
\item \Idef{org1115}Sezione INFN, Bari, Italy
\item \Idef{org1286}Sezione INFN, Rome, Italy
\item \Idef{org36377}Nuclear Physics Group, STFC Daresbury Laboratory, Daresbury, United Kingdom
\item \Idef{org1258}SUBATECH, Ecole des Mines de Nantes, Universit\'{e} de Nantes, CNRS-IN2P3, Nantes, France
\item \Idef{org35706}Suranaree University of Technology, Nakhon Ratchasima, Thailand
\item \Idef{org1304}Technical University of Split FESB, Split, Croatia
\item \Idef{org1017659}Technische Universit\"{a}t M\"{u}nchen, Munich, Germany
\item \Idef{org1168}The Henryk Niewodniczanski Institute of Nuclear Physics, Polish Academy of Sciences, Cracow, Poland
\item \Idef{org17361}The University of Texas at Austin, Physics Department, Austin, TX, United States
\item \Idef{org1173}Universidad Aut\'{o}noma de Sinaloa, Culiac\'{a}n, Mexico
\item \Idef{org1296}Universidade de S\~{a}o Paulo (USP), S\~{a}o Paulo, Brazil
\item \Idef{org1149}Universidade Estadual de Campinas (UNICAMP), Campinas, Brazil
\item \Idef{org1239}Universit\'{e} de Lyon, Universit\'{e} Lyon 1, CNRS/IN2P3, IPN-Lyon, Villeurbanne, France
\item \Idef{org1205}University of Houston, Houston, Texas, United States
\item \Idef{org20371}University of Technology and Austrian Academy of Sciences, Vienna, Austria
\item \Idef{org1222}University of Tennessee, Knoxville, Tennessee, United States
\item \Idef{org1310}University of Tokyo, Tokyo, Japan
\item \Idef{org1318}University of Tsukuba, Tsukuba, Japan
\item \Idef{org21360}Eberhard Karls Universit\"{a}t T\"{u}bingen, T\"{u}bingen, Germany
\item \Idef{org1225}Variable Energy Cyclotron Centre, Kolkata, India
\item \Idef{org1017687}Vestfold University College, Tonsberg, Norway
\item \Idef{org1306}V.~Fock Institute for Physics, St. Petersburg State University, St. Petersburg, Russia
\item \Idef{org1323}Warsaw University of Technology, Warsaw, Poland
\item \Idef{org1179}Wayne State University, Detroit, Michigan, United States
\item \Idef{org1143}Wigner Research Centre for Physics, Hungarian Academy of Sciences, Budapest, Hungary
\item \Idef{org1260}Yale University, New Haven, Connecticut, United States
\item \Idef{org15649}Yildiz Technical University, Istanbul, Turkey
\item \Idef{org1301}Yonsei University, Seoul, South Korea
\item \Idef{org1327}Zentrum f\"{u}r Technologietransfer und Telekommunikation (ZTT), Fachhochschule Worms, Worms, Germany
\end{Authlist}
\endgroup

\end{document}